\documentclass{easees2013}
\usepackage{graphicx}
\usepackage[authoryear]{natbib}
\usepackage[fleqn]{amsmath}
\usepackage[]{amsfonts}%
\usepackage{xcolor}
\usepackage{amssymb}
\usepackage{url}
\newcommand{\apj}{ApJ}
\newcommand{\apjs}{ApJS}
\newcommand{\apjl}{ApJ Letters}
\newcommand{\aap}{A\&A}
\newcommand{\aaps}{A\&AS}
\newcommand{\mnras}{MNRAS}
\newcommand{\nat}{Nature}

\newcommand{\araa}{ARA\&A}

\newcommand{\ssr}{Space~Sci.~Rev.} 
\newcommand{\apss}{Ap\&SS}
\newcommand{\zap}{Z. f. Astr.}
\newcommand{\nar}{New Ast. Rev.}

\newcommand{\pasp}{Pub. Ast. Soc. Pacific}

\newcommand{\physrep}{Phys. Reports}
\newcommand{\revmodphys}{Rev. Mod. Physics}

\DeclareGraphicsExtensions{.pdf}

\newcommand{\Myr}{Ma}
\newcommand{\Gyr}{Ga}

\TitreGlobal{The Ages of Stars}
\begin{document}
%%-----------------------------
%%      the top matter
%%-----------------------------
\title{How accurate are stellar ages\\[0.3cm]
 based on stellar models ?\\[0.6cm]
I. The impact of stellar models uncertainties} 

\runningtitle{Y. Lebreton, M.J. Goupil \& J. Montalb\'an: 
 Stellar Models Uncertainties}%

\author{Y. Lebreton}\address{Observatoire de Paris, GEPI, CNRS UMR 8111, 
 5 Place Jules Janssen, 92195 Meudon, 
France, and Institut de Physique de Rennes, Universit\'e de Rennes 1, 
CNRS UMR 6251, 35042 Rennes, France. 
Email: \url{yveline.lebreton@obspm.fr}}
\author{M.J. Goupil}\address{Observatoire de Paris, LESIA, CNRS UMR 8109, 
92195 Meudon, France.\\ Email: \url{MarieJo.Goupil@obspm.fr}}
\author{J. Montalb{\'a}n}\address{Email: \url{j.montalban@skynet.be}}%

\begin{abstract}
Among the various methods used to age-date stars, methods based on stellar model predictions are
widely used, for nearly all kind of stars in large ranges of masses, chemical compositions and evolutionary stages. 
The precision and accuracy on the age determination depend on both the precision and number of observational constraints, 
and on our ability to correctly describe the stellar interior and evolution.
The imperfect input physics of stellar models as well as the uncertainties on the initial chemical composition of stars are responsible
for uncertainties in the age determination.
We present an overview of the calculation of stellar models and discuss the impact on age of their numerous inputs.
\end{abstract}
\maketitle

%#################################################################
\section{Introduction}

The age of stars cannot be obtained from direct measurements, it can only be estimated or inferred.
As reviewed by \citet{2010ara&a..48..581s}, many methods can be applied to age-date stars, depending on the mass and 
evolutionary stage of the star to be dated, and on whether the star is single or belongs to a group. There are three main 
categories of age-dating methods: quasi direct methods, stellar model dependent methods, and empirical methods. All  of them 
require at some level a knowledge of physical processes. These methods are discussed at different places in this
volume. To set the stage, we just briefly outline them in the following. Firstly, quasi-direct methods are based on 
nucleocosmochronometry, they are applied to the Sun (meteorite analysis, lecture by M. Gounelle at this
School, unpublished chapter) and to halo, very 
metal-poor stars (analyses of the lines of long-lived radio nuclides as Th or U, lecture by D. Valls-Gabaud).  Secondly, 
several empirical methods are currently used, as methods based on the decay of activity (measured from Ca~\textsc{ii} H and K, Mg~\textsc{ii}, 
and $\mathrm{H}\alpha$ lines or from the X-ray luminosity), on the decline of surface lithium abundance, or on the 
relation linking the rotation period and age (gyrochronology, see the lecture by R. Jeffries). 

Finally, several methods rely on stellar internal structure models. Single stars can be age-dated either through their placement on model isochrones 
(hereafter isochrone placement method, lecture by D. Valls-Gabaud) 
or through the fitting by stellar models of some particular stellar observable parameters (hereafter 
``\`{a} la carte'' method):
 
\begin{itemize}
\item \textbf{ Placement on model isochrones (or evolutionary tracks)} requires that, at least, indications of the 
star effective temperature $T_\mathrm{eff}$, luminosity $L$, and surface metallicity [Fe/H] are given by observations. This 
may require conversion from the colour-magnitude diagram to the theoretical Hertzsprung-Russell (HR) diagram (\eg, 
the lecture by S. Cassisi).  In this method, the theoretical isochrone that best matches the stellar observables 
provides the star age $A$, mass $M$, and initial metallicity. The method is widely used to age-date 
large samples of stars, either single or in clusters. For instance, it is applied to A-F stars of masses in the 
range $\approx 1.4- 2.5\ M_\odot$ in the Galactic discs, for Galactic evolution or populations studies and, to K-G 
metal-poor low-mass stars, in the halo or thick disc, for Galactic studies and cosmological applications. When stars belong 
to well-defined groups as open clusters, they can be assumed to be coeval and of the same initial chemical 
composition. In that case, particular features on the isochrones, sensitive to age, like the turn-off (TO) luminosity, are powerful 
tools to age-date the group. 
\item \textbf{\textsl{\`{A} la carte models}} are specific stellar models calculated to 
adjust the observational constraints of a given star, as the oscillation frequencies, interferometric radius, or, mass or radius if the star 
belongs to a binary system. \`{A} la carte models have to be calculated when precise ages are required, for instance to 
constrain the physical state of exoplanets or to better understand physical processes at work in stellar interiors (see Part 2 of these lectures in the chapter on \textsl{The impact of asteroseismology}).
\end{itemize}

In both methods, the accuracy on the inferred ages is impacted by the stellar model calculation 
procedure, in particular by the physical inputs or chemical composition of the models. Furthermore, 
empirical age-dating methods are also affected by stellar model uncertainties since they require calibrations, based on a physical 
knowledge of either stellar atmospheres, or stellar interiors and evolution. 
Other age-dating methods are lithium depletion boundary  which is almost model-independent, 
but only applicable to stars in coeval groups
(see the lecture by R. Jeffries), and the use of white dwarfs cooling sequences, a model-dependent method applicable to old stars (lecture by T. von Hippel).

In the present lectures, we focus on the precision and accuracy of age-dating based on the modeling of stellar interiors and evolution. 
More precisely, here in Lecture 1, we examine the main current uncertainties in the stellar model calculations 
and their impact on the age-dating process, while in lecture 2 we discuss the considerable improvement in age accuracy that results from asteroseismic analysis. 
The present Lecture 1 focuses on stars of masses in the range $0.6-40~M_\odot$, 
of both Population I and II, mainly on the main-sequence (MS). 
We evaluate the impact on age of the main inputs of stellar models (chemical composition, 
energy production, transport of energy and/or chemicals), 
focusing on the processes that have the most significant impact.
In Section~\ref{survey}, we recall some basics of stellar modeling and stellar age-dating.
Section~\ref{microphysics} discusses the impact of the chemical composition and of the microscopic input physics 
on the age of stellar models, while Section~\ref{macrophysics} examines the impact of the macrophysics.
In Section~\ref{conclusions} we provide a synthesis of the weights of the different 
stellar model uncertainties on the error on age.
%#################################################################
\section{Stellar evolution, a brief survey}
\label{survey}

We briefly recall here some basics of stellar structure that will be used in the following. More details can be found in textbooks 
\citep[\eg,][and references therein]{cg68,Maeder09,Kippenhahn13}.

\subsection{Equations of stellar structure and evolution}
\label{equations}

In the general case, the structure of a star can be described with the classical equations of hydrodynamics,
\begin{equation}
\frac{\partial \rho}{\partial t}+\overrightarrow{\nabla}.(\rho \vec{u})\; = \; 0, \hfill \mathrm{continuity}
\end{equation}
\begin{equation}
\rho\left (\frac{\partial }{\partial t}+\vec{u}.\overrightarrow{\nabla}  \right )\vec{u} \; = \; \rho\vec{f}-\overrightarrow{\nabla}P-
\rho\overrightarrow{\nabla}\phi+ \mathrm{div}\mathfrak{S}, \hfill \mathrm{momentum}
\end{equation}
\noindent with
\begin{equation}
\Delta \phi \;  = \; {\nabla}^2 \phi = 4\pi G \rho, \hfill \mathrm{Poisson's\ equation}
\end{equation}
\begin{equation}
\rho T\left (\frac{\partial }{\partial t}+\vec{u}.\overrightarrow{\nabla}  \right ){S}  \; =  \; \rho (\epsilon_\mathrm{nuc}+
\epsilon_\mathrm{visc}) - \overrightarrow{\nabla}.\overrightarrow{F_\mathrm{R}},
 \hfill  \mathrm{energy\ conservation}
\end{equation}
\noindent where $\rho$ is the density, $P$ the pressure, $T$ the temperature, $\vec{u}$ the velocity of the flow, 
$\vec{f}$ the external forces, $\phi$ the gravitational potential, $\mathfrak{S}$ the viscous stress tensor,  
$\overrightarrow{F_\mathrm{R}}$ the heat flux, $S$ the entropy, and $\epsilon_\mathrm{nuc, visc}$ 
the energy produced or lost by nuclear reactions, neutrino loss, viscous heat generation, etc.
Each quantity depends on position in the star and time.

Standard assumptions consist in neglecting the external forces, rotation, magnetic fields, 
the dissipation, and shear instabilities. Even though, solving the 3-D system of the equations of stellar structure and evolution is of a great numerical complexity. 
Therefore, to widely investigate the characteristics of stellar structure and evolution, it is usually assumed that a star, 
during most stages of its evolution, can be treated as a spherically symmetric system, in hydrostatic equilibrium. 
Mass loss is however included in massive stars under a simplified form.
The problem then becomes a 1-D problem; the models are called \textsl{standard} stellar models.
%Furthermore, so-called 1-D \textsl{standard} stellar models also neglect rotation and magnetic fields 

In Lagrangian form, the previous equations are simplified into,
\begin{equation}
\label{mass}
\frac{\partial r}{ \partial m} \; = \; -\frac{1}{4\pi r^2 \boldsymbol{\rho}},\hfill \mathrm{mass\ conservation}
\end{equation}

\begin{equation}
\label{pressure}
\frac{\partial P}{ \partial m} \; = \; -\frac{G m}{4\pi r^4} + \frac{\boldsymbol{\Omega}^2}{ 6 \pi r}, \hfill \mathrm{hydrostatic\ equilibrium}
\end{equation}

\begin{equation}
\label{luminosity}
\frac{\partial L}{ \partial m} \; = \; \boldsymbol{\epsilon_\mathrm{nuc}}-\boldsymbol{\epsilon_\mathrm{grav}}-
\frac{\partial \boldsymbol{U}}{\partial t}+\frac{P}{\boldsymbol{\rho}^2}\frac{\partial \boldsymbol{\rho}}{\partial t},\hfill \mathrm{energy\ conservation}
\end{equation}

\begin{equation}
\label{temperature}
\frac{\partial T}{ \partial m} \; = \; -\frac{G m T}{4\pi r^4 P} \boldsymbol{\nabla} \ \mathrm{with}\ \nabla=
\frac{\mathrm{d} \ln T}{\mathrm{d} \ln P},\hfill \mathrm{energy\ transport}
\end{equation}
where $r$ is the radius of a sphere inside the star and $m$ the mass inside that sphere, 
$L$ is the net luminosity escaping the sphere, $U$ is the internal energy, and 
$\Omega$ is the angular velocity (the related term in Eq.~\ref{pressure} disappears
if rotation is neglected). In these equations as well as in the equations below, the terms in bold require the description of 
physical processes.
Energy transport proceeds either by radiation, convection, or conduction, with for the radiative temperature gradient, 
\begin{equation}
\label{radiation}
\nabla_\mathrm{rad} \; = \; \frac{3}{16\pi acG}\frac{\boldsymbol{\kappa} P}{T^4}\frac{L}{m},
\end{equation}
where $\kappa$ is the mean Rosseland opacity (see Sect.~\ref{opa}). The convective $\nabla_\mathrm{ad}$ 
and conductive $\nabla_\mathrm{cond}$ gradients are discussed in the
following sections.

The temporal evolution of the star is followed by resolving the following equation, written 
for each considered chemical species $i$, of mass fraction $X_{i}$,
\begin{equation}
\left(\frac{\partial X_{i}}{\partial t}\right) \; = \; \left( \frac{\partial X_{i}}{\partial t}\right)_\mathrm{nuc}+
\left(\frac{\partial X_{i}}{\partial t}\right)_\mathrm{transport},
\end{equation}
with, for nuclear evolution,
\begin{equation}
\label{dXdtnuc}
%\left(\frac{\partial X_{i}}{\partial t}\right)_\mathrm{nuc}=\frac{m_{i}}{\rho} \left( \sum_j \boldsymbol{r_{ji}} - \sum_k \boldsymbol{r_{ik}} \right)
\left(\frac{\partial X_{i}}{\partial t}\right)_\mathrm{nuc} \; = \; {\rho A_{i}} \left( \sum_{\mathrm{jk}} \boldsymbol{r^{i}_{jk}} - \boldsymbol{r^{k}_{ij}} \right),
\end{equation}
where $r^{i}_{jk}$ is the reaction rate for a reaction creating the species $i$ from species $j$ and $k$, and  $r^{k}_{ij}$ for reactions destroying the species $i$, 
and $A_i$ is the mass number of species $i$.
For transport processes (convection, diffusion), the chemical evolution equations read,
\begin{equation}
\label{dXidtconv}
\left(\frac{\partial X_{i}}{\partial t}\right)_\mathrm{conv, diff} \; = \; 
\frac{\partial}{\partial m}\left( 4\pi r^2 \rho \boldsymbol{V_{i}} X_{i}\right) + \frac{\partial}{\partial m} 
\left[ \left(  4\pi r^2 \rho\right)^2 \boldsymbol{D} \frac{\partial X_{i}}{\partial m}\right],
\end{equation}
where $V_{i}$ is the diffusion velocity of species $i$ and $D$ the diffusion coefficient whatever the diffusion process is, 
\ie, turbulent and/or diffusive. Some complementary equations describing the transport of angular momentum 
inside the star must be added if rotation is taken into account. This is described later in the lecture.

The resolution of the equations provides values of $m$, $P$, $L$, $T$, and $X_{i}$ throughout the star.
 However, this requires a description of physical processes at work inside the star. Microscopic
  processes (opacities, equation of state, nuclear reactions, neutrino losses, microscopic diffusion) and macroscopic
  processes (mass loss, convective transport,  overshooting and semi-convection, thermohaline
   convection, transport induced by rotation, magnetic field and internal waves,  etc.) intervene in the
   evaluation of the quantities appearing in bold in the equations. Any inadequate  description of these
   processes may contribute to the age uncertainty, at least to some extent that we attempt to quantify in this lecture.

Boundary conditions for the stellar structure equations are to be given in the centre and at the surface. 
In the centre, $m=0$, and $r=0$, $L=0$. At the surface defined at some place where 
$m=M_\mathrm{star}-m_\mathrm{atmosphere}$, the junction has to be made with a model atmosphere calculated 
independently. The model atmosphere provides the stellar model total radius $R_\mathrm{star}$, luminosity $L_\mathrm{star}$ as 
well as the surface pressure $P_S$ and temperature $T_S$. Model atmospheres are discussed in the dedicated  
lecture by F. Martins.

To calculate a stellar evolutionary sequence, one has to provide the initial mass and chemical composition of the star. The 
time starting point can be either the zero-age main-sequence (ZAMS) where the initial model is a homogeneous model starting 
H-burning in the centre, or the pre main-sequence (PMS) where the initial model is a homogeneous, fully convective star, in 
quasi-static contraction \citep{1965ApJ...141..993I}. More sophisticated initial conditions have been explored, where the 
starting point is the birth line and the initial stellar model results from a calculation taking into account the accretion of gas 
onto the star \citep[\eg,][]{1990ApJ...360L..47P}. Furthermore, when rotation is accounted for, one has to provide an initial rotation 
profile, usually assumed to correspond to a solid body rotation.

%-----------------------------------------------
\begin{figure}[!hpt]
\begin{center}
\includegraphics[width=\textwidth]{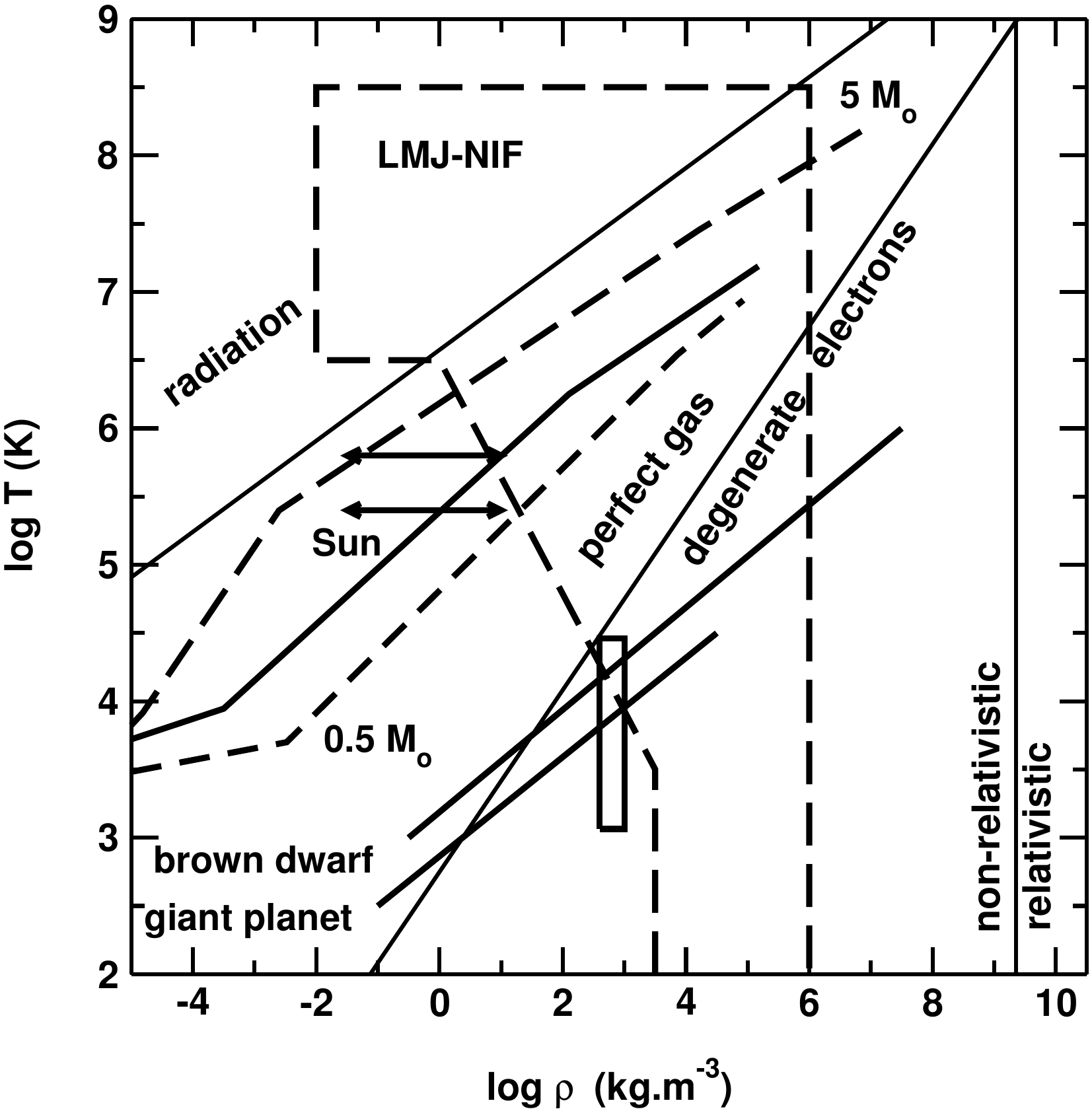}
%\resizebox{0.5\hsize}{!}{\includegraphics{Figs/centre_logrho_logT}}
\caption{Stars in the density-temperature plane. [From \citet{2005ESASP.576..493L}.] 
Interior profiles of stellar models 
of masses $0.5, 1.0$ and $5~M_\odot$ on the MS as well as a brown dwarf and a giant planet profile are shown, 
from the surface (low $T$, low $\rho$) 
to the centre. The processes governing the equation of state are indicated.
The arrows delimit
the region of stellar envelopes where the iron opacities
have been derived from experiments with high-power
lasers, the rectangle corresponds to the region of the interiors
of brown dwarfs and giant planets where the equation
of state of dense matter has been studied with high
pressure experiments using intense lasers, and the large
region delimited by dashed lines is the region 
accessible to the next generation of intense lasers like the
LMJ or the NIF.
}
\label{logrho_logT}
\end{center}
\end{figure}
%-----------------------------------------------
%-----------------------------------------------
\begin{figure}[!hpt]
\begin{center}
\includegraphics[width=\hsize]{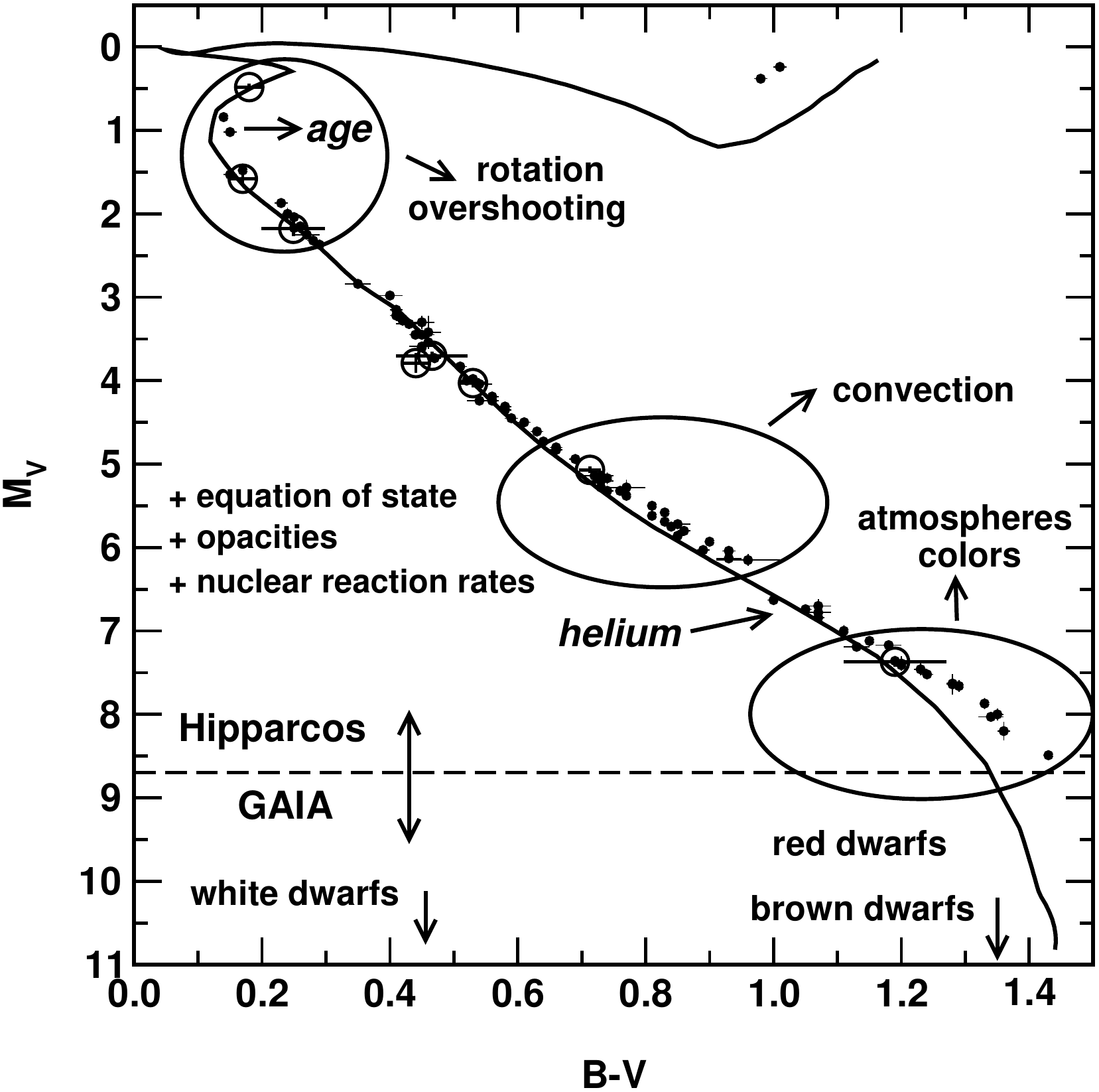}
\caption{The Hyades color-magnitude diagram. [From \citet{2005ESASP.576..493L}.]  Observational
data from Hipparcos \citep{2001A&A...367..111D} are
compared to a model isochrone of $625$~\Myr~\citep{2001A&A...374..540L}.
Different regions are indicated where the impact of the
various physical inputs is crucial. The magnitude limit
of Hipparcos - which will be pushed up by Gaia - is
indicated by the horizontal dashed line.}
\label{Hyades}
\end{center}
\end{figure}
%-----------------------------------------------
\subsection{A diversity of physical processes can affect age-dating}

As mentioned in Sect.~\ref{equations}, a stellar plasma is characterized by its density, temperature, and the individual 
abundances of chemical elements. In Fig.~\ref{logrho_logT}, 
%left panel, 
we show the internal $\rho-\mathrm{T}$ profiles 
--from surface to centre--  
of stars of different masses ($0.5$, $1$, and $5~M_\odot$)
on the MS, as well as those of a brown dwarf and a giant gaseous planet, 
together with the zones indicating the regimes of the equation of state  (see also Sect.~\ref{EoS}).
At a given evolutionary stage, stars of different masses are found in different locations in the $\rho-\mathrm{T}$ plane.
This location changes when evolution proceeds on the MS and beyond.
The physical processes at work in the interior vary from the centre to the surface and change
with the mass and evolution of the star.
As discussed later, those physical processes are sometimes not well understood or their description 
is affected by uncertainties. Since the speed at which a star evolves depends on many physical processes,
the age-dating process is complex and merely uncertain.

In Fig.~\ref{Hyades}, we show the observed position of the best-known stars in the nearer open cluster, 
the Hyades, located at $46$ pc, together with a model isochrone that best fits these observations, at the metallicity of the 
cluster stars ([Fe/H]=0.14 dex). An age of 625~\Myr~is inferred from stellar modelling \citep{1998A&A...331...81P, 2001A&A...374..540L}. 
All along the isochrone, the variety of processes that dominate the uncertainty of the modelling are indicated. 

One important and thorny point comes from the fact that stars of masses higher than $\approx 1.2~M_\odot$ 
develop convective cores that mix material during the MS. 
The heavier the convective core, the larger amount of hydrogen fuel available and therefore the longer the MS lifetime. 
As discussed in the following sections (mainly in Sect.~\ref{macrophysics}), the determination of the convective core extent 
(and of the possible extension of mixing beyond this core by overshooting or rotationally-induced mixing) 
is a caveat that heavily impacts stellar age-dating.
 
\subsection{Time-scales}

\begin{figure}[!htbp]
\begin{center}
\includegraphics[width=\textwidth]{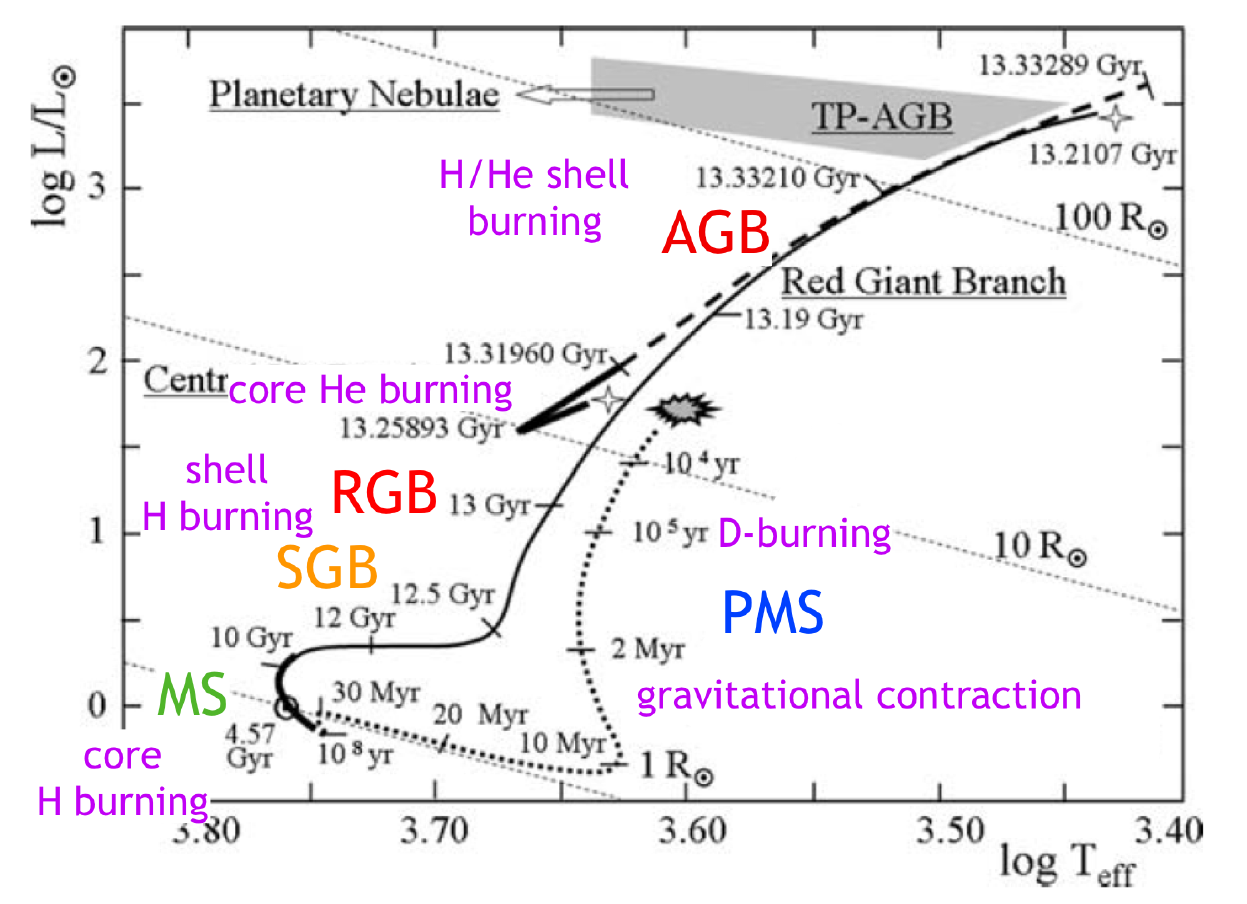}
\caption{The evolution of a star of $1~M_\odot$ from the 
PMS to the tip of the AGB.
[Adapted from Fig. 25.11 by \citet{Maeder09}.]}
\label{stellar_evol1}
\end{center}
\end{figure}

\begin{figure}[!htbp]
\begin{center}
\includegraphics[width=\textwidth]{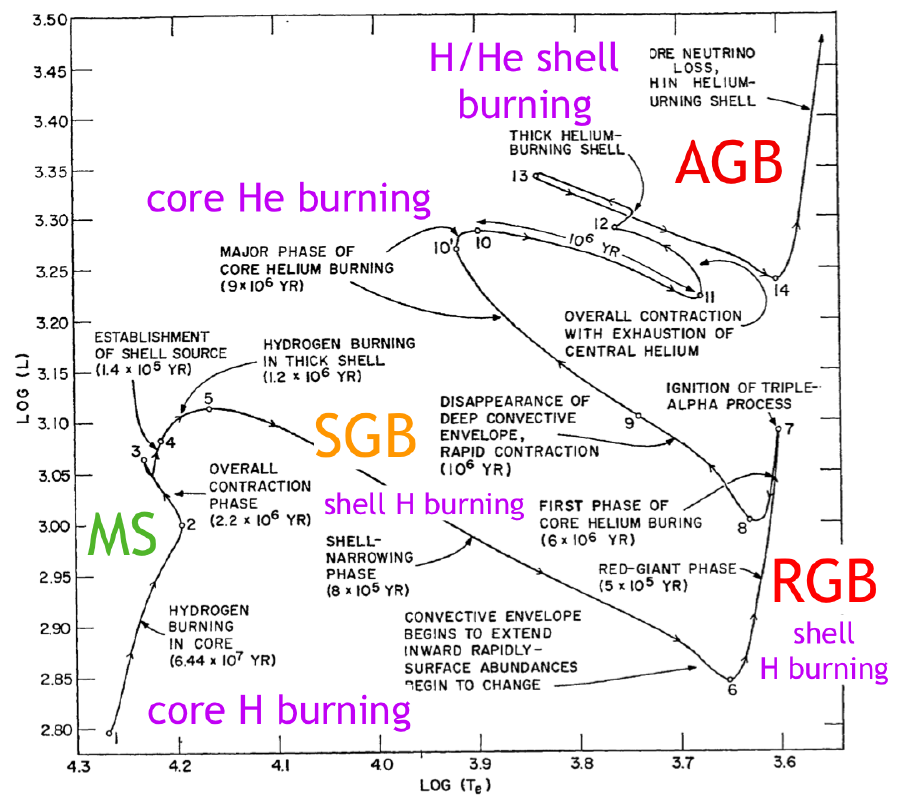}
\caption{The evolution of a star of $5~M_\odot$ from the MS to  AGB.
[Adapted from \citet{1967ARA&A...5..571I}.]}
\label{stellar_evol5}
\end{center}
\end{figure}

In Fig.~\ref{stellar_evol1} and \ref{stellar_evol5}, 
the different stages of the evolution of a $1~M_\odot$ and $5~M_\odot$ star, 
from the PMS to the asymptotic giant branch (AGB), 
are shown together with the corresponding time-scales. 
Different physical processes occur during each phase of evolution which result in different time-scales. 
As a consequence, in the age-dating process, 
the value of the age of the star and its uncertainty depend on mass, 
evolutionary stage, and chemical composition.

%-----------------------------------------------
%-----------------------------------------------
%-----------------------------------------------
%-----------------------------------------------
During the PMS, gravitational contraction is the dominant energy production process. In standard stellar models without accretion, the PMS phase  proceeds on a Kelvin-Helmholtz time scale, $t_\mathrm{KH}$: 
\begin{equation}
t_\mathrm{KH} \; \approx \; \frac{E_\mathrm{int}}{\langle L\rangle} \approx \frac{G M^2}{R \, \langle L\rangle } \, \approx \, t_\mathrm{KH, \odot} \left(\frac{M}{M_\odot}\right)^2 \left(\frac{R}{R_\odot}\right)^{-1} \left(\frac{L}{L_\odot}\right)^{-1},
\end{equation}
%which scales as
%\begin{equation}
%t_\mathrm{KH}\approx t_\mathrm{KH, \odot} \left(\frac{M}{M_\odot}\right)^2 \left(\frac{R}{R_\odot}\right)^{-1} \left(\frac{L}{L_\odot}\right)^{-1}
%\end{equation}
where  $t_\mathrm{KH, \odot}\approx 3.1\times 10^7$ years is the solar value.
However, it has been shown that if accretion of material on to the star is considered during the stellar 
formation and PMS phases, as seen in observations, 
the  time scales are modified \citep[\eg,][and references therein]{2000A&A...359.1025N}. 
Also, the morphology of the evolutionary tracks in the HR diagram in the PMS phase is modified
 when accretion is accounted for.
 As shown in Table~\ref{pmstime}, accretion reduces the duration of the PMS by a factor of three at $3~M_\odot$. 
However, the ratio $t_\mathrm{PMS}/t_\mathrm{MS}$ of the PMS  to the MS 
   lifetime is in the range $0.004-0.02$, which is very short. For evolved stars, the age uncertainties prior to the MS are therefore  negligible in the error budget. For this reason, in the following, we do not consider the PMS
     phase.
   
%-----------------------------------------------
\begin{table}
\setlength{\abovecaptionskip}{0pt}
\setlength{\belowcaptionskip}{10pt}
\begin{center}
\caption{Time scales of star formation (collapse phase) and of PMS, 
after \citet{Maeder09}, as estimated from models including accretion.}
\label{pmstime}
\begin{tabular}{ccccc}
\hline\hline
Final mass & $t_\mathrm{formation}$  & $t_\mathrm{PMS}$  & $t_\mathrm{formation} / t_\mathrm{KH}$ & $t_\mathrm{PMS} / t_\mathrm{KH}$ \\
$(M_\odot)$ &(a) & (a) & -& -\\
\hline
$0.8$ & $7.15\times 10^7$ & $7.15\times 10^7$ & $1.05$ & $1.05$ \\
$1.0$ & $3.82\times 10^7$ & $3.81\times 10^7$ & $0.98$ & $0.98$ \\
$1.5$ & $3.10\times 10^7$ & $3.08\times 10^7$ & $0.87$ & $0.87$ \\
$2.0$ & $1.17\times 10^7$ & $1.15\times 10^7$ & $0.50$ & $0.49$ \\
$3.0$ & $2.68\times 10^6$ & $2.42\times 10^6$ & $0.37$ & $0.34$ \\
$5.0$ & $0.80\times 10^6$ & $0.41\times 10^6$ & $0.69$ & $0.36$ \\
\hline
\end{tabular}
\end{center}
\end{table}
%-----------------------------------------------

%-----------------------------------------------
\begin{figure}[!htbp]
\begin{center}
%\resizebox{0.65\hsize}{!}{
\includegraphics[width=\textwidth]{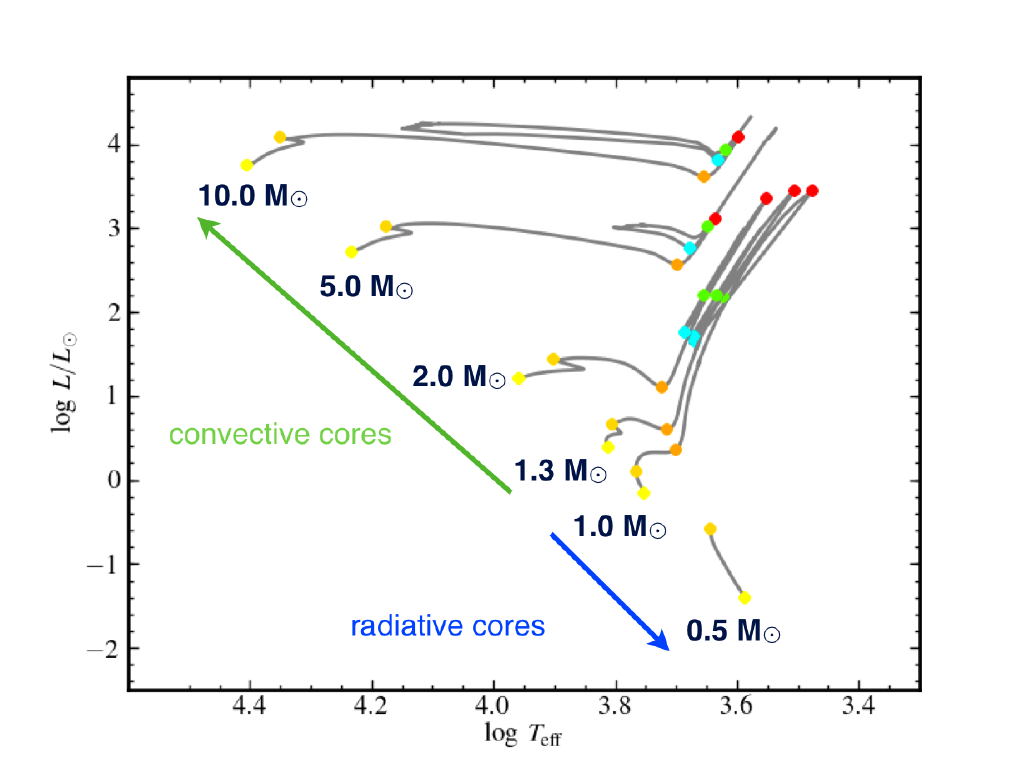}%}
\caption{Evolutionary tracks in the HR diagram generated from \textsl{BaSTi} grids at solar metallicity, and for a solar calibrated 
mixing-length parameter \citep{2004apj...612..168p}. The phases listed in Table~\ref{evoltime} are pinpointed. The 
MS (H-burning) phase is in-between the two consecutive yellow points,  the subgiant branch (SGB, H-shell burning) phase is
between the second (yellow) point and the third point (orange), 
the red giant branch  (RGB, H-shell burning) phase is between the orange and the red points, 
and core He burning occurs beyond the red point.}
\label{tracks}
\end{center}
\end{figure}
%-----------------------------------------------

The different phases of evolution on the MS and beyond occur either on a nuclear time scale $t_\mathrm{nuc}$ or on $t_\mathrm{KH}$. The nuclear time scale reads
\begin{equation}
t_\mathrm{nuc} \, \approx \, \frac{\mathrm{available\ fuel}}{\mathrm{power}}  \, \approx  \, \frac{M_\mathrm{fuel} \, c^2}{\langle L\rangle} \, ,
  \hspace{0.3cm} \mathrm{and\ for\ the\ MS\ phase:} \hspace{0.3cm} 
t_\mathrm{nuc} \; \propto  \;\frac{M_{\mathrm{core}} \, X_\mathrm{H} \, c^2}{ L},
\label{tnucMS}
\end{equation}
where $M_\mathrm{core} X_\mathrm{H}$ is the total mass amount of hydrogen burned during the MS.

Table~\ref{evoltime} provides a summary of the time elapsed in the different phases of the evolution of stars of 
different masses and initial chemical compositions. 
The evolution phases from the MS to core He burning are pinpointed in Fig.~\ref{tracks}. 

%----------------------------------------------------------------------------------
\begin{table}
\setlength{\abovecaptionskip}{0pt}
\setlength{\belowcaptionskip}{10pt}
\begin{center}
\caption{Time scales of the different stages of evolution for Pop I stars and Pop II metal deficient stars of different masses, after \citet{Stein66}.}
\label{evoltime}
{\small 
\begin{tabular}{ccccc}
\hline\hline
$M$ & PMS  & core H-fusion  & shell H-fusion & core He-fusion \\
$(M_\odot)$ &(a) & (a) & (a)&  (a)\\ 
& Pop I/II & Pop I/II &Pop I/II &Pop I/II \\
\hline
$0.7$ & $2.\times 10^8/2.\times 10^8$ & $4.\times 10^{10}/5.\times 10^{10}$ & $ 1.\times 10^7/2.\times 10^7$ & $6.\times 10^7/4.\times 10^7$ \\
$1.0$ & $4.\times 10^7/4.\times 10^7$ & $8.\times 10^{9}/9.\times 10^{9}$ & $6.\times 10^6/9.\times 10^6$ & $3.\times 10^7/2.\times 10^7$ \\

$2.0$ & $3.\times 10^6/3.\times 10^6$ & $5.\times 10^{8}/7.\times 10^{8}$ & $2.\times 10^6/3.\times 10^6$ & $1.\times 10^7/9.\times 10^6$ \\

$5.0$ & $3.\times 10^5/1.\times 10^6$ & $3.\times 10^{7}/8.\times 10^{7}$ & $3.\times 10^5/4.\times 10^5$ & $2.\times 10^7/2.\times 10^7$ \\

$7.0$ & $2.\times 10^5/7.\times 10^5$ & $1.\times 10^{7}/4.\times 10^{7}$ & $2.\times 10^5/1.\times 10^5$ & $8.\times 10^6/8.\times 10^6$ \\

$10.0$ & $ 1.\times 10^5/3.\times 10^5$ & $7.\times 10^{6}/2.\times 10^{7}$ & $7.\times 10^4/4.\times 10^4$ & $3.\times 10^6/3.\times 10^6$ \\

$15.6$ & $6.\times 10^4/2.\times 10^5$ & $3.\times 10^{6}/1.\times 10^{7}$ & $2.\times 10^4/2.\times 10^4$ & $1.\times 10^6/1.\times 10^6$ \\
\hline
\end{tabular}
}
\end{center}
\end{table}
%----------------------------------------------------------------------------------

\subsection{Isochrone placement and main sequence turn-off}

\subsubsection{Evolutionary tracks and isochrones}

To age-date large ensembles of stars, grids of stellar evolutionary tracks are calculated for a given range of mass, metallicities, and evolutionary stages.
 Furthermore, these grids are based on a given set of input physics and parameters. 
 Input parameters (\eg, initial helium abundance or $\Delta Y/\Delta Z$, mixing-length parameter for convection and overshooting parameter, etc.) 
 are discussed later in the lecture. We recall that, along an evolutionary track, the age varies 
 (the initial mass and composition are fixed,  but their actual value can change due to mass loss and/or diffusion and mixing processes inside the star). 
 From grids of evolutionary tracks, grids of isochrones (fixed age and initial chemical 
 composition, increasing mass), are then built by interpolation.

In the age-dating process, a star with given observed values of $L$, $T_{\mathrm{eff}}$, and surface
 metallicity is placed in the HR 
diagram and its age, mass, and initial chemical composition are inferred by inversion in the isochrone grids.
As discussed in, \eg~\citet{2004mnras.351..487p} and \citet{2005a&a...436..127j}, 
such inversion does not provide precise ages in some regions of the 
HR diagram, either because isochrones are very  close to each other and cannot be disentangled (case of low mass stars, not 
evolved and close to the ZAMS), or, because of the complex morphology of isochrones, several evolutionary stages can be 
assigned to the same star (case of the MS turn-off region, RGB, and He-burning regions). Bayesian inversion, considering 
priors, like the initial mass function (IMF) of the stellar sample,  has been shown to improve the age-dating results in the 
degeneracy regions. Nevertheless, some problems may remain as discussed by
\citet[][]{2004mnras.351..487p, 2005a&a...436..127j}, and in the lectures of D. Valls-Gabaud and T. von Hippel
in the present volume. Note that the \texttt{PARAM} 
Web tool \citep{2002a&a...391..195g,2006a&a...458..609d} allows to determine the age of a given star
with this technique.

%-----------------------------------------------
\begin{figure}[!htbp]
\begin{center}
\includegraphics[width=0.9\textwidth]{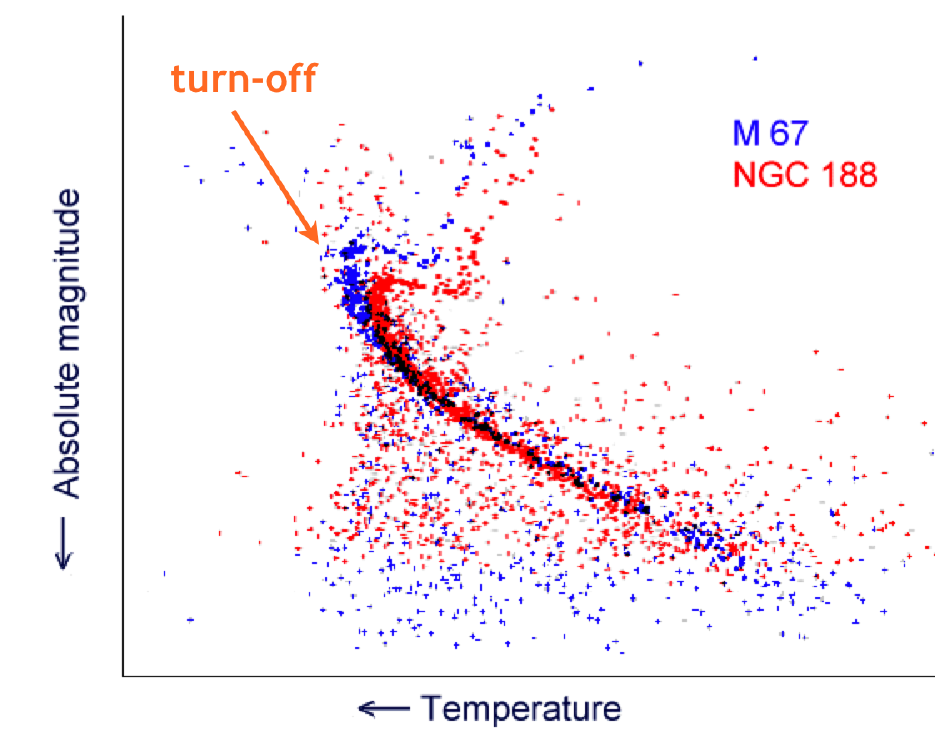}
\caption{Position in the HR diagram of stars members of two open clusters and their turn-off position. The M67 cluster is 
younger than NGC188, its turn-off is bluer and brighter.
[Adapted from Wikipedia].}
\label{M67NGC188}
\end{center}
\end{figure}
%-----------------------------------------------

\subsubsection{Cluster main-sequence turn-off}

Stellar clusters (open clusters) are very interesting case-studies since they are constituted of stars that can reasonably be assumed to originate 
from the same molecular cloud and therefore to have same age and initial chemical composition, but different masses.
Stars members of open clusters therefore draw an isochrone in the HR diagram (Fig.~\ref{M67NGC188}). 
A particular feature of cluster isochrones is the turn-off point 
which marks the end of the main-sequence. %(see the turn-off point of the Hyades cluster in Fig.~\ref{Hyades}). 
As illustrated in Fig.~\ref{M67NGC188}, the younger the cluster, the brighter and bluer its turn-off point.
The luminosity at turn-off is a robust age indicator, as explained below.
The case of globular clusters is more complicated since it is now accepted that they are
multi-population structures \citep[\eg,][]{2009IAUS..258..233P}.

\subsubsection{Theoretical relation between the turn-off luminosity and turn-off age}
\label{grids}
%\subsubsection{Stellar models to examine the dependencies of the turn-off luminosity}

In order to evaluate the impact of the parameters of stellar models on age-dating based on the TO luminosity, we have calculated several grids of stellar models of masses 0.6, 0.7, 0.8, 0.9, 1.0, 1.1, 1.2, 1.3, 1.4, 1.5, 1.75, 2.0, 2.5, 3.0, 
4.0, 5.0, 7.5, 10., 20., 30., 40. $M_\odot$, and evolutionary stages covering the evolution from the
 ZAMS to the beginning of the SGB. Each grid corresponds to a given set of model parameters or input physics, as
 is described later.  
 We have used the \texttt{cesam2k} code \citep{2008ap&ss.316...61m}.
  The reference grid corresponds to models calculated with the input physics listed below.
\begin{itemize}
 \item {Opacities:}\ \textsl{OPAL96} opacities \citep{1996apj...464..943i} complemented at low temperatures by \textsl{WICHITA} tables \citep{2005apj...623..585f}. 
    \item {Equation of state:}\ \textsl{OPAL05}  \citep[][]{2002apj...576.1064r}.
    \item {Nuclear reaction rates:}\ \textsl{NACRE} data
    \citep[][]{1999nupha.656....3a} except for the $^{14}N(p,\gamma)^{15}O$ reaction where 
    we adopted the revised \textsl{LUNA} rate \citep{2004phlb..591...61f}.
   \item {Convection:} \textsl{CGM} convection theory of \citet{1996apj...473..550c} with 
   a solar mixing-length parameter {$\alpha_\mathrm{conv}=\ell/H_P=0.688$} ($\ell$ is the mixing-length 
   and $H_P$ the pressure scale height) resulting from the
    calibration of the radius and luminosity of a solar model with the same input physics 
    \citep[see \eg,][]{2008ap&ss.316...61m}.  
   \item {Atmospheric boundary condition:} grey model atmospheres with the classical Eddington T-$\tau$ law.
  \item {Solar mixture:}  
  \textsl{GN93} mixture  \citep{1993pavc.conf..205G}, which corresponds to  $(Z/X)_\odot=0.0245$. 
  \item {Stellar chemical composition:} The initial $Z/X$ is solar. The initial helium abundance
  is derived from   $(Y_0 - Y_\mathrm{P})/{(Z- Z_\mathrm{P})}{=}{\Delta Y}/{\Delta Z}$, where $Y_\mathrm{P}$ 
   and $Z_\mathrm{P}$ are the primordial abundances. 
   We adopted $Y_\mathrm{P}{=}0.245$  \citep[\eg,][]{2007apj...666..636p}, $Z_\mathrm{P}{=}0.$ 
   and, ${\Delta Y}/{\Delta Z}{\approx}2$. This latter roughly corresponds to the solar $(\Delta Y/\Delta Z)_\odot$
   obtained from the solar model calibration.
    \item {Microscopic diffusion and convective core overshooting:} are not included in the reference grid.
    \end{itemize}

%-----------------------------------------------
\begin{figure}[!hptb]
\begin{center}
\includegraphics[width=0.895\textwidth]{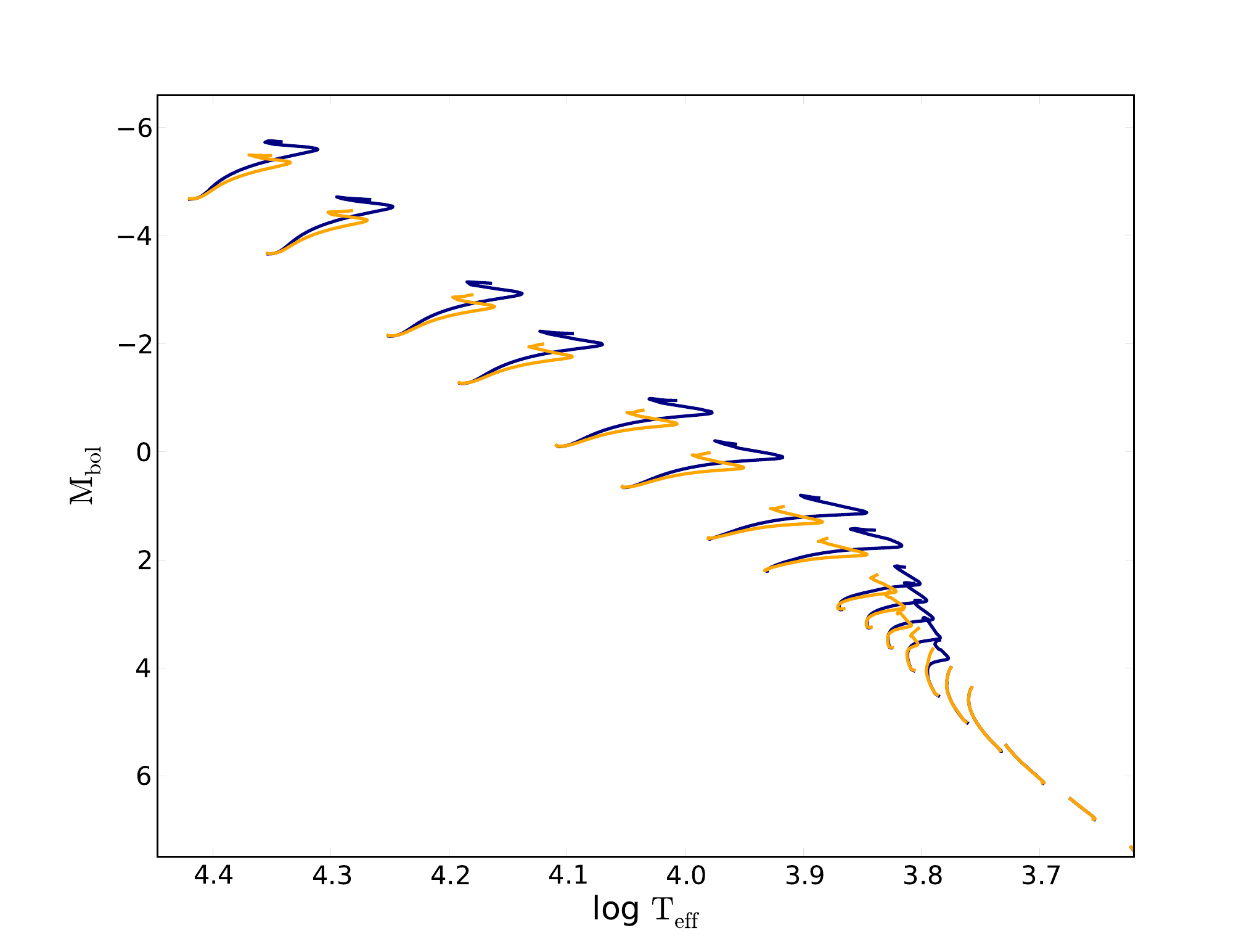}
\includegraphics[width=0.895\textwidth]{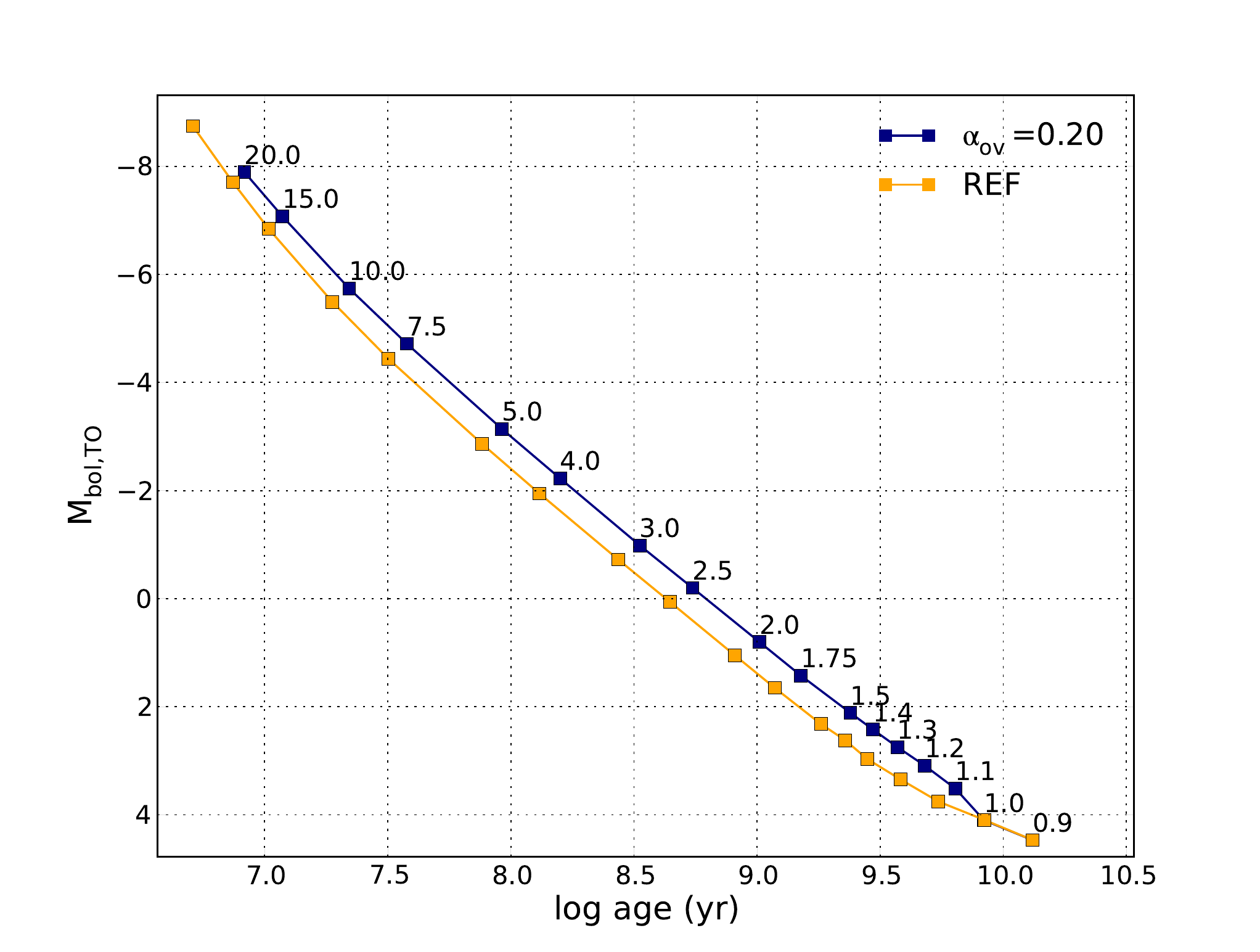}
\caption{\textsl{Top:} evolutionary tracks in the HR diagram for our reference grid (orange) and for the
grid including overshooting in blue (see Sect.~\ref{macrophysics}). Stellar masses are indicated in the right figure. 
\textsl{Bottom:} bolometric magnitude at TO as a function of age for both grids.
}
\label{turnoff}
\end{center}
\end{figure}
%-----------------------------------------------
%\subsubsection{Turn-off luminosity vs. age}

For each mass in each grid, we extracted the value of the luminosity and age at turn-off, which we defined for convenience as the
stage where the central hydrogen abundance drops to $X_c< 10^{-4}$.
In Fig.~\ref{turnoff}, left panel, we plot the evolutionary tracks in the HR diagram. 
In Fig.~\ref{turnoff}, right panel, we plot the bolometric magnitude at turn-off as a function of the TO age, 
at different masses, for the reference grid and for a grid including 
overshooting of the convective core (see Sect.~\ref{macrophysics}).
The figure shows that, \textsl{for fixed input physics and free parameters of stellar models}, 
a precise observational measure of the turn-off luminosity allows to infer the age of a cluster quite precisely: 
for instance, the $\log\mathrm{age}- M_{\mathrm{bol}}$ relation is about linear between $1.2$ and $5~M_\odot$, 
with a slope $\mathrm{d} M_\mathrm{bol}/\mathrm{d(age)}\sim 1.7\ 10^{-3}\ \mathrm{mag} \, \mathrm{Ma}^{-1}$. 
As a result,  an error of $0.01$ mag on $M_\mathrm{bol}$ would imply an error on the age of  $\approx 6$~\Myr. 
However, as discussed in the following, the theoretical TO luminosity is very sensitive to 
imperfections in stellar models as well as to badly known stellar parameters. 
Also, it is important to recall that a precise and accurate determination of the stellar luminosity requires precise 
distances and apparent magnitudes, as well as bolometric corrections. 
While distances and magnitudes will be exquisitely precise when the Gaia mission 
delivers its data \citep{2012mnras.426.2463l}, improved bolometric corrections will require to go on 
progressing on model atmospheres (\textsl{cf.} the lecture by F. Martins).
 
\subsection{Homologous stars}

Homology provides simple scaling relations that help to grasp the internal structure of a star and its sensitivity to parameter 
changes, along its evolution. Homology relations are established and commented with a lot of details in the text book by \citet{cg68}.
We briefly recall some  relations that will be useful in the framework of the present lecture.

Let us consider a star (hereafter star $1$) with total mass $M_0$ and radius $R_0$ divided into concentric 
spherical shells
We denote by $x=r_0/R_0$ the fractional distance to the centre for the shell located at radius $r_0$, and 
by $M_0(x)$ the mass inside the sphere of radius $r_0$. 
A second star (star $2$), with mass $M$ and radius $R$, is said to be homologous to star $1$ if $M(x)/M=M_0(x)/M_0$.

For homologous stars, starting from Eqs.~\ref{mass} and \ref{pressure}, 
it can be shown that 
\begin{equation}
P(x)\propto P_\mathrm{c} \; \propto  \; \frac{M^2}{R^4}\hspace{0.75cm} \mathrm{and }\hspace{0.75cm} T(x)
\propto T_\mathrm{c} \; \propto  \;\mu \frac{M}{R}\hspace{0.5cm} \mathrm{for\ an\ ideal\ gas,}
\end{equation}
where $\mu$ is the mean molecular weight.

To get an expression for the luminosity, one has to make assumptions on the opacity $\kappa$.
Opacity can roughly be approximated by a power law, the Kramers' law that reads,
\begin{equation}
\label{eqhom1}
\kappa \; = \; f(X,Z) \, \rho^n \, T^{-\alpha} \, ,
\end{equation}
where $X$ and $Z$ are the hydrogen and metal mass fractions. 
In low mass stars, one can assume that opacity is roughly dominated by bound-free (bf) and free-free (ff) transitions 
with $n\sim 1$ and $\alpha\sim3.5$, while in high mass stars, opacity is dominated by Thomson electron scattering (es) with $n=\alpha=0$.
This yields
\begin{equation}
\label{kramersthomson}
\kappa\sim \kappa_\mathrm{bf, ff} \; \propto \; f(Z) \, (1+X) \, \rho \, T^{-3.5} \, ; \\
\kappa\sim \kappa_\mathrm{es} \; \simeq \; 0.02 (1+X) \, . \\
\end{equation}

Similarly, the nuclear energy production rate can be approximated by
\begin{equation}
\epsilon \; = \; g(X,Z) \, \rho^{u-1} \, T^{s} \quad .
\end{equation}
For the proton-proton ($p-p$) chain $s\sim 4$ and $u\sim 2$, while for the CNO cycle $s\sim 20$ and $u\sim 2$.

For homologous stars on the MS, from Eq.~\ref{temperature} and \ref{luminosity}, one gets simple scaling relations, expressing the behaviour of the total luminosity.
For instance, for a high mass, fully radiative star, in which one can roughly assume that the opacity is governed 
by electron scattering, one finds,
\begin{equation}
L  \; \sim  \; \frac{\mu^{4}}{\kappa_{0}} M^3 \quad ,
\end{equation}
where no assumption on the mode of energy generation or on thermal equilibrium has to be made.
Conversely, for a low mass, fully radiative star, dominated by Kramers opacity,
\begin{equation}
\label{homol_lum}
L  \; \sim \; \frac{\mu^{7.5}}{\kappa_{0}} \, \frac{M^{5.5}}{R^{0.5}} \quad .
\end{equation} 

In this latter relation, there is a slight dependence on the mode of energy generation inside the star, through the radius dependency.
Using Eqs.~\ref{eqhom1} to \ref{homol_lum}, we obtain
\begin{equation}
R_\mathrm{pp}  \; \propto \;  {\mu^{-0.43}}{M^{0.14}} \quad , 
\mathrm{\, and \quad } R_\mathrm{CNO} \propto {\mu^{0.55}}{M^{0.73}} \quad ,
\end{equation}
for stars in which hydrogen fusion is dominated by the p-p chain and CNO cycle, respectively.

The luminosity mainly depends on how efficiently energy can be transported by radiation. For a star in thermal equilibrium (\eg, on the MS), 
$L_{\mathrm{nuc}}$ and therefore $ T_{\mathrm{c}} $ adapt themselves to the surface luminosity.
In some cases, the dependence of luminosity on stellar models input parameters can be understood by homology relations. 
The impact on age can then be deduced via the nuclear time-scale (see Eq.~\ref{tnucMS}).
%#################################################################
\section{Impact of chemical composition and microphysics uncertainties on stellar ages}
\label{microphysics}

In this section we examine the impact on age-dating of the chemical composition and of microscopic input physics entering stellar models.

\subsection{Chemical composition}

The initial chemical composition is, after the initial mass, the second main input of stellar models. 
It is usually expressed in mass fraction.
The abundances in mass fraction of H, He, and metals (\ie, of all elements heavier than helium) 
are denoted respectively by $X$, $Y$, and $Z$. The abundance of a given element ${i}$ is denoted by $X_{i}$. 
Since observations generally provide abundances relative to hydrogen, 
very often, the global abundance of metals is expressed by the ratio $Z/X$ (see below).

All elements do not intervene at the same level in stellar model calculation.
On one hand, some elements directly enter the calculation of the stellar structure via the physical 
processes (nuclear reaction rates, opacity, equation of state, diffusion, etc.), and thus, their abundances impact the structure of the star.
For instance,
\begin{itemize}
\item the nuclear reaction rates are dominated by H (on the MS), then by He, C, O, etc.,
\item the mean Rosseland opacity is governed by some leading elements, mainly H, He, Fe, O, Ne, etc.,
\item the equation of state requires the global amount of metals $Z$ that intervenes in the pressure calculation, 
as well as the individual abundances $X_{i}$ that intervene in the calculation of ionization equilibria,
\item the microscopic diffusion --and more generally transport processes-- concern all the elements.
\end{itemize}
On the other hand, some elements are tracers of transport processes and their abundances do not impact much the star structure 
(for instance $\ ^{6, 7}\mathrm{Li}$, $\ ^{9}\mathrm{Be}$, $\ ^{13}\mathrm{C}$, etc.).
The nature of leading elements, for a given process, depends on the physical conditions inside the star ($T$, $P$), and therefore on the mass and evolution state.

\subsubsection{Heavy elements}

Observations provide present surface abundances, not initial abundances nor inner abundances. 
In the case of the Sun, observations in the photosphere, meteorites, and interstellar medium 
provide individual abundances of all elements $X_{i}$, isotopic ratios, and the global $Z/X$ 
\citep[see \eg,][]{2009ara&a..47..481a}. Stellar data are sparser. 
Generally, one has access to the abundance in number of metals relative to hydrogen [M/H] or to [Fe/H] 
(if only iron is measured) and sometimes to a few individual abundances like those of C, N, O, Ca, or $\alpha$-elements (see below).

In the modelling, one uses the ratio $Z/X$ of abundances in mass fraction, which is related to the observed abundances in number by the following relation,
\begin{equation}
\mathrm{[M/H]}=\log (Z/X)- \log (Z/X)_\odot \quad ,
\end{equation}
where a value for the solar $(Z/X)_\odot$ has to be chosen (see below), 
and where $\mathrm{[M/H]}$ is often taken to be equal to $\mathrm{[Fe/H]}$.
One also has to choose a mixture of heavy elements, \ie,  the abundances of individual metals. Usually, it is assumed that 
$(X_{i}/Z)=(X_{i}/Z)_\odot$ unless individual abundances are measured (for instance $\alpha$-elements enhanced mixture, CNONa in globular clusters, etc.).

In the following sections, we examine the impact of the abundances of heavy elements on age. For that purpose, we consider 
as examples, the solar mixture ([Fe/H]=0), a depleted mixture with [Fe/H]=$-1.0$ dex (representative of some halo or thick disc stars), 
and an $\alpha$-elements enhanced mixture.

\subsubsection{Helium}

The helium abundance cannot be inferred directly from the spectra of tepid stars because of the lack of lines.
In the Sun, the helium abundance in the convective envelope (CE) has been inferred  from helioseismology (see Lecture 2 on \textsl{The impact of asteroseismology})).
The helioseismic solar value is $Y_\mathrm{CE, \odot}= 0.2485\pm 0.0034$ \citep{2004ApJ...606L..85B}. 
 Because of diffusion processes that occurred during the solar lifetime, $Y_\mathrm{CE, \odot}$ is expected to be different from 
 the initial helium abundance in the molecular cloud where the Sun formed. From the calibration of the solar model, \ie,  from the requirement that 
 a model of $1\ M_\odot$ reaches at solar age $t_\odot \sim 4.57$~\Gyr, the observed solar luminosity, radius, and surface metal abundance $(Z/X)_\odot$,
  one derives the solar initial helium abundance $Y_\mathrm{0, \odot}$ and metal to hydrogen ratio $(Z/X)_\mathrm{0, \odot}$. 
  The solar model calibration also provides the convection parameter $\alpha_\mathrm{conv}$ (see Sect.~\ref{conv}).
More details about the solar model calibration are given in Lecture 2 (\textsl{The impact of asteroseismology}). 
 
 The initial helium abundance $Y_0$ is therefore usually a free parameter of stellar models. 
The main hypothesis/choices that are currently made for this quantity are listed below.
\begin{itemize}
\item $Y_0$ can be set to the solar calibrated value $Y_\mathrm{0, \odot}$, which depends on the input physics of the 
associated solar model.
\item $Y_0$ can be derived from the relation $Y_0= Y_\mathrm{p} + Z\times (\Delta Y/\Delta Z)$, where $  Y_\mathrm{p} $ 
is the primordial helium abundance, and  $\Delta Y/\Delta Z$ the helium-to-heavy elements enrichment ratio.
This relation accounts for the enrichment of helium and heavy elements in the interstellar medium resulting 
from Galactic evolution. The value of the primordial helium abundance is quite secure today. 
For instance, on one hand, 
\citet{2013JCAP...11..017A} got $Y_\mathrm{p}=0.2534 \pm 0.0083 $ from observations in H~\textsc{ii} regions. 
On the other hand, from the observations of the Cosmic Microwave Background by the  WMAP and 
Planck missions and standard Big Bang nucleosynthesis, \citet{2008JCAP...11..012C}  
inferred $Y_\mathrm{p}=0.2487 \pm 0.0002$ (WMAP), while \citet{2013arXiv1307.6955C} 
inferred $Y_\mathrm{p} = 0.2463 \pm 0.0003$ (Planck). 
 Conversely, $\Delta Y/\Delta Z$ is imprecise and can vary from place to place in the Galaxy.
Stellar modellers currently use values in the range $\Delta Y/\Delta Z=2 \pm 1$, resulting from solar calibration. However, 
as reported by \citet{2010A&A...518A..13G}, a large dispersion is found in the literature  with 
$\Delta Y/\Delta Z$ values that vary from $0.5$ to $5$ at least.
\item In most favourable cases, where precise and numerous observational constraints are available for the considered star, 
the initial helium content of the star can be inferred from modelling. This is 
\textsl{\`{a} la carte} modelling, 
thoroughly described in lecture 2.
\end{itemize}

In the following, to estimate the impact of the choice of $Y$ on age-dating, we consider models with 
$Y=0.25, 0.28$, and $0.31$, and  $\Delta Y/\Delta Z= 2$ and $5$.

\subsubsection{Solar mixture}
\label{solarmixture}

In most stellar models, stellar mixtures are assumed to be similar to the solar mixture, \ie,  $(X_{i}/Z)_\mathrm{star}=(X_{i}/Z)_\odot$.
However, the choice to make on the solar mixture is still subject to discussion. In the years from 1993 to now, 
there have been several revisions of the solar photospheric mixture. A major revision took place in 2003, 
when 3-D solar model atmospheres including non local thermodynamical equilibrium effects 
as well as improved atomic data were used to infer solar photospheric abundances \citep[see \eg,][]{2009ara&a..47..481a}. 
The unexpected result has been a decrease of the abundances of C, N, O, Ne, Ar, and $\left(Z/X\right)_\odot$. 
In Table~\ref{solar_global} below, we list some of the $\left(Z/X\right)_\odot$ determinations.

\begin{table}[!ht]
\setlength{\abovecaptionskip}{0pt}
\setlength{\belowcaptionskip}{10pt}
\begin{center}
\caption{Values of the solar $\left(Z/X\right)_\odot$ ratio from 1993 to 2010 obtained successively by 
\citet{1993pavc.conf..205G} (\textsl{GN93}), \citet{1998SSRv...85..161G} (\textsl{GN98}),
\citet{2005A&A...435..339A} (\textsl{AGS05}), \citet{2008A&A...488.1031C} (Caff08), \citet{2009ara&a..47..481a} (\textsl{AGSS09}), and \citet{2009LanB...4B..712L} (Lod09). }
\label{solar_global}
\centering
\begin{tabular}{ l c c cccc}
\noalign{\smallskip}
\hline\hline
\noalign{\smallskip}
 & \textsl{GN93} & \textsl{GN98} & \textsl{AGS05} & Caff08 & \textsl{AGSS09} &	 Lod09  \\
\noalign{\smallskip}
\hline
\noalign{\smallskip}
 $\left(Z/X\right)_\odot$ & 0.0245 & 0.0229 & 0.0165& 0.0209 & 0.0181 & 0.0191  \\
\noalign{\smallskip}
\hline
\end{tabular}
\end{center}
\end{table}

From the \textsl{GN93} to the \textsl{AGSS09} results, the solar oxygen abundance decreased by $\sim 34$ per cent. 
This impacted the total solar metallicity $\left(Z/X\right)_\odot$, 
which decreased by $\sim 25$ per cent. One of the main consequences is a degradation of the agreement between the 
helioseismic solar model and observations \citep[\eg,][]{2009ara&a..47..481a}. 
The decrease of the O abundance induces a decrease of opacity, 
which leads to a convective envelope shallower than the seismically inferred value. 
The $\mu$-decrease degrades the agreement between solar model structure and helioseismology observations. 
It has been suggested that an increase of the Ne abundance (non directly measured in the solar photosphere)  
could compensate for the oxygen decrease. 
However, while the  increase in opacity due to Ne improves 
the agreement with helioseismology for the location of the base of the convective zone and the He abundance in it, 
the density and sound speed profiles still do not match the seismic estimates 
\citep[for a review, see][]{2008PhR...457..217B}.
 In the following, we consider the effects on age-dating
of a change from the \textsl{GN93} mixture to the \textsl{AGSS09} one. 

\subsubsection{$\alpha$-elements}

In stars, the $\alpha$-elements (O, Ne, Mg, Si, S, Ar, Ca, Ti) are synthesized by $\alpha$ particles (\eg, helium nuclei) capture reactions that  proceed as,
\begin{equation}
\ _{6}^{12}\mathrm{C} (\alpha, \gamma) \ _{8}^{16}\mathrm{O} (\alpha, \gamma) \ _{10}^{20}\mathrm{{Ne}} (\alpha, \gamma) \ _{12}^{24}\mathrm{{Mg}} (\alpha, \gamma) \, \cdots
\end{equation}
and so on up to the synthesis of Si, S, Ar, Ca, and Ti. In the early Galactic life, nucleosynthesis was dominated by massive, 
short-living stars ending as type \textsc{ii} supernovae (SN), 
which produced $\alpha$-elements together with iron-peak elements.
Later, SN \textsc{i}a resulting from the accretion of gas from a stellar companion onto a white dwarf also contributed to the enrichment of the interstellar medium, 
providing again iron-peak elements, but only little  amounts of $\alpha$-elements \citep[see \eg,][]{1979ApJ...229.1046T}.
As a result, metal-poor, old stars in the halo and thick disc show $\alpha$-elements enhancements with respect to younger, thin disc stars
(see the lecture by M. Haywood). There is a trend for $\alpha$-elements to increase 
when [Fe/H] increases with similar trends observed in disc, bulge, and halo
\citep{2010A&A...513A..35A}. 
The impact of $\alpha$-elements enhancements on stellar models is through opacity changes.
In the following, to estimate how the choice of $\alpha$-elements enhancement affects age-dating, 
we consider models with $\mathrm{[\alpha/Fe]}=0.0$ (\ie,  no enhancement with respect to the Sun) and $0.4$ dex 
(corresponding to the important enhancement observed in old population stars).

\subsection{Nuclear reactions}
\label{nuc}
%-----------------------------------------------
\begin{figure}[!htpb]
\begin{center}
\resizebox{0.95\hsize}{!}{\includegraphics{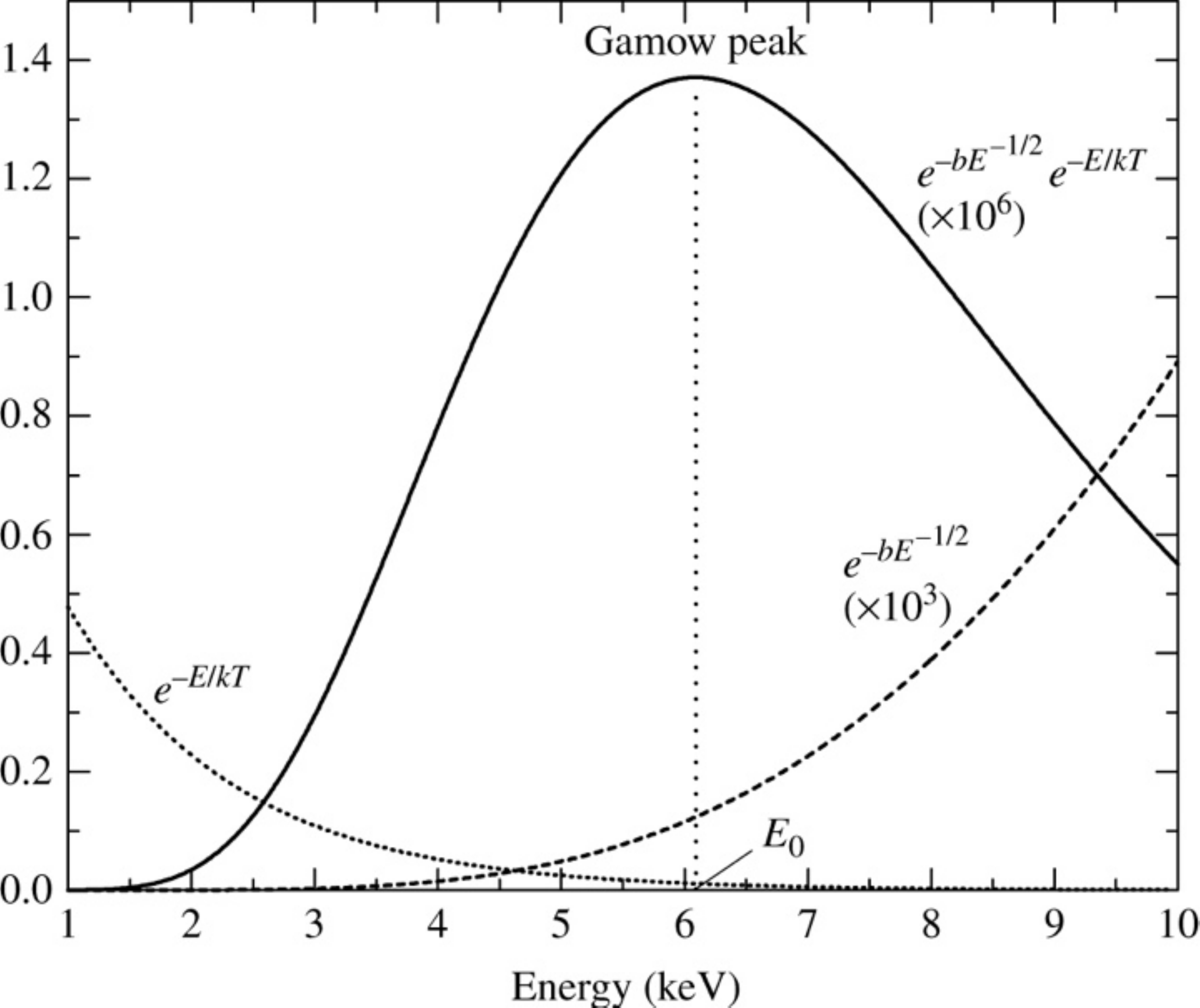}}
\caption{The Gamow peak. Dotted line is the high energy tail of the Maxwell-Boltzmann distribution
of particles velocities. Dashed line is the probability of penetration of the Coulomb barrier by the tunnel
effect. The product of the two functions is the continuous line, showing the maximum probability of 
fusion \citep[for a detailed description, see][]{1968psen.book.....C}.}
\label{gamow}
\end{center}
\end{figure}

%-----------------------------------------------
\subsubsection{Nuclear reactions rates}

A very comprehensive presentation of stellar nucleosynthesis can be found in the text book by \citet{1968psen.book.....C}. 
We briefly recall a few points here.

\begin{description}
\item[Reaction rates. ] 

The temporal evolution of a species ${i}$ (mass number $A_{i}$, charge number $Z_{i}$) 
under the effect of nuclear reactions is expressed by Eq.~\ref{dXdtnuc}. 
The reaction rate $r^{k}_{ij}$, \ie,  the number of reactions per second and per gram for a reaction of the type, $i+j \Rightarrow k+\cdots$, reads
\begin{equation}
r^{k}_{ij} = N_{A}^{2} \langle \sigma v \rangle_{ij} \frac{X_{i}}{A_{i}} \frac{X_{j}}{A_{j}},
\end{equation}
where $N_{A}$ is the Avogadro number and where the effective cross-section of a non resonant
 nuclear reaction $ \langle \sigma v \rangle_{ij} $ reads
\begin{equation}
\langle \sigma v \rangle_{ij} \propto \frac{\left(Z_{i} Z_{j} /A_\mu\right)^{1/3}}{T^{2/3}} S_0 
\exp \left[ -C  \left(\frac{Z_i^2 Z_j^2 A_\mu}{T}\right)^{1/3} \right] \, 
\left(1+f(T)\right) \, ,
\end{equation}
where $S_0$ is the astrophysical factor (S-factor), $A_\mu$ is the nucleon number of the reduced particle, 
$C$ is a constant,
and $f(T)$ is a correction to the Gaussian (Gamow peak, see below).
The S-factor has to be evaluated theoretically or experimentally. It is the source of uncertainty in the rate.
Note that the effective cross-section has to be corrected for electron screening, implying that 
$\langle \sigma v \rangle_s= f_s \langle \sigma v \rangle$ where $f_s$ is the screening factor (see below).

%-----------------------------------------------
\begin{figure}[!htbp]
\begin{center}
\resizebox{0.95\hsize}{!}{\includegraphics{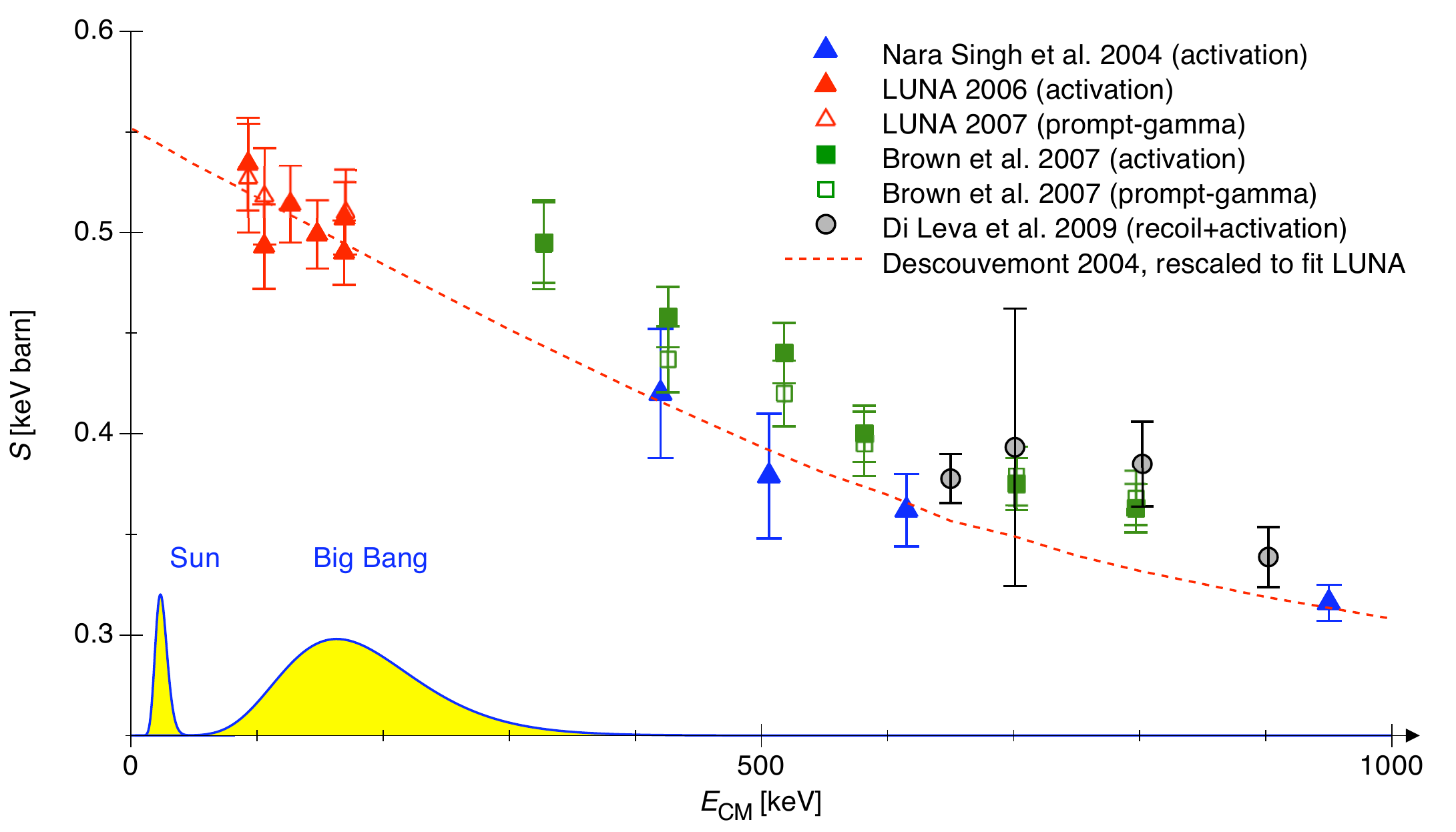}}
\caption{Astrophysical S(E)-factor for $\ _{2}^{3}\mathrm{He} (_{2}^{4}\mathrm{He}, \gamma) _{4}^{7}\mathrm{Be}$ reaction, 
after \citet{2010ARNPS..60...53B}. For this reaction, \textsl{LUNA} laboratory measurements closely approach the Gamow peak at the temperatures in the solar centre (in yellow are the Gamow windows for solar centre and Big Bang nucleosynthesis
temperatures).}
\label{He3He4Broggini}
\end{center}
\end{figure}
%-----------------------------------------------

\item[Gamow peak.] 

For thermonuclear fusion to take place between charged particles in stellar interiors, a Coulomb barrier has to be crossed
by interacting nuclei. 
A nuclear reaction rate depends \textsl{(i)} on the energy of particles, therefore on the temperature, 
and \textsl{(ii)} on the probability of penetration of the Coulomb barrier by the tunnel effect.
The probability for a nuclear reaction to occur shows a maximum (the Gamow peak, Fig.~\ref{gamow}) 
resulting from the  combined contribution of the Maxwell-Boltzmann 
high energy tail and of the Coulomb barrier penetration probability. 

\item[Astrophysical S-factor. ] 

The S-factor can be derived either from theory or from experimental data. 
The experimental measurement of S-factors is difficult because nuclear reactions take place in stars at low energies
(typically from a few keV to less than 0.1 MeV), while in the laboratory nuclear reactions are produced at higher energy.
Getting the astrophysical S-factor therefore requires extrapolation of laboratory measurements to low energies, 
implying the risk to omit unknown resonances, etc.
Progress has been accomplished in the last ten years with the advent of low energy, high intensity underground accelerators, 
which have begun to give access to the low energy domain, down to energies in the solar Gamow window, 
as illustrated by Fig.~\ref{He3He4Broggini} \citep[see][for a review on the \textsl{LUNA} experiment capabilities]{2009RPPh...72h6301C}.

\item[Energy production. ] 

The energy production reads,
\begin{equation}
\epsilon_\mathrm{nuc, j} = r^{k}_{ij} Q_{j} \frac{\rho}{N_{A}} = N_{A} \rho \frac{X_{i}}{A_{i}} \frac{X_{j}}{A_{j}} \langle \sigma v \rangle_{ij} Q_{j},
\end{equation}
where $Q_{j}$ is the energy released by one, $i+j \Rightarrow k+\cdots$, nuclear reaction.

\end{description}

%-----------------------------------------------
\begin{figure}[!htbp]
\begin{center}
%\resizebox{0.6\hsize}{!}{
\includegraphics[width=0.9\textwidth]{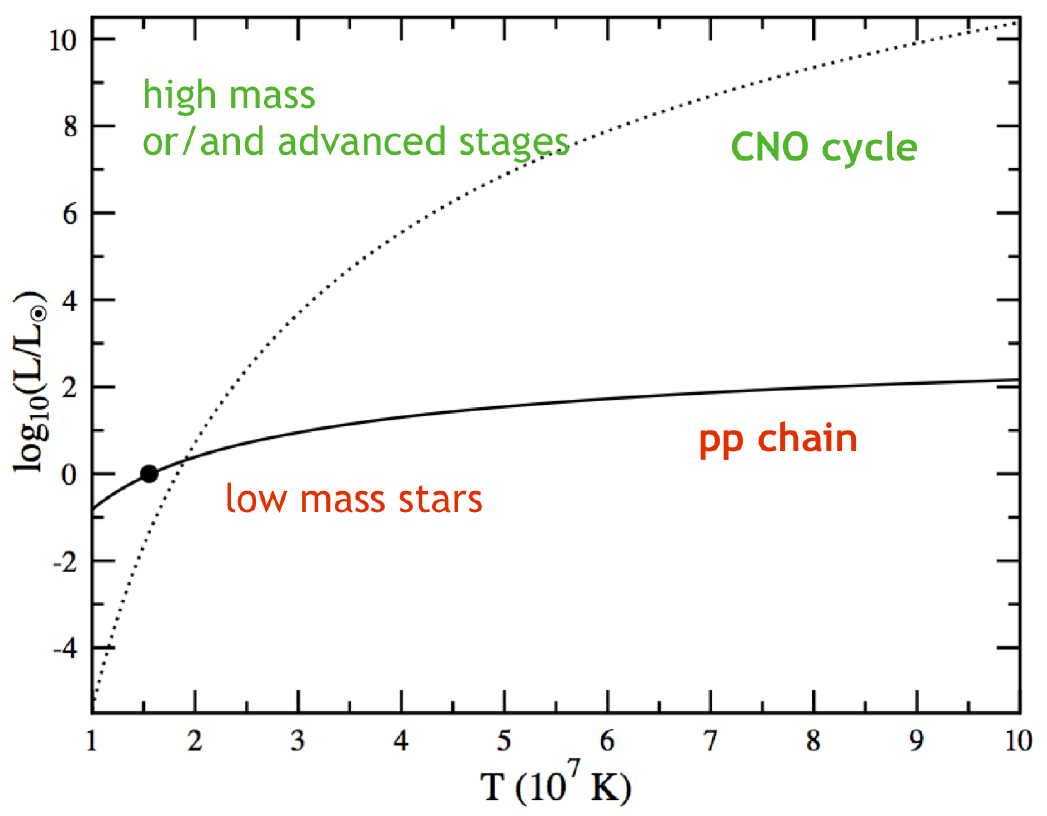}
\caption{The respective weights of the $p-p$ chain and CNO cycle in the global stellar luminosity 
for different central temperatures. [From \citet{2011RvMP...83..195A}.]}
\label{ppcnoweights}
\end{center}
\end{figure}
%-----------------------------------------------

\subsubsection{Impact on age-dating of hydrogen burning leading reactions}

As it is well-known, in stars, hydrogen burning proceeds either by the proton-proton chain ($p-p$) in low-mass stars 
with low central temperatures, or by the CNO cycle in high-mass stars and/or for advanced evolutionary stages (see Fig.~\ref{ppcnoweights}).
We examine the impact on age-dating of the two leading nuclear reactions for hydrogen burning.
The $p-p$ chain is led by its slowest reaction, $p(p,e^+\nu)d$, whose rate is obtained from theory, while the CNO cycle is led by the 
$^{14}\mathrm{N} (p, \gamma) ^{15}\mathrm{O}$ reaction, whose rate is inferred from laboratory experiments.

%-----------------------------------------------
\begin{figure}[!hptb]
\begin{center}
%\resizebox{0.5\hsize}{!}{
\includegraphics[width=0.85\textwidth]{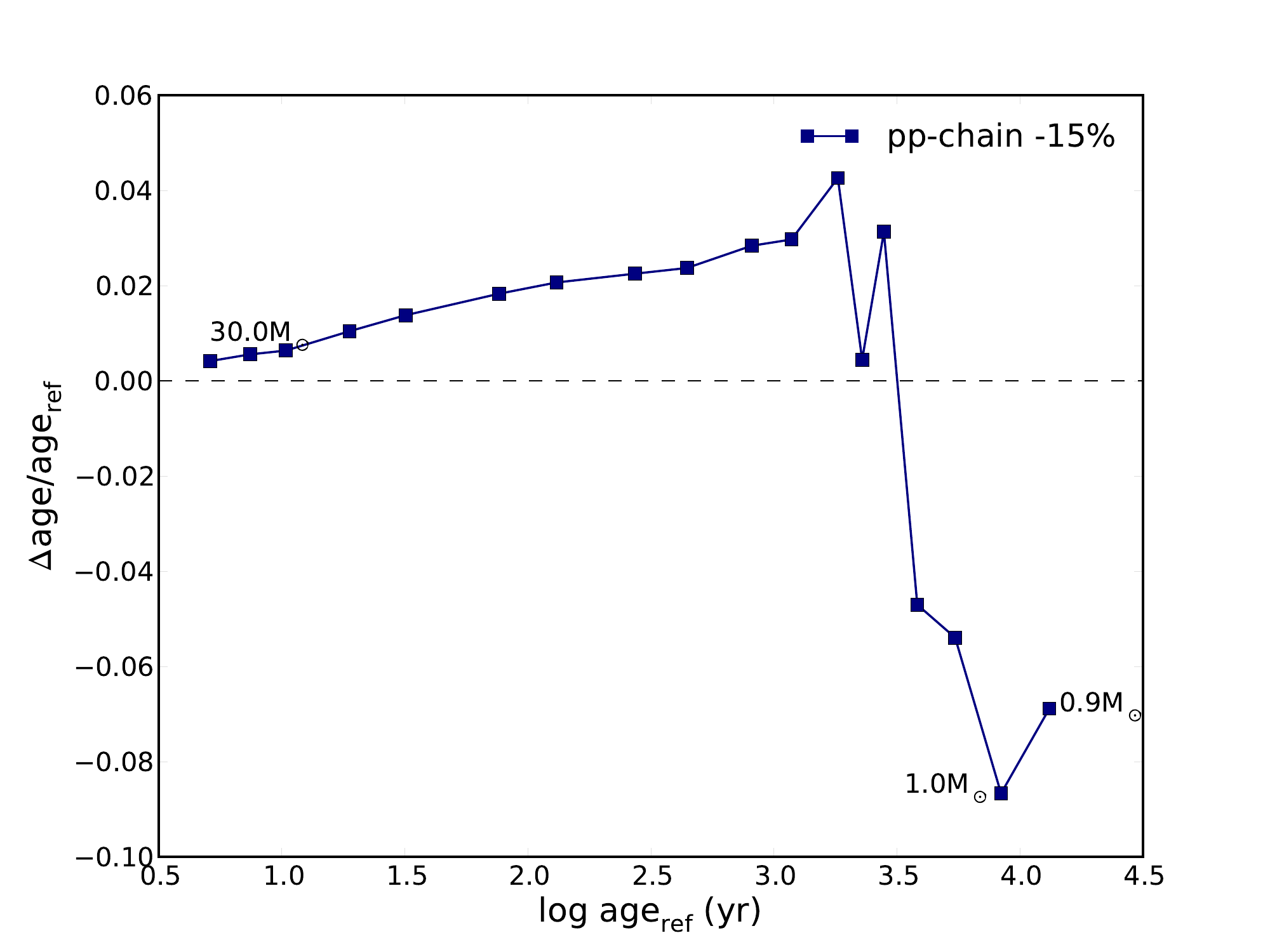}
%\resizebox{0.5\hsize}{!}{
\includegraphics[width=0.85\textwidth]{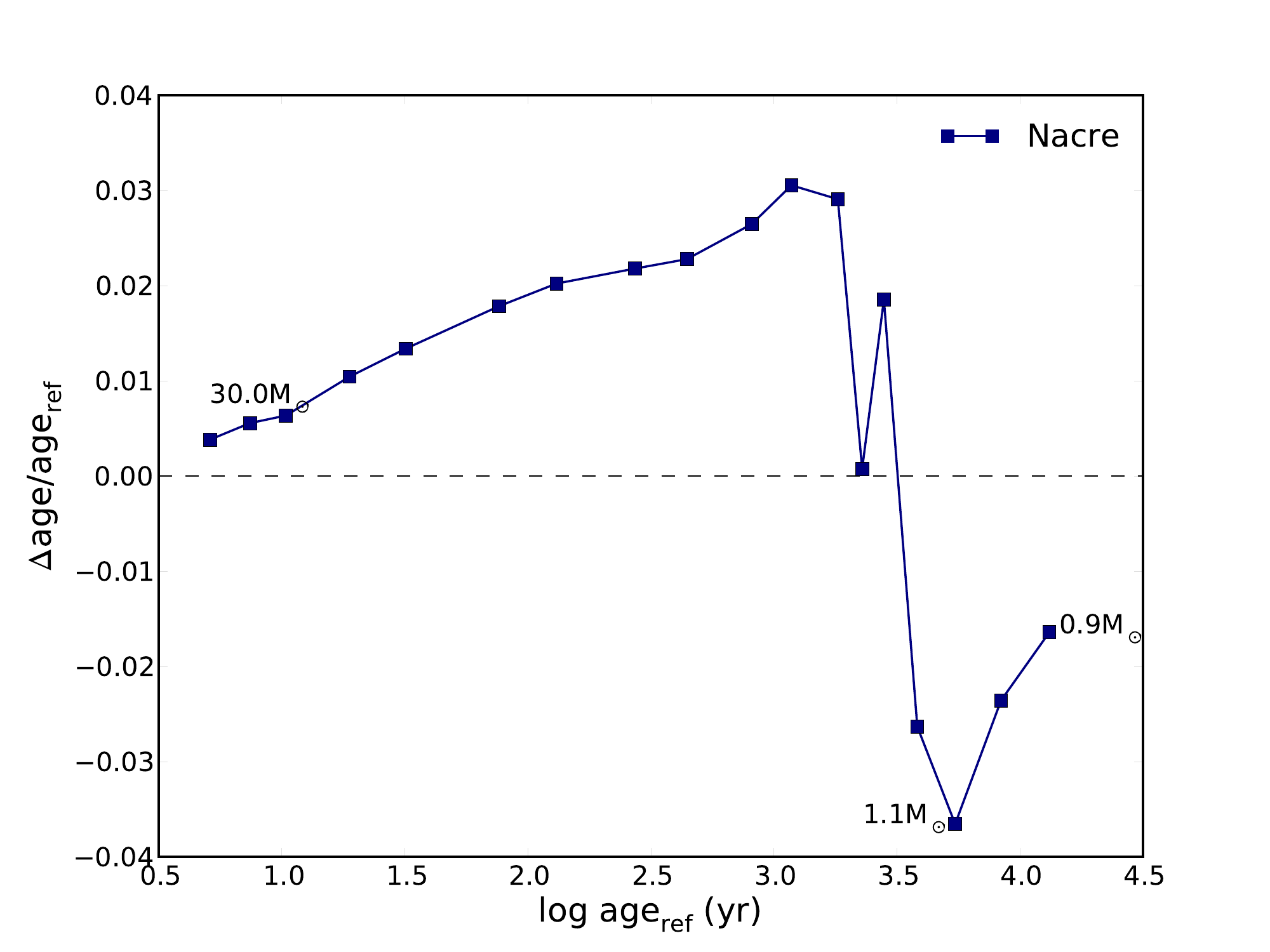}
\caption{{\sl Top:} relative difference of the TO age between models where the p-p reaction rate has been 
decreased by $15$ per cent and the reference model. Each point corresponds to a model of different mass, 
as listed in Sect.~\ref{grids}. 
The extreme values of the considered mass range are pinpointed as well as the mass at 
which the relative age difference is maximum.
{\sl Bottom:} same for the comparison between \textsl{NACRE} and \textsl{LUNA} rates for 
the $^{14}\mathrm{N} (p, \gamma) ^{15}\mathrm{O}$ reaction.
}
\label{errorpp}
\end{center}
\end{figure}
%-----------------------------------------------

\begin{description}
\item[p-p chain: $\mathbf{p(p,e^+\nu)d}$ reaction.]
The rate of this reaction is too small to be measured in the laboratory. It is derived from the theory of weak interaction \citep[see \eg,][]{2011RvMP...83..195A}. 
On the other hand, \citet{1998PhLB..416..365D} have estimated that this rate is constrained by helioseismology at a level of $\pm 15$ per cent. 
In the following, we take this value as an error bar for the p-p reaction rate.

In Fig.~\ref{errorpp}, left panel, we show the effect on the TO age of a decrease of $15$ per cent 
of the $p(p,e^+\nu)d$ reaction rate.
It shows that the maximum effect at turn-off occurs for masses $\approx 1\ M_\odot$ (age $\approx 10$~\Gyr), 
where the age difference is of ${\sim}9$ per cent.
More specifically, at low mass where the $p-p$ chain dominates, a decrease of the rate causes an increase 
of central density resulting in a more compact core. 
The age, which roughly varies as indicated by Eq.~\ref{tnucMS} is smaller. At moderate mass, where both p-p and CNO operate, 
the $E_\mathrm{pp}/E_\mathrm{CNO}$ ratio is smaller, 
resulting in a lower central density and a higher age (maximum effect of 
$\approx 4$ per cent at $\approx 1.5\ M_\odot$). 
At high mass, since the CNO cycle dominates, the effect of a decrease of the $p-p$ reaction rate is very small. 

\item[CNO cycle: $\mathbf{^{14}\mathrm{N} (p, \gamma) ^{15}\mathrm{O}}$ reaction.]
The rate of this reaction has been measured with the \textsl{LUNA} device 
\citep[see \eg,][]{2002nuas.conf..111F,2008PhRvC..78b2802M}. 
Impressively, the reaction rate is now measured down to centre of mass energies of $70$ keV, 
approaching the physical conditions at the centre of a RGB star of $1\ M_\odot$. 
Extrapolation of the rate down to energies relevant for a $1\ M_\odot$ on the MS is still needed 
(see Fig.~\ref{ppCNO1} and \ref{ppCNO2}). 
From these new measurements, a major revision of the reaction rate followed, 
leading to a reduction of the S-factor by $\sim 50$ per cent. 

%-----------------------------------------------
\begin{figure}[!htbp]
\begin{center}
%\resizebox{0.95\hsize}{!}{
\includegraphics[width=0.99\textwidth]{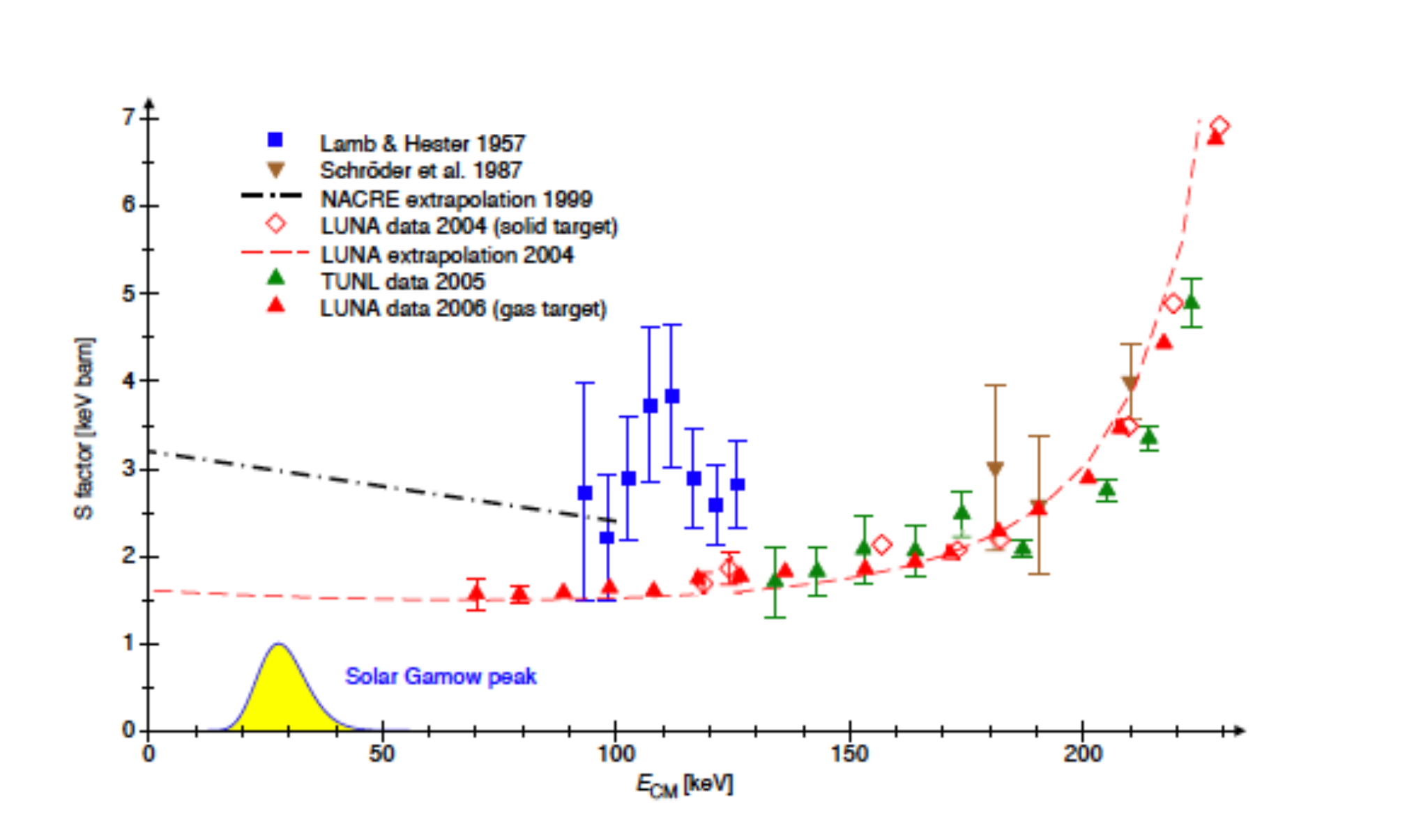}
\caption{Same as Fig.~\ref{He3He4Broggini} for 
the $^{14}\mathrm{N} (p, \gamma) ^{15}\mathrm{O}$ reaction rate.
[From \citet{2010ARNPS..60...53B}.]}
\label{ppCNO1}
\end{center}
\end{figure}
%-----------------------------------------------
%-----------------------------------------------
\begin{figure}[!htbp]
\begin{center}
%\resizebox{0.95\hsize}{!}{
\includegraphics[width=0.99\textwidth]{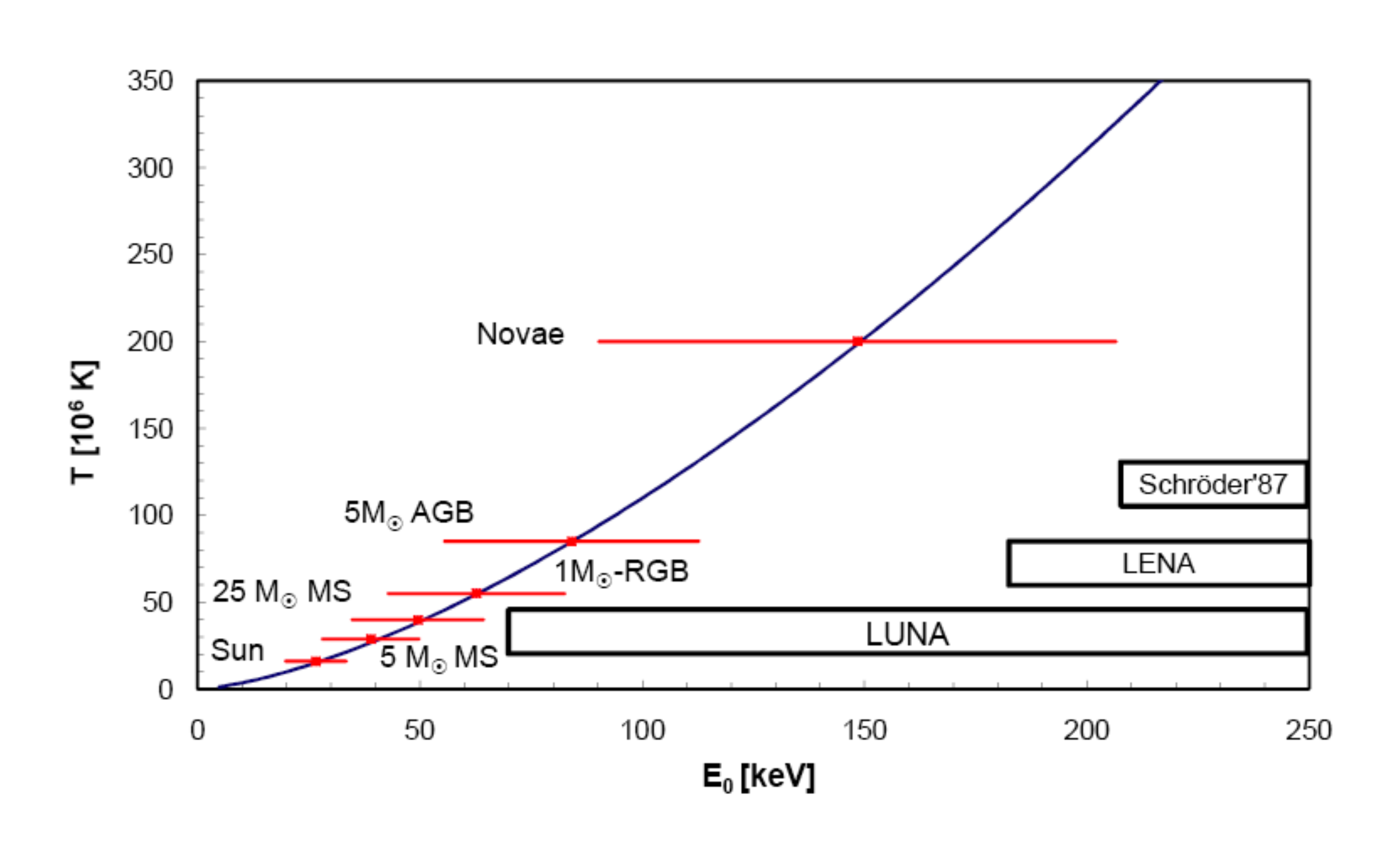}
\caption{The $^{14}\mathrm{N} (p, \gamma) ^{15}\mathrm{O}$ reaction rate: 
the leap towards low temperatures accomplished by the \textsl{LUNA} experiment.
[From  \citet{2009RPPh...72h6301C}.]}
\label{ppCNO2}
\end{center}
\end{figure}
%-----------------------------------------------

In turn, in a calibrated solar model, the $p-p$ \textit{vs} CNO balance is drastically modified
 ($E_\mathrm{CNO}/E_\mathrm{tot}$ decreases from $1.6$ to $0.8$ per cent when changing from 
\textsl{NACRE} to \textsl{LUNA} rate). 
Furthermore, the decrease of the nuclear energy produced at given density and temperature 
affects the onset of convective cores in solar-like stars: a
convective core first appears at 
higher mass, or equivalently, the convective core is less massive at a given mass (see Fig.~\ref{14NNACRELUNA}).

This has indeed consequences for stellar age-dating. \citet{2004A&A...420..625I} 
examined the impact of the reduced rate on the isochrones of metal poor ([Fe/H]$=-2.0$ dex) globular clusters 
and found that the turn-off was brighter and bluer with an age reduction of 0.7 to 1~\Gyr~ 
(see Fig.~\ref{diff_nuc}). 
On the other hand, we show in Fig.~\ref{errorpp}, right panel, that at solar metallicity the age impact is rather small,
with a maximum difference in the range 3-4 per cent.

\end{description}

%==========================
\begin{figure}[!hptb]
\begin{center}
%\resizebox{0.5\hsize}{!}{\includegraphics{FIGURES/task13_nuc_cc}}\resizebox{0.5\hsize}{!}{\includegraphics{FIGURES/hrformi}}
\resizebox{0.5\hsize}{!}{\includegraphics{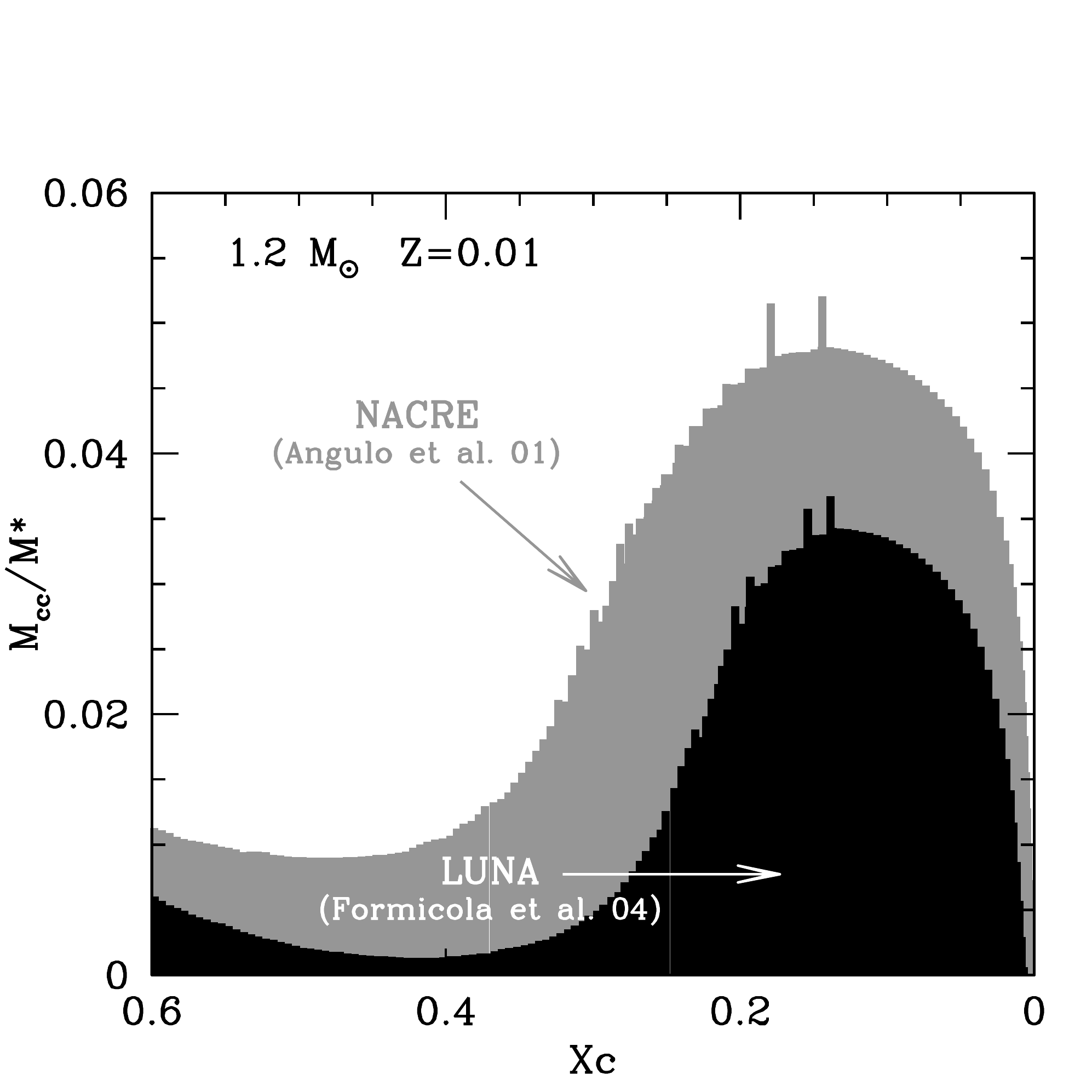}}\resizebox{0.5\hsize}{!}{\includegraphics{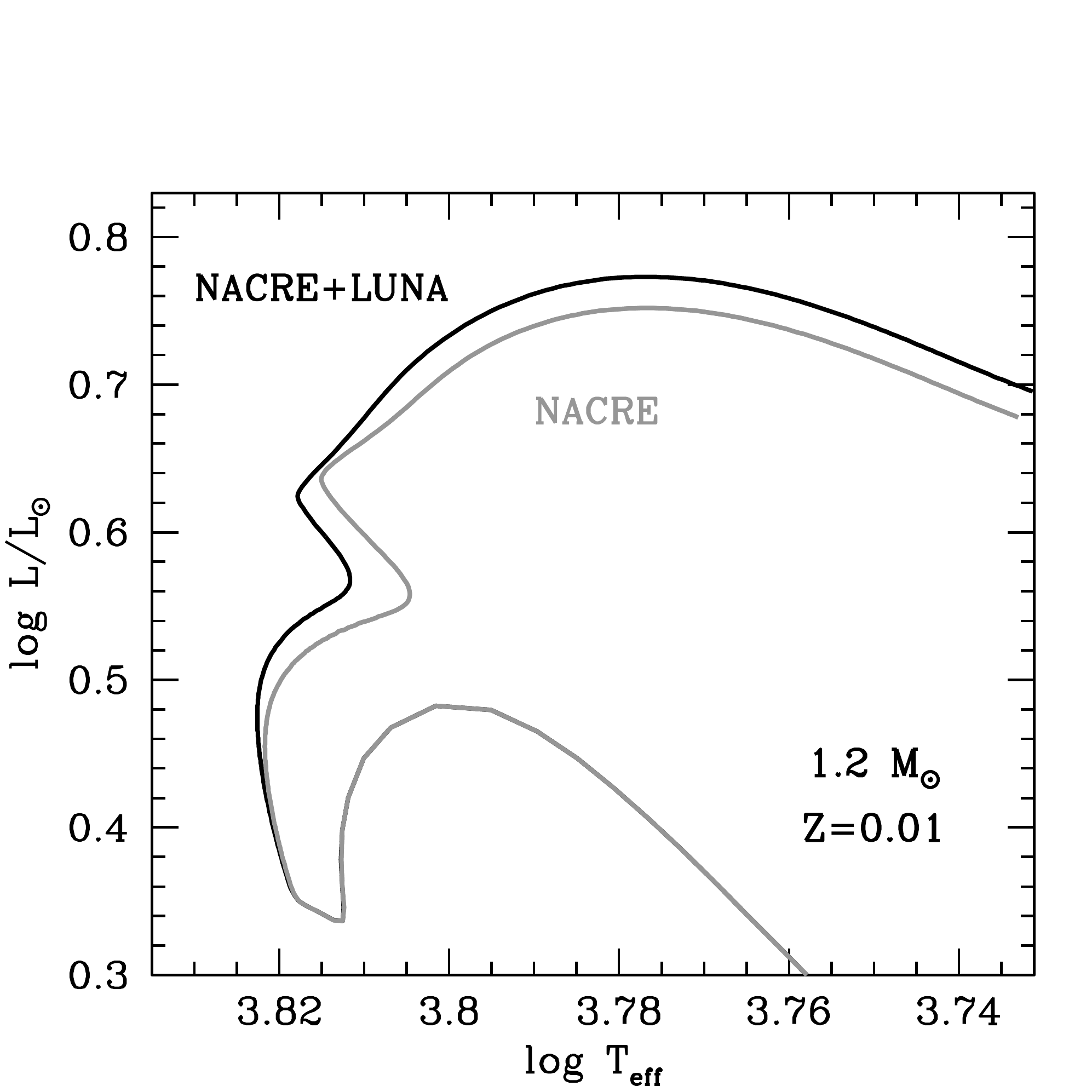}}
\caption{Impact of the revision of the $^{14}N(p,\gamma)^{15}O$ reaction rate in {CL\'ES} models
\citep{2008ap&ss.316..149s} of $1.2 M_\odot$ and $Z=0.01$, \ie,  at a mass close to the mass of apparition 
of a convective core on the {MS}. 
{\sl Left:} evolution of the convective core mass on the {MS} with in grey the core obtained with the \textsl{NACRE} rate for $^{14}N(p,\gamma)^{15}O$ and in black the one obtained with the \textsl{LUNA} rate. 
{\sl Right:} comparison of the tracks in the HR diagram of models with \textsl{LUNA} or \textsl{NACRE} 
reaction rates. [After \citet{2010Ap&SS.328...29L}.]}
\label{14NNACRELUNA}
\end{center}
\end{figure}
%==========================

%-----------------------------------------------
\begin{figure}[!htbp]
\begin{center}
\resizebox{0.85\hsize}{!}{\includegraphics{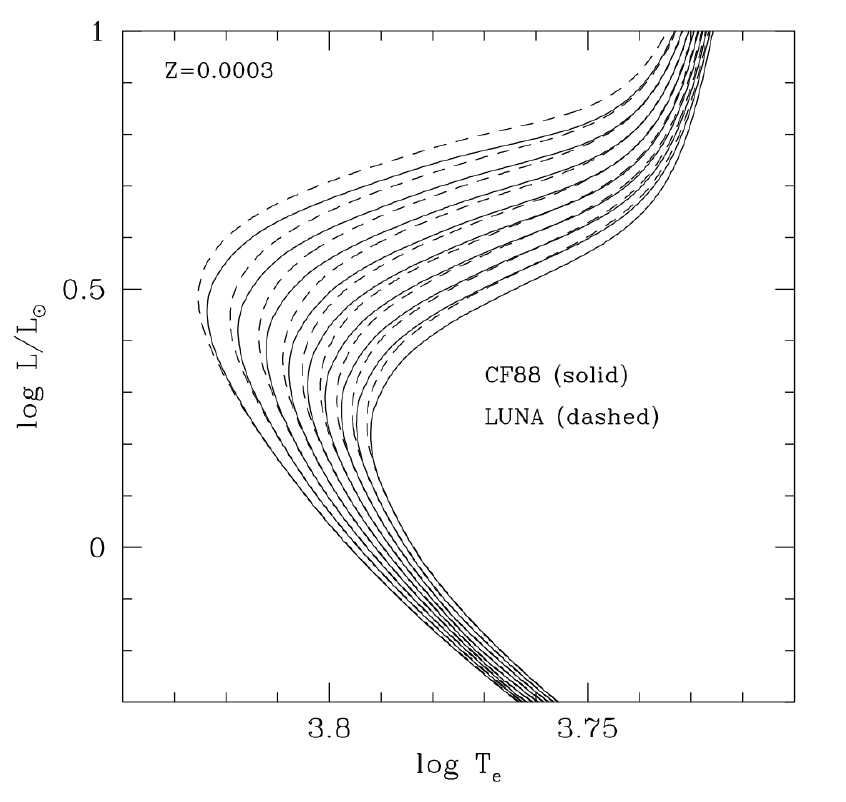}}
\caption{Effect on globular cluster isochrones of a decrease in the 
$^{14}\mathrm{N} (p, \gamma) ^{15}\mathrm{O}$ rate. [From \citet{2004A&A...420..625I}.]}
\label{diff_nuc}
\end{center}
\end{figure}
%-----------------------------------------------

\subsubsection{Screening factor}

In theory, the nuclear reactions rates are calculated for bare nuclei, where a positively charged nucleus $i$ 
collides with another,  positively charged, target nucleus $j$. 
In stars, the interactions between nuclei occur in presence of electrons that are
negatively charged.  The electron cloud surrounding the nuclei reduces the Coulomb barrier between them.
In this picture, the nuclear reaction rate is expected
to be enhanced in the presence of electrons.
Thus, one has to correct the non screened reaction rate $\langle \sigma v\rangle$ into a screened rate 
$\langle \sigma v\rangle_s=f _\mathrm{s}\langle \sigma v\rangle$, where $f_\mathrm{s}$ is a screening factor.

In the case where the screening is weak -which is suitable for MS stars considered here- first estimations of the screening factor have been obtained by 
\citet{1954ApJ...119..464S,1954AuJPh...7..373S, 1973ApJ...181..439D,1977ApJ...212..513M}. 
These authors treated the screening in a static case where they neglected the displacement of the interacting nuclei
within the plasma. On the other hand, 
astrophysical constraints on the screening were derived by \citet{2001A&A...371.1123W}, 
who obtained a range of allowed values of $f$ in the range $0.98-1.10$ using the constraint on the solar model coming from the seismic solar sound speed. 
More recently, \citet{2011ApJ...729...96M} developed a new approach, the dynamical screening, 
where they considered that the interaction energy of a pair of nuclei depends on the relative velocity of the pair.
% as illustrated in Fig.~\ref{screening}. 
The slower the velocity, the higher the screening. 
\citet{2011ApJ...729...96M} estimated that in the 
solar case the dynamic screening factor is $f_\mathrm{d}\sim 0.996$, while in the static case it is $f_\mathrm{s}=1.042$.

Concerning the age-dating, we have compared the turn-off age of models including the classical static weak 
screening with the one of models without screening (which mimics dynamic screening). The age differences never 
exceed 5 per cent. 

%-----------------------------------------------
%\begin{figure}
%\begin{center}
%\resizebox{0.7\hsize}{!}{\includegraphics{Figs/screening.png}}
%\caption{Screening effects in nuclear reaction rates (illustration).}
%\label{screening}
%\end{center}
%\end{figure}
%-----------------------------------------------

%\input opacity.tex

\subsection{Opacities}
\label{opa}
\subsubsection{Opacities in stellar models}

%-----------------------------------------------
\begin{figure}[!htbp]
\begin{center}
\resizebox{0.86\hsize}{!}{\includegraphics{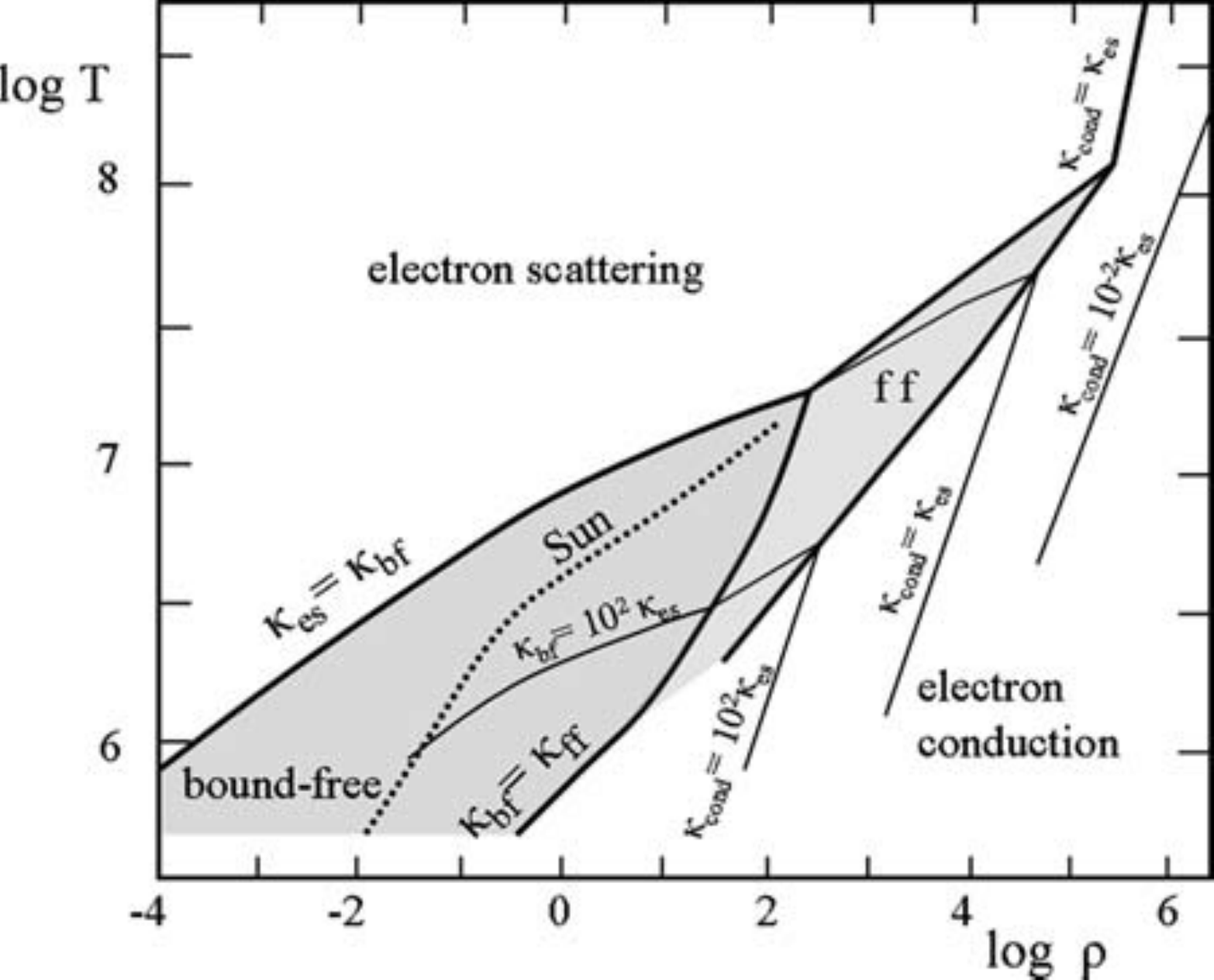}}
\caption{Opacity regimes in the density-temperature plane. [From \citet{Maeder09}.] 
}
\label{kappa_rhoT}
\end{center}
\end{figure}
%-----------------------------------------------

Radiative opacity, i.e. the ability of a medium to block radiation, is one of the main physical inputs of
stellar models. More generally, inside a star, opacity controls the transport of energy by photons (radiative opacity) or by
particles (the so-called conductive opacity). It  therefore tunes the stellar luminosity.
In the following, we only consider the radiative opacity, conduction being important -only- in dense regions of stars
like the centre of very low mass or very evolved stars \citep{2003ApJ...582L..43C}.

In the general case, the opacity (absorption coefficient) of a plasma depends on the frequency of the radiation. It is denoted
by $\kappa_\nu$ and its unit is $\mathrm{m^2 \, kg^{-1}}$. In stellar interiors (not in the atmosphere), the radiative transport can be treated in the
diffusion approximation \citep[see the text book by][]{1978stat.book.....M}. In the equation for energy transport (Eqs.~\ref{temperature} 
and \ref{radiation}), the opacity enters 
as an harmonic mean on frequency, weighted over the temperature derivative of the Planck function 
$B_\nu(T)$,
\begin{equation}
\frac{1}{\kappa_{\mathrm{R}}}=\frac{\pi}{a \, c \, T^3} \int\limits_{0}^{+\infty} \frac{1}{\kappa_\nu} \frac{\partial 
B_\nu(T)}{\partial T} \mathrm{d}\nu.
\end{equation}
The Rosseland mean opacity $\kappa_{\mathrm{R}}$, hereafter denoted by $\kappa$, is a function of $\rho$, T, and chemical composition.

%-----------------------------------------------
\begin{figure}[!htbp]
\begin{center}
%\resizebox{0.5\hsize}{!}{\includegraphics{FIGURES/OP_OPAL}}
\resizebox{0.95\hsize}{!}{ \rotatebox{270}{\includegraphics{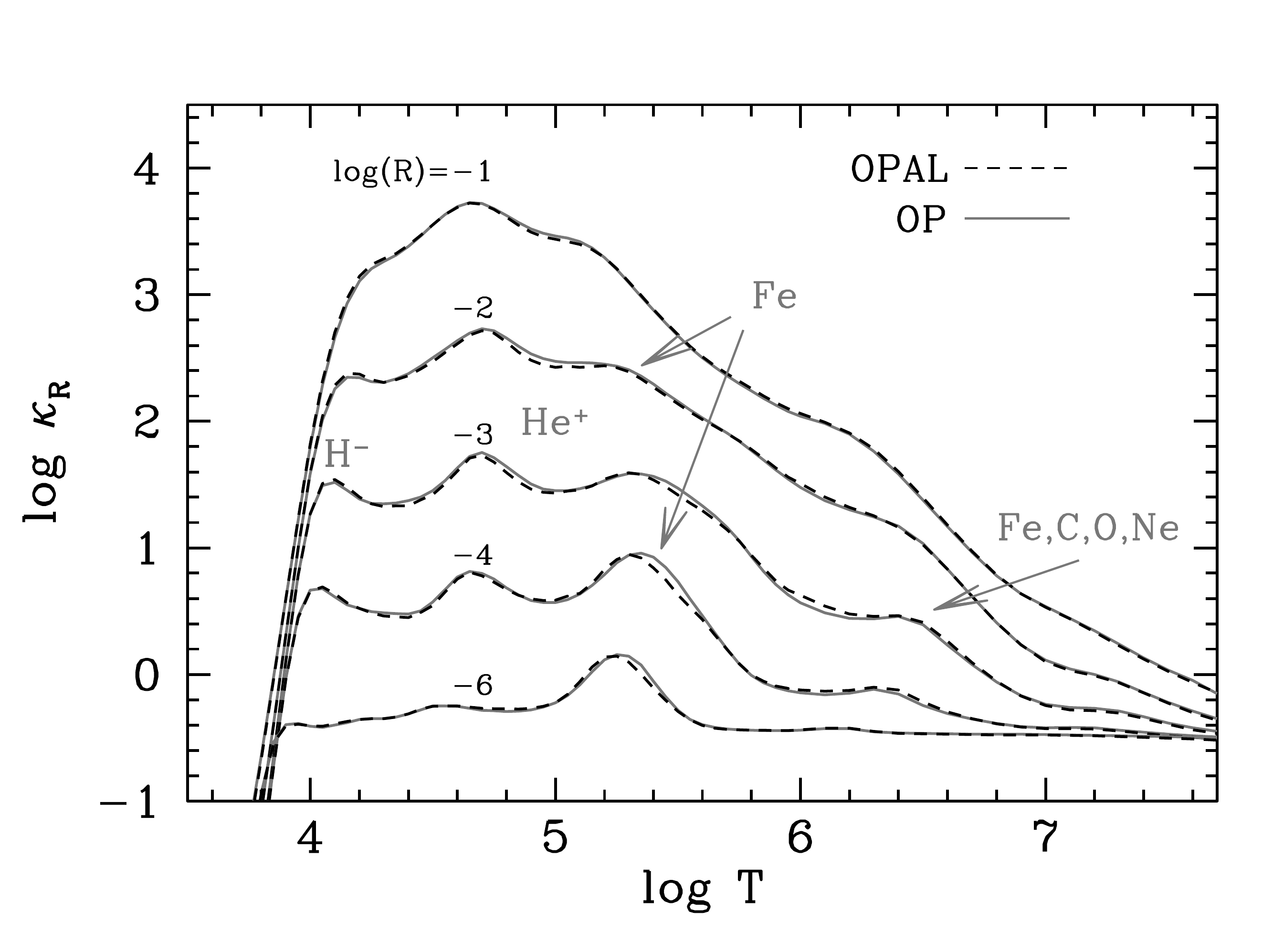}}}
\caption{Comparison of \textsl{OPAL} and \textsl{OP} opacities for X=0.70, Z=0.02,
and \textsl{AGSS09} solar mixture. See \citet{2005MNRAS.360..458B} for the same comparison, but for the
\textsl{GN93} mixture.
}
\label{OPOPAL}
\end{center}
\end{figure}
%-----------------------------------------------

In stellar models, different contributions to opacity have to be taken into account depending on the temperature and density of the  
plasma (see Fig.\ref{kappa_rhoT}). In very high density regions, opacity is dominated by the conduction by degenerate 
electrons. In high temperature, low density regions, the opacity is dominated by photon diffusion on electrons (electron 
scattering) and is approximately given by $\kappa_\mathrm{es}\approx 0.02 (1+X)$. In the regions of intermediate 
temperature and density, photon absorption related to ionization (bound-free processes) or photon scattering by ions 
(free-free transitions) can roughly be described by a Kramers' law with
$\kappa_\mathrm{bf, ff}\approx f(Z) (1+X) \rho T^{-3.5}$ (see also Eq.~\ref{kramersthomson}).
In low temperature, low density regions, the opacity is dominated by photon absorption in bound-bound transitions.
In these regions, the calculation of opacity is difficult because it implies all species in all accessible energy levels.
A census over the properties of these levels is therefore needed from atomic and molecular physicists.

Modern opacities currently used in stellar models were independently obtained by the \textsl{OPAL}  
\citep{1996apj...464..943i} and the \textsl{OP} \citep{2005MNRAS.360..458B} groups. 
For low temperatures, the Wichita group \citep{2005apj...623..585f} has provided opacities accounting for the contribution 
of molecules and grains. Practically, opacities are delivered as tables listing the opacity as a function of 
the temperature $T$,  
the quantity $R=\rho/T^3_6$ where $T_6=T/10^6$, and chemical composition ($X, Y, Z$). 
In these tables, the opacity calculation is based on millions of 
transitions for 21 chemical elements, constituting ions, atoms, molecules, and grains.

%-----------------------------------------------
\begin{figure}[!hptb]
\begin{center}
\includegraphics[width=0.87\textwidth]{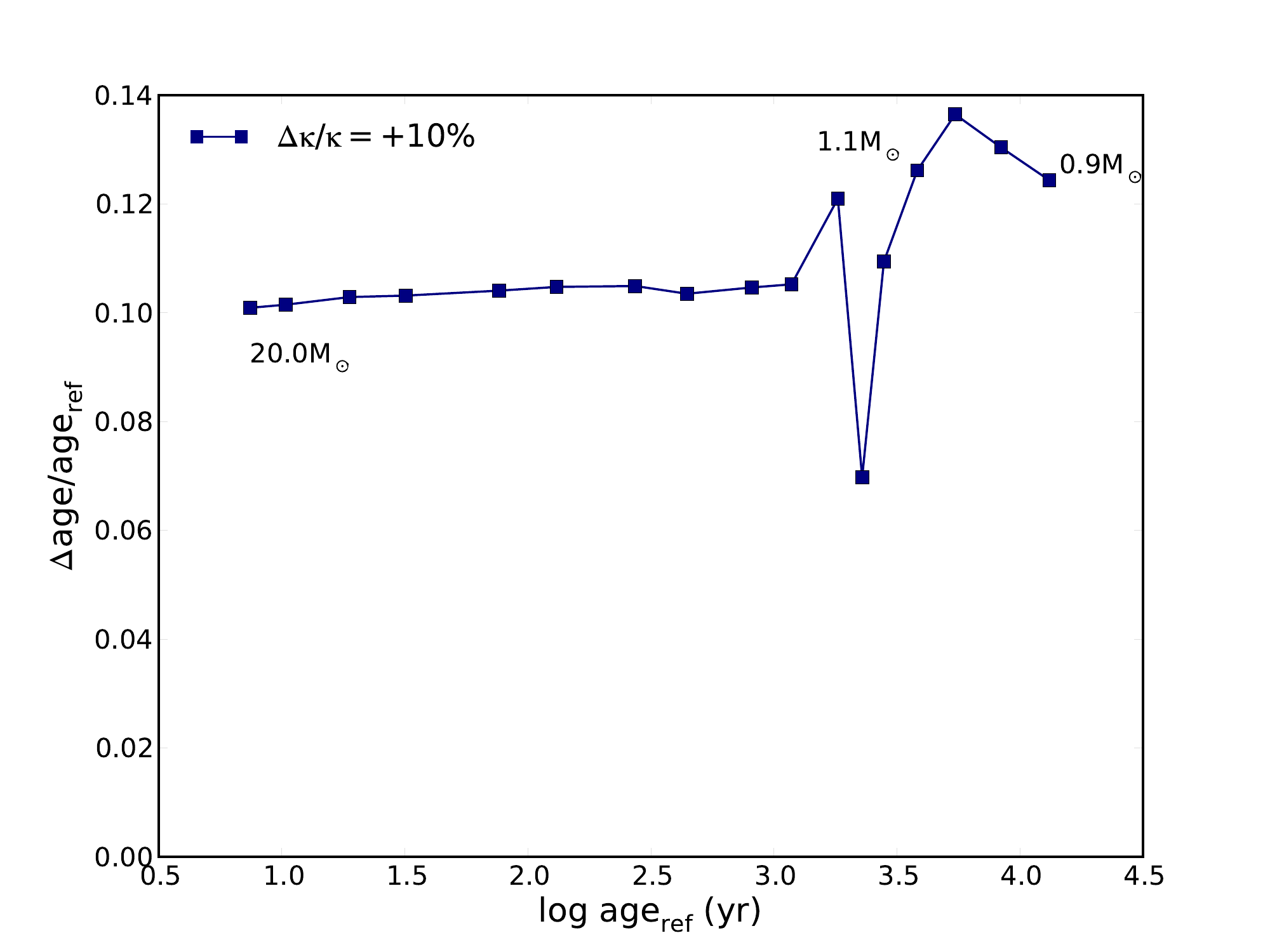}
\includegraphics[width=0.87\textwidth]{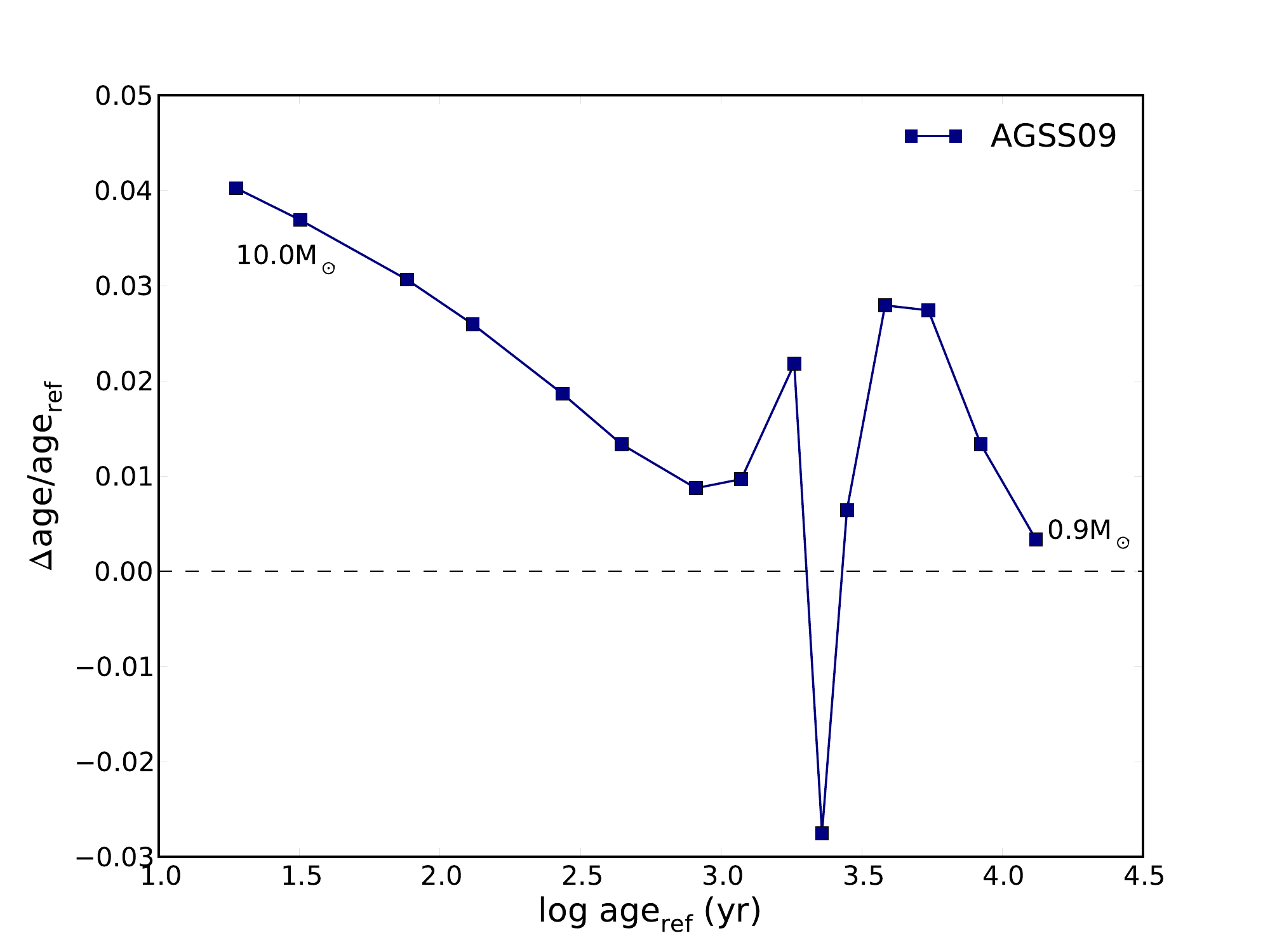}
\caption{Same comparison as in Fig.~\ref{errorpp}. {\sl Top:} effect on turn-off age of an increase of $10$ per cent of the opacity. 
{\sl Bottom:} effect on turn-off age of changing the \textsl{GN93} to the \textsl{AGSS09} solar mixture.}
\label{diff_kap_solmix}
\end{center}
\end{figure}
%-----------------------------------------------

Thorough comparisons of \textsl{OP} and \textsl{OPAL} opacities \citep[see for instance][]{2005MNRAS.360..458B} have shown
a very good agreement between the two groups with differences in opacities which do not exceed
$5-10$ per cent (Fig.~\ref{OPOPAL}) except locally in the so-called $Z$-bump\footnote{The $Z$-bump corresponds to the sudden increase of opacity related 
to the ionisation of heavy elements like iron.}, where differences can still reach $30$ per cent. 

\subsubsection{Impact on age resulting directly or indirectly from opacities}

Opacities affect stellar age-dating in different manners. First, the uncertainties and 
shortcomings in the opacity calculation directly impact the age-dating. Furthermore, 
since the net opacities in a model depend on the chemical composition adopted in the modelling, any
uncertainty on the abundances indirectly impacts the age-dating through opacity changes.
We examine below the effect on age of changes of opacity resulting from different sources.

\begin{itemize}
\item {\sl Uncertainty in the radiative opacity.}

In Fig.~\ref{diff_kap_solmix}, we show that, in case of the Rosseland opacity were 10 per cent higher, 
stellar models ages at turn-off would be increased by $6$ to $14$ per cent. This is due to the fact 
that larger opacity implies lower luminosity, and therefore higher lifetime (see Eqs.~\ref{homol_lum} and \ref{tnucMS}).

\item {\sl Change of opacity due to uncertainty on the solar mixture.}

As discussed in Sect.~\ref{solarmixture}, solar models based on the \textsl{AGSS09} solar
 mixture of heavy elements \citep[][]{2009ara&a..47..481a} do not reproduce
the helioseismic observations as well as models based on the canonical 
\textsl{GN93} mixture \citep{1993pavc.conf..205G} do. 
The \textsl{AGSS09} mixture is deficient in O and C, N, 
Ne, and Ar
with respect to the \textsl{GN93} mixture. For the \textsl{AGSS09} mixture, 
$(Z/X)_{\mathrm{\odot, AGSS09}}=0.0181$, while for the \textsl{GN93} mixture 
$(Z/X)_{\mathrm{\odot, GN93}}=0.0245$. 
As illustrated in Fig.~\ref{AGS05GN93}, below the convection zone of a calibrated solar model,
the opacity is $20$ per cent smaller when the \textsl{AGSS09} mixture is used instead of the \textsl{GN93} one.
%\citep[see e.g.][]{2004ApJ...614..464B}.

As a case study, we compare stellar models based on the two solar mixtures \textsl{GN93} and \textsl{AGSS09}
 and assuming the same $\Delta Y/\Delta Z$ value.
The smaller $(Z/X)$ in the \textsl{AGSS09} case implies a smaller value of $Z$ and $Y$, and a higher value of $X$ in these models.
As shown in Fig.\ref{diff_kap_solmix}, as a consequence of a higher value of $X$, 
the age at turn-off is higher in most models
(Eq.~\ref{tnucMS}). 

%-----------------------------------------------
\begin{figure}[!htbp]
\begin{center}
%\resizebox{0.5\hsize}{!}{\includegraphics{Figs/kappa_AGS05_GN93.png}}
%\resizebox{\hsize}{!}{\includegraphics{Figs/dkmix1.eps}\includegraphics{Figs/dkmix12.eps}}
%\caption{Opacities differences at fixed given structure and same value of $Z$ ($Z=0.02$) for a $1 M_\odot$ model 
%at solar age (left) and a $12 M_\odot$ model in the middle of the {MS} (right) calculated with the 
%{\small {\small GN93}} or the {\small AGS05} solar mixture. The boundaries of convective regions are indicated by vertical lines. After %\cite{2008CoAst.157..160M}}
%\resizebox{0.85\hsize}{!}{
\includegraphics[width=0.9\textwidth]{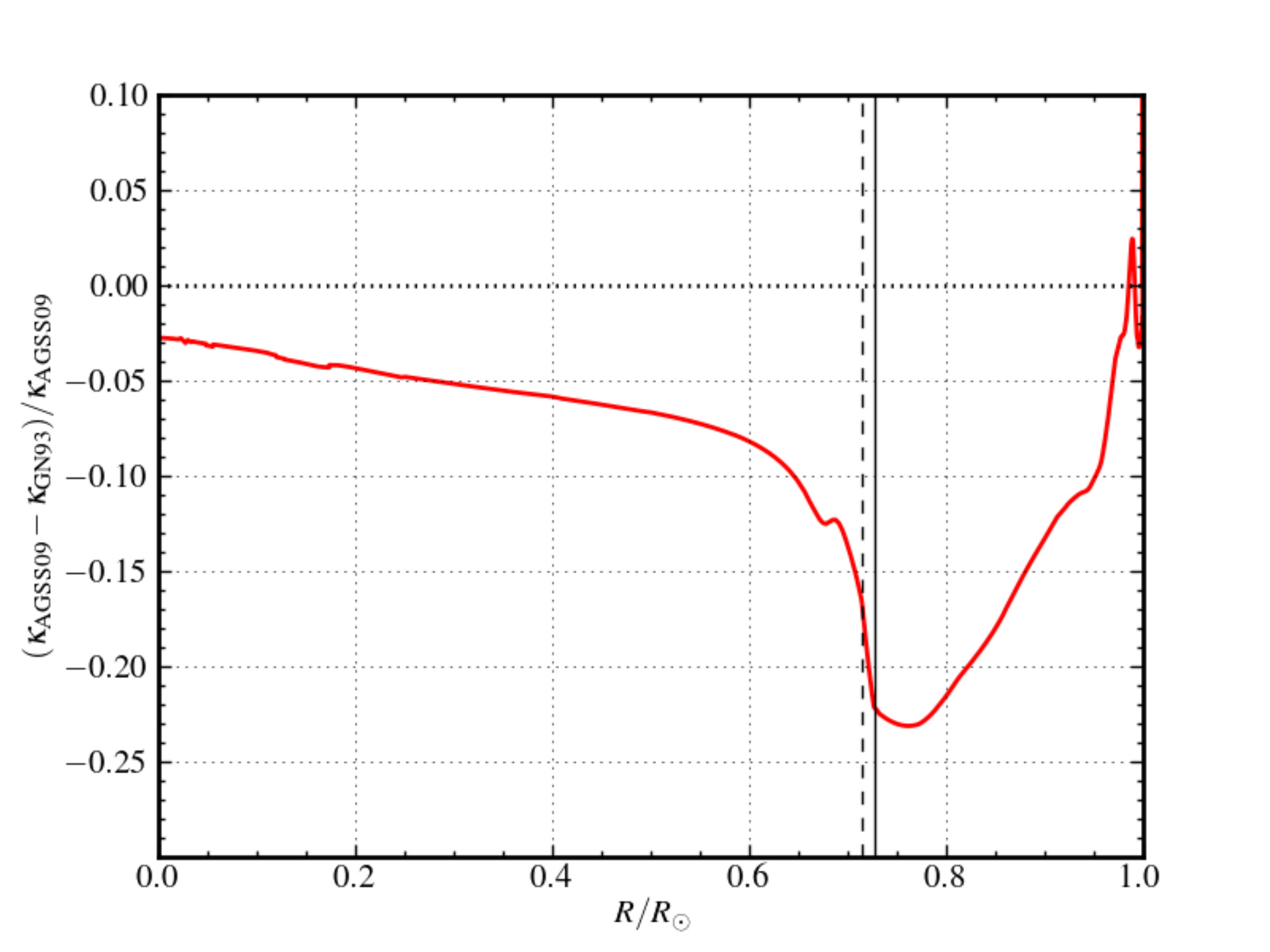}
\caption{Difference in the opacity as a function of radius between two calibrated solar models, calculated with the \texttt{cesam2k} code 
with either the \textsl{AGSS09} or the \textsl{GN93} solar mixture.
Vertical lines indicate the locus of the base of the convective envelope (\textsl{AGSS09}: continuous line, \textsl{GN93}: dashed line).
}
\label{AGS05GN93}
\end{center}
\end{figure}
%-----------------------------------------------

\item {\sl Change of opacity due to $\mathbf{\alpha}$-elements enhancement.}

The effect of an $\mathbf{\alpha}$-elements enhancement on the age of globular clusters at very low [Fe/H] has been studied 
in several papers \citep[see for instance][and references therein]{2002HiA....12..439V,2012ApJ...755...15V}.
As an illustration, Fig.~\ref{kappa_alpha} shows that an enrichment in oxygen or in the other $\mathbf{\alpha}$-elements produces 
cooler and fainter tracks in the HR diagram, which in turn induces a decrease of the age at turn-off. \citet{2012ApJ...755...15V}
have shown that the impact of oxygen is overwhelming in the age decrease, 
with at $\mathrm{[Fe/H]}=-2.27$ dex, a decrease of 1~\Gyr~per step of $+0.3$ dex in [O/Fe].

In Fig. \ref{diff_alpha_elts}, left panel, we have compared the turn-off age of stars of different masses with heavy elements 
mixtures of different [Fe/H] values ($-1.0$ and $0.0$ dex), and including either an
$\mathbf{\alpha}$-elements enhancement  of $[\alpha/\mathrm{Fe}]=0.4$ dex
 or a solar -non enhanced- value $[\alpha/\mathrm{Fe}]=0.0$ dex. We used the BaSTI grids 
of stellar evolutionary tracks calculated for a constant value of ${\Delta Y/\Delta Z}$
 \citep{2004apj...612..168p}.
 There is a general decrease in age with a maximum of 20 per cent for $\mathrm{[Fe/H]}=0.$ (solar) and 5 per cent at 
$\mathrm{[Fe/H]}=-1.$ dex. The models enriched in $\alpha$-elements have a higher luminosity and the same initial hydrogen abundance,
which turns into a smaller age. 

%-----------------------------------------------
\begin{figure}[!htbp]
\begin{center}
\resizebox{0.85\hsize}{!}{\includegraphics{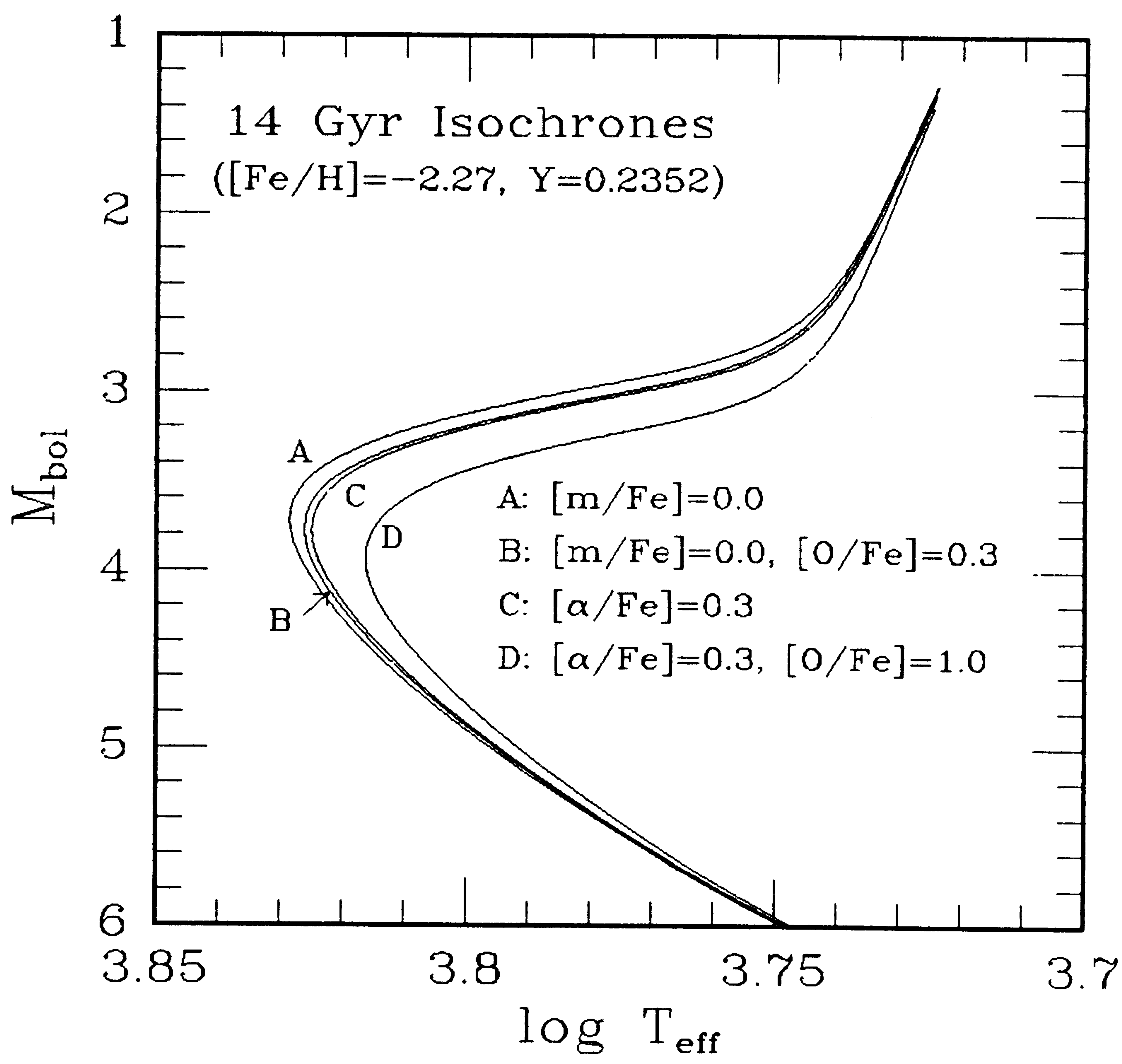}}
\caption{Effect of an enrichment of oxygen and $\mathbf{\alpha}$-elements on 
the MS turn-off. [From \citet{2001NewAR..45..577V}.]
}
\label{kappa_alpha}
\end{center}
\end{figure}
%-----------------------------------------------
\item {\sl Change of opacity due to uncertainty on the metallicity.}

In Fig.~\ref{diff_alpha_elts}, right panel, we show that in case of the error on the metallicity 
[Fe/H] were of $\pm 0.1$ dex the TO ages would differ by up to $8$ per cent.
A change of [Fe/H], at constant ${\Delta Y/\Delta Z}$, in a stellar model has two main competing effects: 
\textsl{(i)} the helium abundance and therefore the mean molecular weight $\mu$ increases 
which tends to increase the luminosity, and
\textsl{(ii)} the opacity increases which tends to reduce the luminosity. 
A smaller luminosity corresponds to an increase of age.
In low-mass stars, the bound-bound and bound-free opacities, which play an important role, increase
a lot when [Fe/H] increases. As a result, the luminosity is smaller and the TO age is higher.
In high mass stars, where free-free opacities and scattering are more important, the opacity is
less affected by an increase of metals. In these stars, due to the change of helium resulting from the [Fe/H]
increase, the luminosity is higher and the turn-off age decreases.

\item {\sl Change of opacity due to uncertainty on the He abundance or on ${\Delta Y/\Delta Z}$.}

To quantify the effect of changing the initial helium abundance of stellar models, we have compared the
ages at turn-off of models calculated with initial helium contents of $Y=0.25, 0.28$, and $0.31$. 
We find that a decrease of $Y$ from 0.28 to 0.25 induces a decrease of the turn-off age in the range 10 to 35 per cent for the interval
of mass we considered (see Fig.~\ref{diff_deltaFeH_Y}, left panel). This is due to the fact that increasing $Y$ also increases the mean molecular weight and in turn the luminosity. With a higher luminosity the age is smaller.
Similarly, an increase of the $\mathrm{\Delta Y/\Delta Z}$ ratio, from 2 to 5, 
produces a decrease of the turn-off age (see Fig.~\ref{diff_deltaFeH_Y}, right panel).

\end{itemize}

%-----------------------------------------------
%\begin{figure}
%\begin{center}
%\resizebox{0.5\hsize}{!}{\includegraphics{Figs/diffFeH0.pdf}}\resizebox{0.5\hsize}{!}{\includegraphics{Figs/diffFeHm1.pdf}}
%\caption{Same comparison as in Fig.~\ref{errorpp} for the effect of an enrichment of $\alpha$ elements with %$\mathrm{\alpha/Fe}=0.4$ dex at  $\mathrm{Fe/H}=0.$ (left) and $\mathrm{Fe/H}=-1.0$ (right). Models were 
%taken from the BaSTI grid \citep{2004apj...612..168p}.
%\orange{mass range max mass}
%}
%\label{diff_alpha_elts}
%\end{center}
%\end{figure}

%-----------------------------------------------
\begin{figure}[!hptb]
\begin{center}
%\resizebox{0.5\hsize}{!}{
\includegraphics[width=0.87\textwidth]{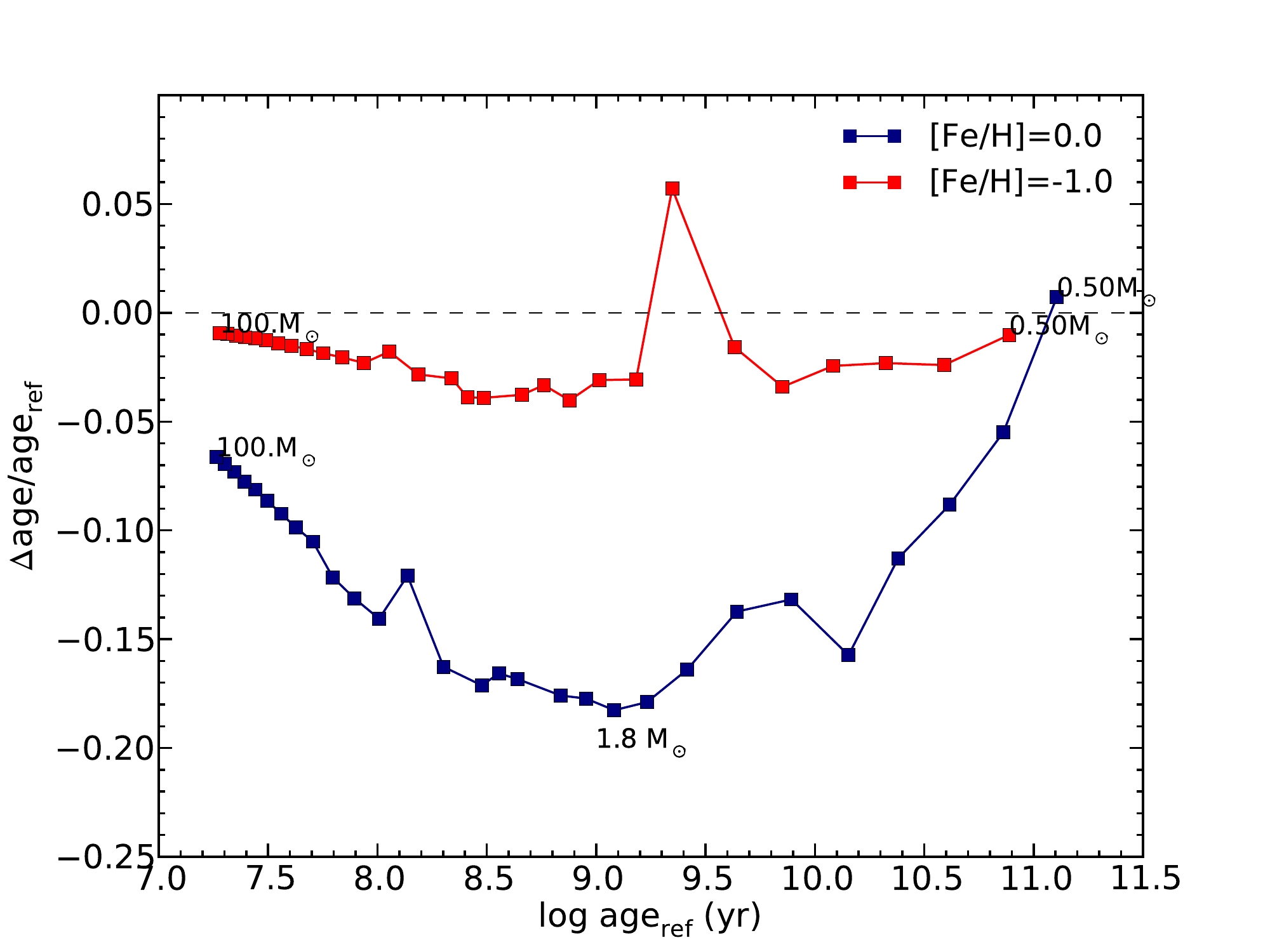}
%\resizebox{0.5\hsize}{!}{
\includegraphics[width=0.87\textwidth]{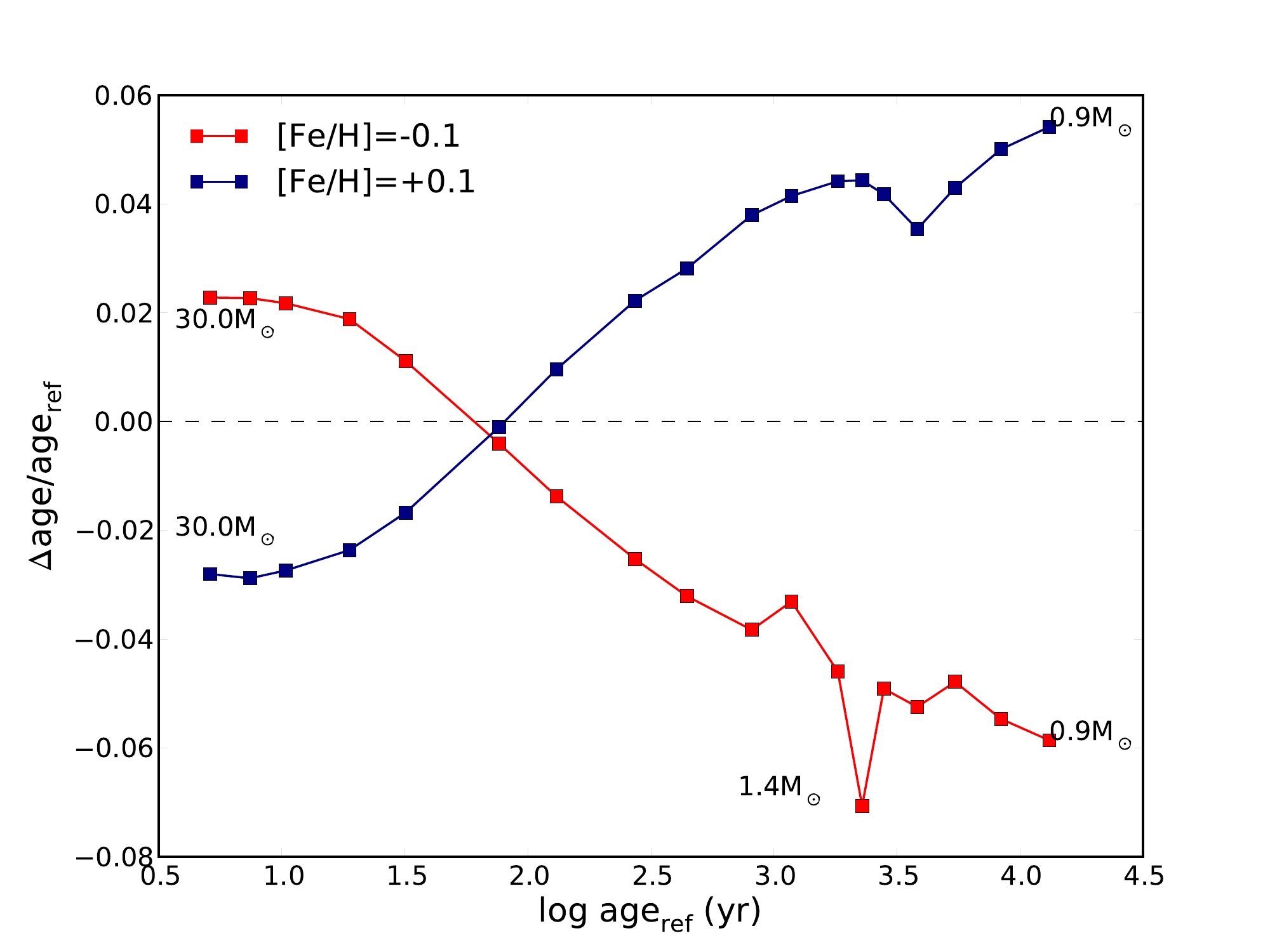}
\caption{Same comparison as in Fig.~\ref{errorpp}. {\sl Top:} the effect on TO age of an enrichment of $\alpha$-elements of $[\mathrm{\alpha/Fe}]=0.4$ dex at  $[\mathrm{Fe/H}]=0.$ (navy) and $[\mathrm{Fe/H}]=-1.0$ (red). Models were 
taken from the \textsl{BaSTI} grid \citep{2004apj...612..168p}. 
{\sl Bottom:} effect of a change of  $[\mathrm{Fe/H}]$ by $\pm 0.10$ dex.
}
\label{diff_alpha_elts}
\end{center}
\end{figure}
%-----------------------------------------------

\begin{figure}[!hptb]
\begin{center}
%\resizebox{0.5\hsize}{!}{
\includegraphics[width=0.87\textwidth]{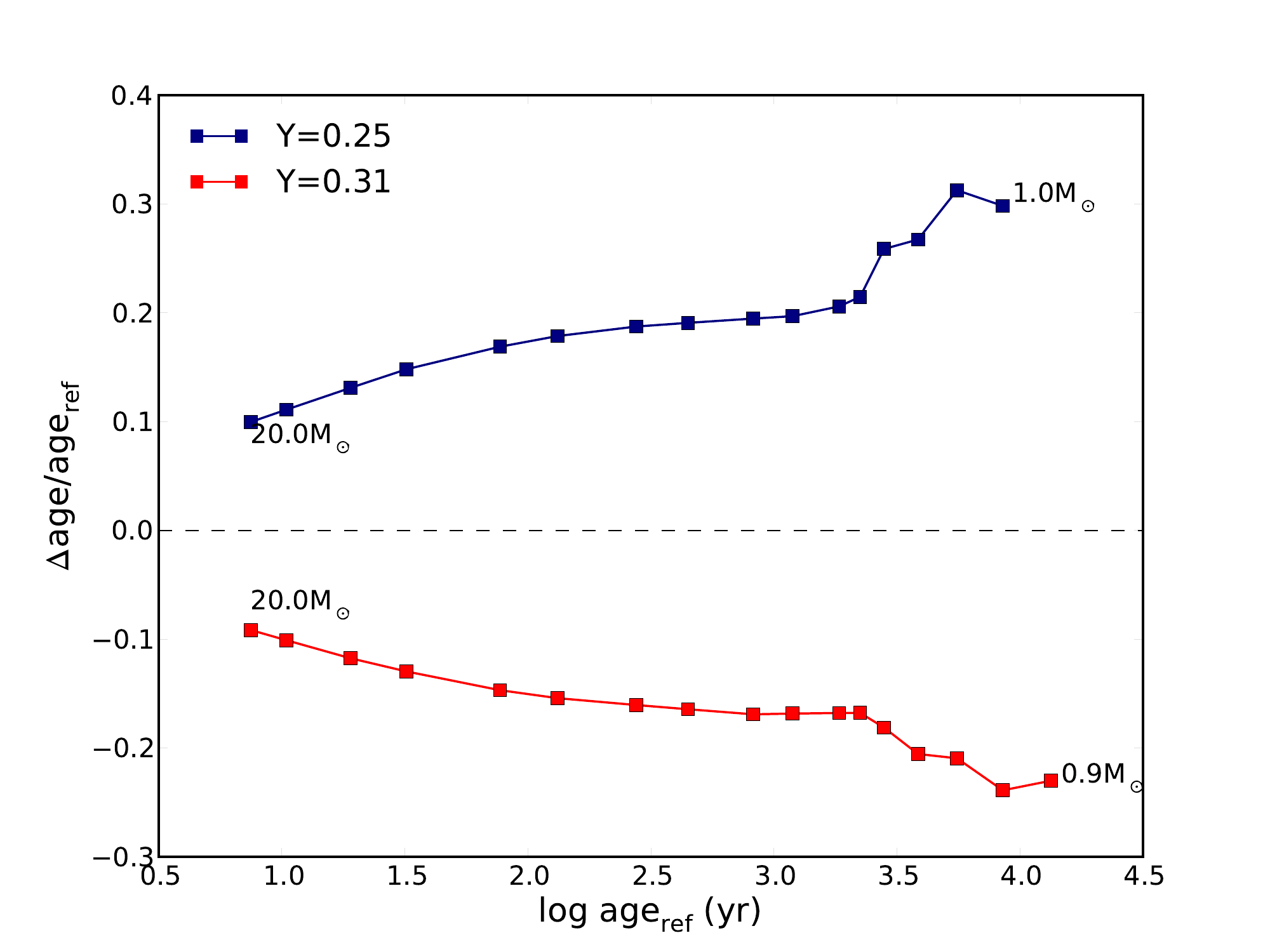}
%}\resizebox{0.5\hsize}{!}{
\includegraphics[width=0.87\textwidth]{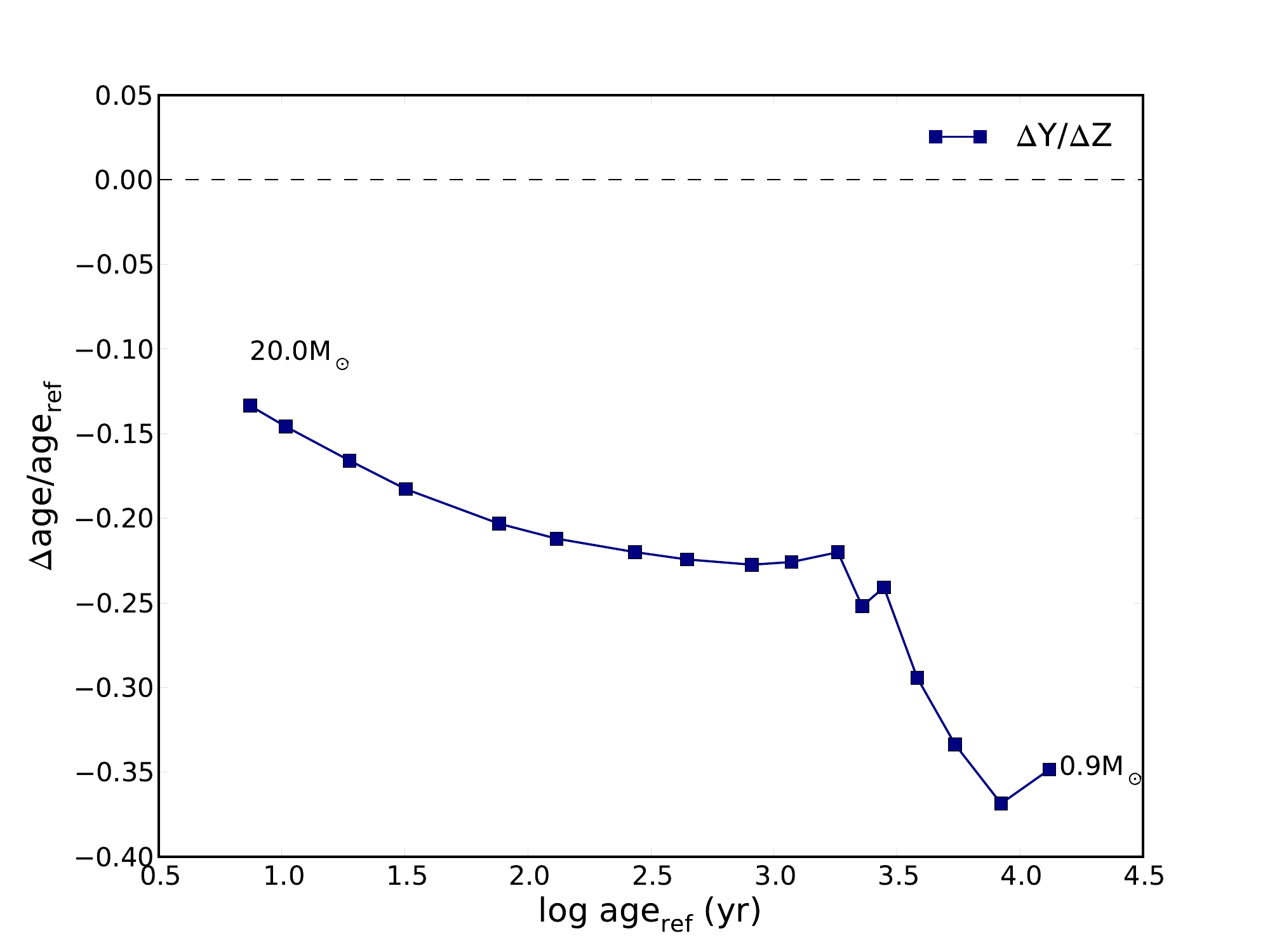}
\caption{Same comparison as in Fig.~\ref{errorpp}. {\sl Top:} effect of a change of $Y$ by 0.03 (with respect to a reference value $Y=0.28$).
 {\sl Bottom:}  effect of a change of $\Delta Y/\Delta Z$ from $2$ to $5$.
%\orange{mass range max mass}
}
\label{diff_deltaFeH_Y}
\end{center}
\end{figure}

%-----------------------------------------------
%\begin{figure}
%\begin{center}
%\resizebox{0.5\hsize}{!}{\includegraphics{Figs/diff_REF_DYDZ.pdf}}
%\caption{Same comparison as in Fig.~\ref{errorpp}. Effect of a change of $\Delta Y/\Delta Z$ from $2$ to $5$.
%\orange{mass range max mass}}
%\label{diff_DYDZ_diff}
%\end{center}
%\end{figure}

%\input eos.tex

\subsection{Equation of state}
\label{EoS}

\begin{figure}[!hptb]
\begin{center}
%\resizebox{0.5\hsize}{!}{
\includegraphics[width=0.87\textwidth]{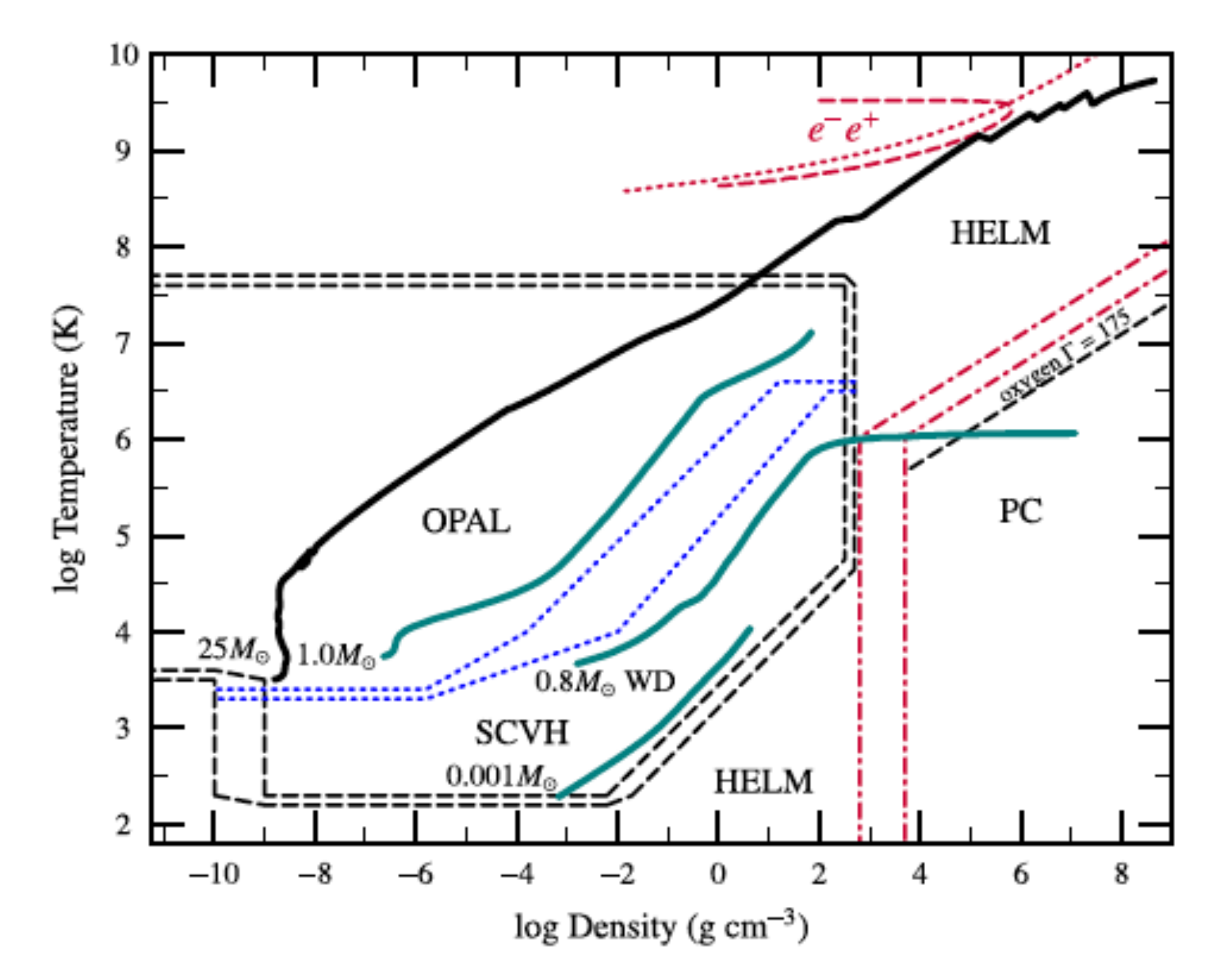}
%}\resizebox{0.5\hsize}{!}{
\includegraphics[width=0.7\textwidth]{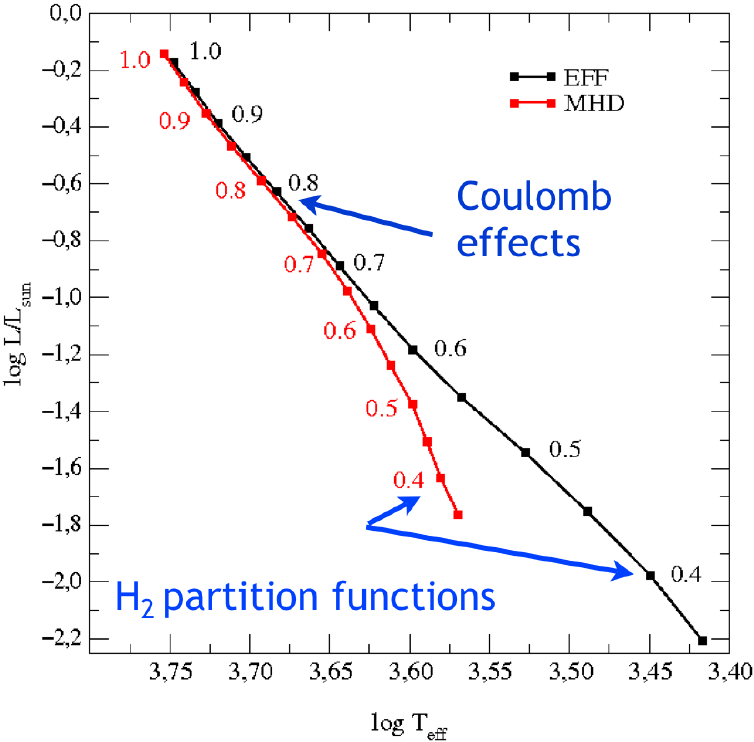}
%\resizebox{!}{0.4\vsize}{\includegraphics{Figs/EOS_MESA.png}}\resizebox{!}{0.4\vsize}{\includegraphics{Figs/EOS_HR.png}}
\caption{Equation of state.
{\sl Top:} the patchwork used in the \texttt{MESA} code to cover the whole stellar $\rho-T$
plane with available EoS, after \citet{2011ApJS..192....3P}. {\sl Bottom:} comparison of EFF 
\citep{1973A&A....23..325E} and MHD \citep{1999ApJ...526..451N} EoS.
It shows the impact on the ZAMS position of Coulomb effects and of $H_2$ partition functions included in MHD EoS,
after \citet{1988ESASP.286..661L}.
}
\label{FIGSEoS}
\end{center}
\end{figure}

Depending on the location of the star or region of a star in the temperature-density plane, different contributions to the
equation of state (EoS) have to be considered (top panel, Fig.~\ref{FIGSEoS}). Prior to 1990, the equation of state used to calculate
stellar models usually only included contributions from the ideal gas, degenerate electron gas, and radiation.
Then in the early 90s, a leap forward has been accomplished, in the context of the work dedicated to the improvement of opacity,
and more sophisticated EoS including the departures from ideal gas were made available. Both the \textsl{OPAL} EoS \citep{2002apj...576.1064r} and
the MHD EoS \citep[part of the \textsl{OP} Opacity Project,][]{1999ApJ...526..451N} include the Coulomb effects, volume effects, and $H_2$ partition functions.
We point out that during the last ten years, the numerical accuracy of these EoS has been improved.

Currently, stellar evolution codes use either the \textsl{OPAL05} or the MHD EoS, which have been 
compared by \citet{2006ApJ...646..560T} 
and by \citet{1999ApJ...518..985B}, this latter in the context of helioseismology.
When necessary, for the modelling of dense very low mass stars, the dedicated EoS of \citet{1995ApJS...99..713S} is used.
Furthermore, several packages of EoS tables make a patchwork of the previous EoS, in order to cover the temperature-density plane as widely as possible. 
This is the case of Irwin's FreeEos used in
\citet[][]{2003ApJ...588..862C}, and of  the EoS used in the \texttt{MESA} code \cite[]{2011ApJS..192....3P}, see the top panel, in Fig.~\ref{FIGSEoS}.

 Taking into account the non ideal effects in the EoS changes the location of  stellar models of low mass
  in the HR diagram
(Fig.~\ref{FIGSEoS}, bottom panel). More importantly the effects of the EoS can be probed by helioseismology through the
modification they imply for quantities as the sound speed or the adiabatic index $\Gamma_1$. 
Several EoS, among which the \textsl{OPAL} and MHD EoS,
have been discussed and probed in the context of helioseismology 
\citep[see for instance][and references therein]{1997ApJ...491..967G,1997A&A...322L...5B, 2001ApJ...546.1178G}.

The impact on the turn-off age of using two different EoS (\textsl{OPAL} and \textsl{FreeEOS}) has been evaluated by \citet{2013A&A...549A..50V} for a $0.9~M_\odot$
star with $Z=0.006$ (metal rich globular cluster in the Large Magellanic Cloud). Their Table D.1 shows that the difference in age is lower than 1 per cent.
Moreover, we considered the \textsl{OPAL01} and \textsl{OPAL05} versions of the \textsl{OPAL} EoS at solar metallicity and different stellar masses and found differences in the turn-off age
that are lower than 1.5 per cent.

%-----------------------------------------------
%-----------------------------------------------

%\input micdiff.tex

\subsection{Microscopic diffusion}
\label{micdiff}

%-----------------------------------------------
\begin{figure}[!htbp]
\begin{center}
%\resizebox{0.5\hsize}{!}{\includegraphics{Figs/micdiff_Y.png}}\resizebox{0.5\hsize}{!}{\includegraphics{Figs/micdiff_ZX.png}}
\resizebox{0.5\hsize}{!}{\includegraphics{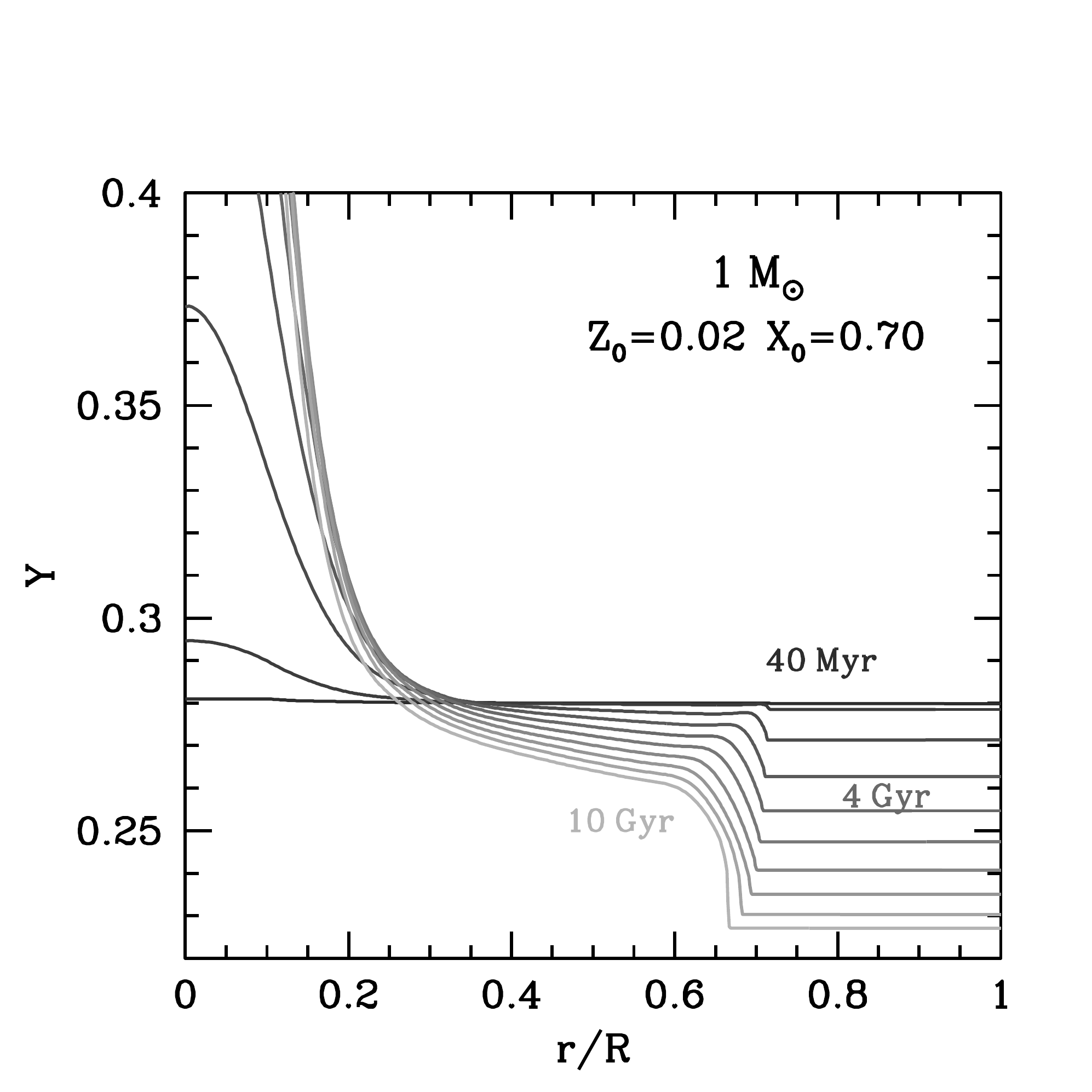}}\resizebox{0.5\hsize}{!}{\includegraphics{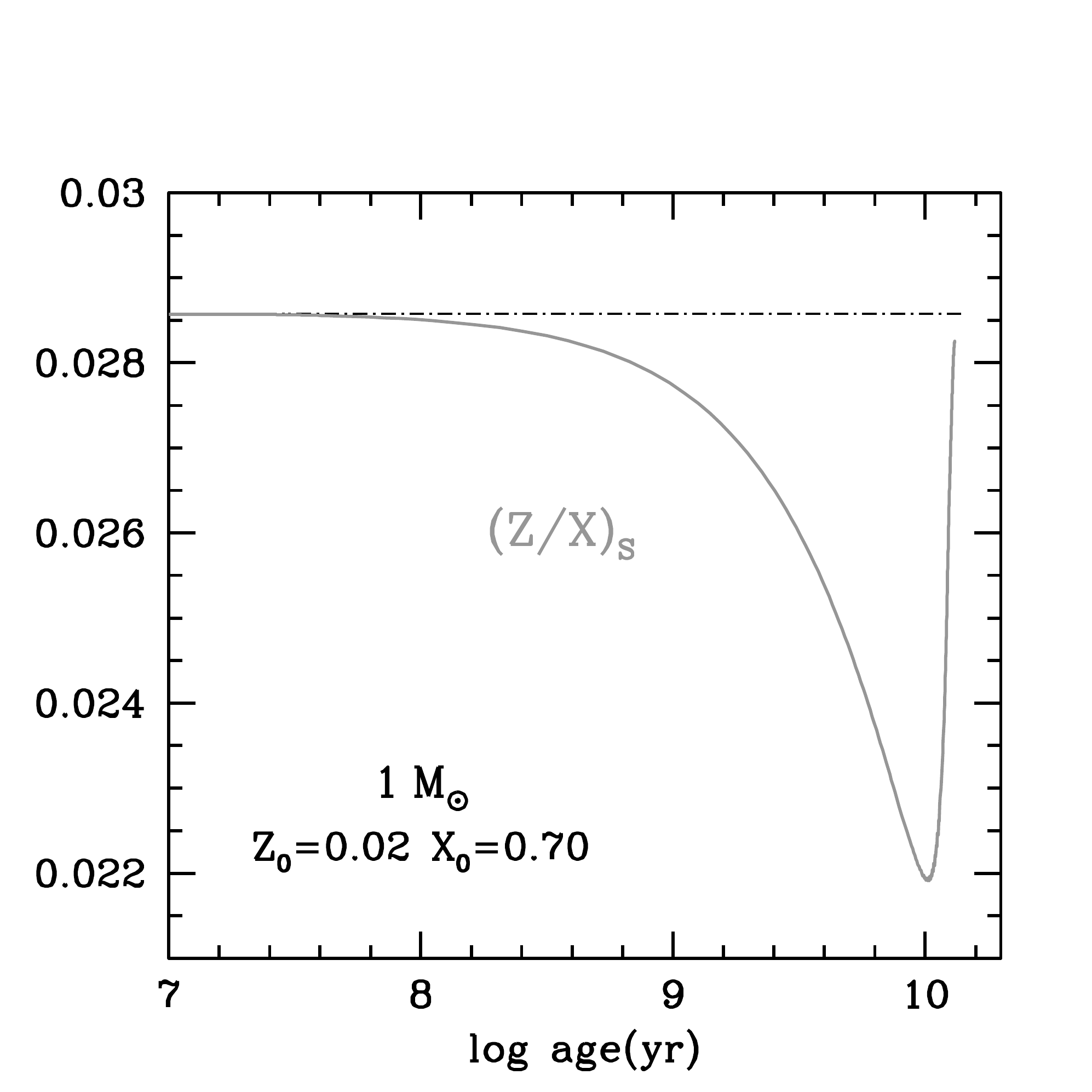}}
\caption{{\sl Left:} evolution of the helium profile in a solar model along the MS, the greyer, the older. 
As evolution proceeds, microscopic diffusion depletes helium at the surface, while nuclear reactions enrich the core in helium.  
{\sl Right:} evolution of the surface $Z/X$ ratio in a solar model along the evolution from the ZAMS to the RGB. 
Microscopic diffusion depletes the surface $Z/X$ during the MS, but the first dredge-up in the RGB brings $Z/X$ back to its initial
value. [From Lebreton and Montalb{\'a}n, EES2009, unpublished.]
}
\label{micYZ}
\end{center}
\end{figure}
%-----------------------------------------------

Microscopic (atomic) diffusion is the transport of chemical elements inside stars by 
different diffusion processes. In low mass K-G stars, transport by pressure (gravitational settling), 
temperature, and concentration
gradients are dominant processes and the diffusion velocity of a species $i$ with respect to protons 
reads, 
\begin{equation}
\label{allchap}
v_{i/p}= D_{i/p} \Bigg[ - \frac{1}{c_{i}}\frac{\partial c_i}{\partial r} 
+ \frac{1}{P} \left(2A_{i}-Z_{i}-1\right)\frac{\partial P}{\partial r} 
+ \frac{1}{T} \left(2.65 Z_{i}^2 +0.805(A_{i}-Z_{i})\right)\frac{\partial T}{\partial r}\Bigg],
\end{equation}
see \citet{1960ApJ...132..461A}. In this equation $c_i$ is the relative concentration of ion $i$ in 
the mixture.

In hotter A-F stars, radiative forces have to be taken into account to explain abundance anomalies
\citep{1970ApJ...160..641M,1998ApJ...504..559T,2007EAS....26...37A,2012A&A...546A.100T}.
This leads to add a term in Eq.~\ref{allchap}  of the form
\begin{equation}
v_{i/p}= D_{i/p} \Bigg[ \dotsm  
+ \frac{A_{i} m_p}{k T} (g_{i, \mathrm{rad}}-g)\Bigg],
\end{equation}
where $g$ is the local gravity and $g_{i, \mathrm{rad}}$ the radiative acceleration.

%-----------------------------------------------
\begin{figure}[!htbp]
\begin{center}
\resizebox{0.85\hsize}{!}{\includegraphics{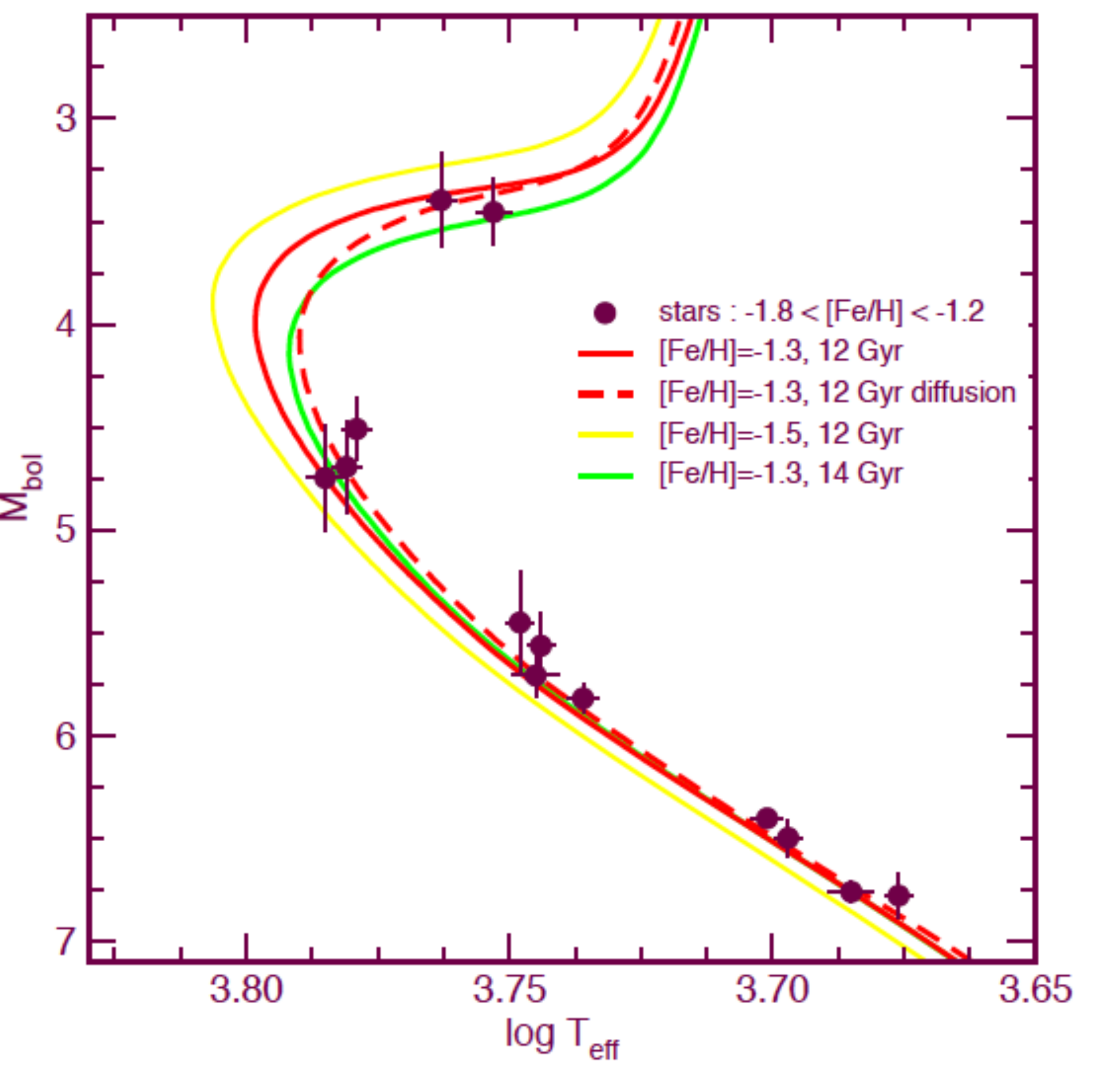}}
\caption{
Effects of microscopic diffusion and/or of a change of [Fe/H] on the age of halo stars observed by Hipparcos.
[From Lebreton,  Gaia Science sheet on 
{\sl `` Stellar Ages, Galactic Evolution \& the Age of the Universe''}, 
{\protect\url{http://www.cosmos.esa.int/web/gaia/science-topics}}.]
}
\label{micdiffhalo}
\end{center}
\end{figure}
%-----------------------------------------------

%-----------------------------------------------
\begin{figure}[!htb]
\begin{center}
%\resizebox{0.85\hsize}{!}{
\includegraphics[width=0.9\textwidth]{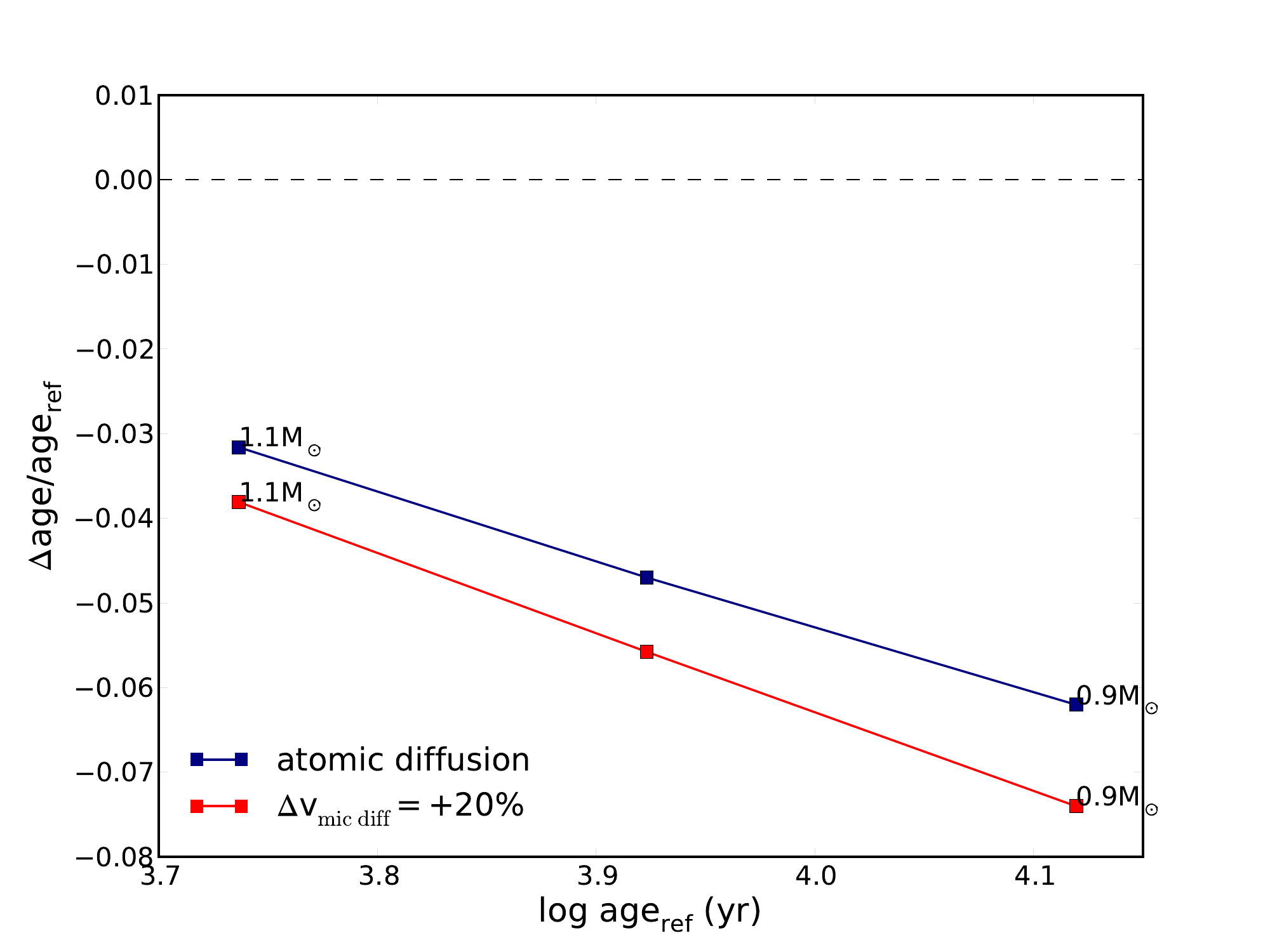}
\caption{
Same comparison as in Fig.~\ref{errorpp}. In blue the effect of microscopic diffusion, 
in red the effect of an increase of the microscopic diffusion velocity by 20 per cent. 
}
\label{velmic}
\end{center}
\end{figure}
%-----------------------------------------------

In low mass stars, on the MS,  atomic diffusion transports helium and metals towards the centre 
(and depletes them in the envelope), while hydrogen is pushed up from the centre towards the envelope (see 
Fig.~\ref{micYZ}).
On the other hand, in post SGB phase, during the first dredge up, the convection zone extends deep in the star and 
the resulting mixing kind of restores the initial abundance of metals at the surface, as confirmed by spectroscopic 
observations \citep[][]{2006Natur.442..657K}.

In models including microscopic diffusion, the envelope opacity increases due to the enhancement of hydrogen in the envelope. 
In turn, the envelope is deeper and the effective temperature is smaller as seen in the HR diagram of Fig.~\ref{micdiffhalo}. 

Atomic diffusion is a very slow process. \citet{1976ApJ...210..447M} estimated the surface abundances depletion time scale,
\begin{equation}
\frac{X_i(t)}{X_i(0)}= \exp\left(-\frac{t}{\tau}\right) \hspace{0.5cm} \mathrm{where} \hspace{0.5cm} \tau \propto \frac{M_\mathrm{CE}}{T_\mathrm{BCE}^{\frac{2}{3}}} \quad ,
\end{equation}
and where $M_\mathrm{CE}$ is  the mass in the convective envelope and $T_\mathrm{BCE}$ is the temperature at its bottom.
Table~\ref{taudiff} lists the variation of $\tau$ with the mass of the star: the lower the stellar mass, the deeper the convective envelope, and the slower the process.

The increase of the central helium abundance (and therefore of the mean molecular weight), leads to an increase of the luminosity and therefore to a decrease of the duration of the MS. This is illustrated in Fig.~\ref{micdiffhalo}, where we show that atomic diffusion reduces the age at turn-off of low-mass stars by a few per cent. This has consequences for the age-dating of globular clusters, the ages of which are reduced by ${\sim}1$~\Gyr~when diffusion is accounted for.

\begin{table}
\setlength{\abovecaptionskip}{0pt}
\setlength{\belowcaptionskip}{10pt}
\begin{center}
\caption{Time scales for surface abundance depletion due to atomic diffusion.}
\label{taudiff}
\begin{tabular}{ccccc}
\hline\hline
$M (M_\odot)$ & 1.0  & 1.1  & 1.2 & 1.4 \\
\hline
$\tau$ (\Gyr) & 5.4  & 1.5  & 0.11 & 0.0043 \\
\hline
\end{tabular}
\end{center}
\end{table}

Furthermore, there are several possible formalisms to take atomic diffusion into account in stellar models, via
the first term expressing the variation of chemical composition in Eq.~\ref{dXidtconv} 
\citep[see e.g.][this latter provides coefficients for collisions]{1969fecg.book.....b, mp93, 1994ApJ...421..828T, 1986ApJS...61..177P}.
\citet{2007EAS....26...25T} showed that with these different formalisms the diffusion velocities may change by up to $20$ per cent. As a consequence, the effect on age-dating is reinforced if the diffusion velocities are higher 
(Fig.~\ref{velmic}).

Finally, the gravitational settling efficiency increases when mass increases because the convective zones are thinner.
As a result, for masses higher than $\gtrsim 1.2 M_\odot$, there is a rapid quasi-total depletion of helium and metals at 
the surface of those stars, which is not observed. To properly model these stars, it is necessary to account for radiative forces in the calculation. 
Up to now, only 
two stellar evolution codes include radiative accelerations, the Montr\'eal code \citep{ 2000ApJ...529..338R} and the 
\texttt{TGEC} code \citep{2012A&A...546A.100T}. Other codes use recipes to prevent the full depletion (turbulence, mass loss, rotation).
%#################################################################
\section{Impact of stellar hydrodynamics (macrophysics) uncertainties on stellar ages}
\label{macrophysics}

\subsection{Convection}
\label{conv}

Heat and chemical element transport  by convection play an important role in
stellar evolution.  When integrating the  1D  equations for stellar structure, 
one only needs to determine where the medium is convective and how  
the temperature gradient is modified  in the convective regions 
and their surrounding layers. The way these pieces of information are obtained 
is described in all text books of stellar structure and evolution. 
We provide here a brief overview. %without detailing the derivations. 

 \subsubsection{Onset of convection }
\begin{figure}[!htb]
\begin{center}
\resizebox{0.85\hsize}{!}{\includegraphics{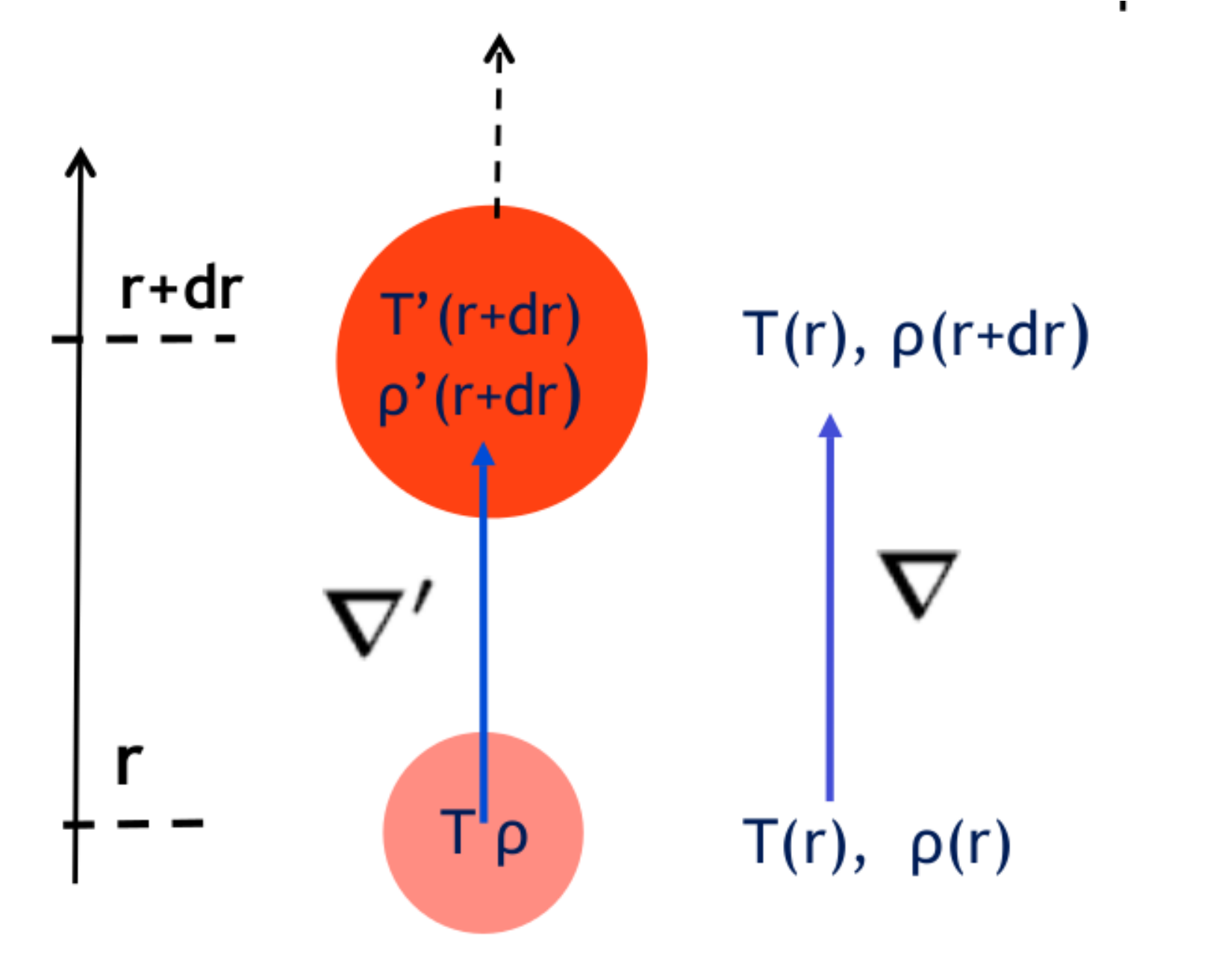}}
\caption{Convective instability scheme.}
\label{des1}
\end{center}
\end{figure}

  The onset of convection originates from a thermal
instability due to buoyancy. Convection takes place whenever 
the radiative gradient 
is not able  to transfer  the energy efficiently enough. 
Let us consider a gravitationally stratified medium with both temperature and density  
$T(z)$ and $\rho(z)$ decreasing outwards.  A blob 
 of gas, which is  slightly less dense (hotter) than the
surrounding medium, rises up from its equilibrium position  due to buoyancy. 
The Mach number  of the medium, $Ma$, \ie,  
 the ratio of the convective velocity over the sound speed $v/c_\mathrm{s}$,  
 is small ($Ma \sim 10^{-4}-0.3$ from the
 bottom to the top of the solar convective region). One then assumes  that pressure
equilibrium is maintained between the ascending bubble and its 
surroundings. Then at a given level, say $\delta r$,
above its initial position, the blob keeps on rising  
if it remains less dense than the surrounding medium (unstable
stratified medium). In contrast, gravity pulls the blob back if  
it becomes denser (cooler) than the surrounding environment 
(stably stratified medium, see Fig.~\ref{des1}). The rising blob 
(density $\rho^\prime$, temperature $T^\prime$)
remains less dense (hotter) {than} the cooling
medium ($\rho,T$) if the blob density decreases faster than that of the medium
 (its temperature decreases slower). This condition reads:
$$ \rho^\prime(r+\Delta r)< \rho(r+\Delta r)  \mathrm{\ \ or,\  equivalently\ \ }  T^\prime(r+\Delta r)> T(r+\Delta r) \quad . $$

 The condition for convective instability then is 
  $ \nabla^\prime < \nabla$ 
  with $\nabla=\mathrm{d}\log T/\mathrm{d}\log P$ for the medium and  $\nabla^\prime$ for  the blob. 

If the blob moves rapidly enough that its motion  can be assumed  adiabatic,  
$\nabla^\prime=\nabla_\mathrm{ad}=(\gamma-1)/\gamma$, 
where $\gamma=c_P/c_V$ is the ratio of the specific heat at constant pressure to the specific heat at
constant volume and $\gamma=5/3$ for an  ideal monoatomic gas. 
Then the condition for the onset of convection becomes
  $ \nabla_\mathrm{ad} < \nabla$. 

If the convection is inefficient,
 the temperature gradient of the medium 
remains nearly radiative hence $ \nabla \approx \nabla_\mathrm{rad}$. 
The reality lies in-between. The blob radiates energy during its motion, 
then $  \nabla^\prime>\nabla_\mathrm{ad} $. 
Convective heat transport decreases the temperature gradient of the medium,
 then $ \nabla < \nabla_\mathrm{rad}$. 
As a result,  the medium is usually characterized by 
\begin{equation} 
\label{gradall}
 \nabla_\mathrm{ad}<\nabla^\prime < \nabla  <\nabla_\mathrm{rad} \quad .
 \end{equation}
The Schwarzschild criterion for convective instability in a homogeneous medium,  can
  then be conveniently stated as 
\begin{equation}\label{schwar}
 \nabla_\mathrm{ad}  <\nabla_\mathrm{rad} \quad .
 \end{equation}
Both gradients are known at each level $r$,   
regions which are unstable against convection are then easily 
identified in 1-D stellar evolutionary codes, in the framework
of the mixing-length theory \citep{1932VeGoe...2..220B}.
%The determination of the
%extension of the associated chemically mixed region  is another story (see Sect.\ref{overshooting}). 
Furthermore, in presence of a $\mu$-gradient, the criterion for convective instability
becomes the Ledoux criterion \citep[\eg,][]{Kippenhahn13}.

%-----------------------------------------------
%-----------------------------------------------

 \subsubsection{Location of  convective regions in 1-D stellar models }

The next issue then is where in a stellar model the  criterion  for convective instability 
(Eq.~\ref{schwar}) is satisfied. Let us first consider the radiative gradient, 
$\nabla_\mathrm{rad}$ (defined in Eq.\ref{radiation}). It can also be written as 
$$\nabla_\mathrm{rad}= \frac{H_P}{KT} F_\mathrm{tot} \; \propto \; \kappa \, \frac{L}{m} \quad ,$$ 
where $F_\mathrm{tot} $ is the total heat flux,  $H_P$ the pressure scale height, $L$ the
luminosity, and $K$ the thermal  conductivity, all at level $r$. 
In stars, convection can take place in the envelope, in the core, and in intermediate layers,
mainly because $\kappa$ or $L/m$ become large.

\begin{itemize}
\item {\sl Convective envelopes:} for a given star (with a given total luminosity  $L$
 and total mass $M$),  the opacity $\kappa$ in the envelope 
(where the mass $m$ at level $r$ is $m \sim M $) is large 
 due the presence of  $H^{-}$ ions in partial ionisation zones, hence the radiative gradient is large. 
Moreover, in these regions the value of $\gamma $ drops close to one  and therefore 
$\nabla_\mathrm{ad}$ becomes small. Both properties favour
the onset of convection.
  As a consequence, cool stars do develop
 extended  convective outer layers. The effective temperature, or the radius, depend on
 the properties of outer convection.
Hence uncertainties in the description of
inefficient convection in stellar models may affect the shape of the isochrones and 
 accordingly the ages deduced from isochrone fitting.

\item {\sl Convective  cores:} the ratio $L/m$  is quite large 
in the central regions 
when the nuclear energy rate strongly depends on
 temperature. This happens  when the CNO cycle  significantly operates, 
that is for MS stars of  
mass larger than about $1.1 ~ M_\odot$, depending on the
 chemical composition. 
Uncertainties in the location of the boundary of the central mixed region 
(see Sect.~\ref{overshooting})
involve variations of the lifetime of the central hydrogen burning phase and  
directly affect the ages. 
\end{itemize}

\subsubsection{Efficiency of convection }

The properties of stellar convection are governed by the competition 
between several characteristic time scales: \textsl{(i)} the buoyancy driving time scale 
$t_\mathrm{b}= 2 \pi/N_\mathrm{BV}$, where $N_\mathrm{BV}$, 
the Brunt-V\"{a}is\"{a}l\"{a}  (BV) frequency, 
is the frequency associated to the oscillation of a perturbed 
parcel of a gravitationally stratified fluid (see lecture 2), \textsl{(ii)} the viscous time scale 
$t_\mathrm{visc}=2 \pi/ \omega_{visc}$, 
and \textsl{(iii)} the radiative time scale $t_\mathrm{rad}=2 \pi/\omega_\mathrm{rad}$.
These latter quantities read,
\begin{equation}
\omega_\mathrm{visc}= \frac{\nu}{\ell^2} \quad \, ; \quad \quad 
\omega_\mathrm{rad}= \frac{K}{\ell^2}i \quad ,
\end{equation}
where 
$\nu $ is the kinematic viscosity and $\ell$ a characteristic length scale.
The  Rayleigh number  measures the strength of the instability
$$ Ra = \frac{N_\mathrm{BV}^2}{\omega_\mathrm{visc}\; \omega_\mathrm{rad}} \; . $$
Both viscous and radiative effects inhibit the development of
 the instability.  In stellar conditions such as in the Sun, 
the Rayleigh number ($Ra_\mathrm{\odot} \sim 10^{23}$) is huge and the instability leads to 
a strong driving.
  
On the other hand, the Prandtl number of the 
fluid  measures the ratio of the thermal and viscous time scales:
$$ Pr = \frac{t_\mathrm{rad}}{t_\mathrm{visc}} = \frac{\omega_\mathrm{visc}}{\omega_\mathrm{rad}} \quad .$$
In stellar conditions, $Pr$ is small ($Pr_\mathrm{\odot} \sim 10^{-10}-10^{-3}$) 
   and  the fluid can be considered as inviscid. 
As a consequence of the inviscid nature and of the
 large scales involved, the Reynolds number $Re$, is quite large. 
 With the characteristic length scale $L$,  say $L \sim 10^6-10^9$ m,
and velocity $U$, say  $U \sim 10^2-10^3 ~\mathrm{m \, s^{-1}}$,
the solar Reynolds number is  $$Re = \frac{UL}{\nu} \sim  10^{12}- 10^{14},$$ 
much larger than the critical number ($\approx~2300$) beyond which turbulence sets in.
Stellar convection is highly turbulent with a wide range 
of spatial scales involved  \citep[\eg,][]{2009LNP...756...49K}. %,2009LNP...756..107C}.

%($1\ \mathrm{Mm}$ - $1\ \mathrm{cm}$) 

As shown by 3-D simulations by \eg~\citet{1998ApJ...499..914S}, the
convective motions in stellar envelopes show  narrow 
cool descending plumes and hot rising bubbles, both types of motions penetrating
in the adjacent stably stratified layers. In 1-D stellar models, 
 however, the description of convection, the ``mixing length theory'' (MLT),  
 is based on  a very simple picture. It 
assumes that a blob which is less dense than the surrounding medium rises up  to a  level where 
it dissolves giving back its  energy excess to the
medium. The distance $d_\mathrm{MLT}$ is the mixing length 
and is usually taken as a fraction of the pressure scale height, $H_P$, that is 
  $d_\mathrm{MLT} = \alpha_\mathrm{MLT} \times H_P$. 

When convection takes place somewhere, its impact depends on its efficiency.  
 A measure of the efficiency $S$ is given by the ratio of the thermal time scale 
to the buoyancy time scale.
 $S$ is also the product of the Rayleigh number by the Prandtl number
\citep{1996apj...473..550c}. The quantity
\begin{equation}
\label{eff}
 S= \frac{t_\mathrm{rad}}{t_\mathrm{b}}\; = \; Ra \, \times \, Pr \; ,
 \end{equation}
measures the ability of convection to transport heat. 
The efficiency can then be either large or small. 
An inefficient convection $S\ll1$ however does not mean that the convective flux is small.
   
In stellar envelopes, convection is inefficient ($S\ll1$)
at the top of the convection zone. The
superadiabatic gradient defined as the difference between 
the actual gradient and the adiabatic 
one   is proportional to the squared mixing-length parameter: 
$$\left(\nabla-\nabla_\mathrm{ad}\right)\propto \alpha_\mathrm{MLT}^2 $$

In convective cores, convection is quite efficient and the actual gradient is close to adiabatic
whatever the mixing-length value. On the other hand, non-local effects generate overshooting beyond the
Schwarzschild limit. This adds a new free parameter, the overshooting distance 
$d_\mathrm{ov}=\alpha_\mathrm{ov} \times H_P$. Therefore, the implementation of turbulent convective transport in  
1-D stellar codes  remains one major weak point of stellar
evolution theory.  Over the years, many tentative works  have aimed at extending the  phenomenological
description  proposed by \citet{1958za.....46..108b} after the work of 
\citet{1925Prandtl}. 
Despite these efforts, the MTL including its  improved variants (see below) 
 basically remains  in use in the current stellar evolutionary codes.
 
\subsubsection{Convective  gradient}
In stellar convective regions, one needs to determine the actual temperature gradient, $\nabla$. 
It is derived from the total flux conservation law 
  $F_\mathrm{tot}=F_\mathrm{rad} + F_\mathrm{conv}$, where the total
flux  is  known at each level $r$: 
$$F_\mathrm{tot}= \frac{L}{4\pi r^2} \; . $$
The radiative flux  depends on the unknown temperature gradient $$ F_\mathrm{rad}= -K ~ \nabla ~~~~;
~~~~K=\frac{4ac}{3}\frac{T^4}{\kappa \rho}\frac{1}{H_P} \; .$$
One also needs the convective flux $F_\mathrm{conv}$. Assuming pressure
equilibrium, the convective flux is identified with the enthalpy flux and is therefore 
  defined as $$ F_\mathrm{conv} = \rho~ c_p ~ <w~ \theta> \; , $$
where  
$\theta$ are the temperature deviations from the horizontal mean $T$, and $\omega$ the counterpart for
the vertical velocity. 
   Because of the turbulent nature of the convection, one must
 compute  an ensemble average of the statistical 
 fluctuations of velocity $w$ and temperature $\theta$
with respect to a static background. In order for turbulent convection implementation  to be tractable 
in a 1-D stellar code, several assumptions and approximations, listed belowr, have to be made.

\begin{itemize}
\item Convection can be assumed to be
incompressible because the Mach numbers are small ($Ma \ll1$).  
%The continuity equation  becomes $\nabla \cdot \vec v=0$. 
Actually, pressure and density fluctuations with respect to the averaged background 
are neglected except for the density fluctuation entering
the source of  convective instability, \ie,  the  buoyancy acceleration  $\delta \rho\times g$.

\item The second assumption is that of a stationary flow. 
All quantities are considered as statistical averages. 
This is justified by the fact that the dynamical time scales of
relevance for turbulent convection are much shorter than the evolutionary time
scale for MS stars.

\item The turbulence is assumed to be isotropic and homogeneous. 
The relevant quantities are horizontally averaged.  A better description 
ought to include the horizontal heat exchange between rising, hotter  blobs and cooler, descending plumes.  

\item  $\langle w \, \theta \rangle \; = \; {v} \, \Delta T$,  that is the product of  mean velocity times 
 mean temperature difference between the blob and the surrounding at the time of
 dissolution.
 
\item  Convection in 1-D stellar models is local, 
that is the convective gradient at a given level $r$ 
is written in terms of quantities defined at the same level. 
This is a strong assumption, which is
not justified. One consequence is a non-physical
treatment of the boundaries between radiative and
convective regions. They  are imposed
by the Schwarzschild criterion, which does not allow  for convective penetration into the
neighbouring radiative layers.

\item  The motion of the blob is assumed to stop  after some travel distance $l$ where 
it gives back  its heat excess to the medium. 
The distance is taken to be some fraction of $H_P$. 
This fraction is a free parameter, which
makes the formulation non predictable.  
In the solar case, this distance is derived from
a calibration process because the solar model must match its independently known mass, 
luminosity, and radius, at its current age. The value however depends on the physical inputs used
to build the solar model. There is no reason that the same value applies to another star with a
different mass, chemical composition, and age.  Actually, 3-D numerical simulations of
stellar envelopes show that the mixing length should vary across the HR diagram
 \citep[\eg,][]{2014arXiv1403.1062M}.
This is confirmed by seismic
studies of a few stars \citep[see \eg,][]{2005A&A...441..615M}.

\item   Turbulent pressure, turbulent kinetic energy are  discarded. 3-D simulations 
however show that they are not negligible \citep{1999A&A...351..689R,2003MNRAS.340..923R,2004IAUS..224..155T}.
\end{itemize}

With the above assumptions, using conservation of the flux and of the energy, 
it is possible to  derive a local relation between
the convective flux and the superadiatic gradient ($\nabla - \nabla_\mathrm{ad}$) such that 
$$F_\mathrm{conv}= -K_1 ~ (\nabla - \nabla_\mathrm{ad})  ~ \phi(S) \, , $$
where $\phi$ is a function of the efficiency $S$ (Eq.~\ref{eff}), 
and $K_1$ depends on the
equilibrium stratification properties.

In the formulation of \citet{1958za.....46..108b}, the heat is assumed to be  
transported by one eddy-size blobs (\ie, 
 corresponding to one single  spatial turbulent scale). 
Although this is an unjustified assumption, the resulting formulation 
 was and still is the one
implemented in  most 1-D stellar codes to compute the temperature gradient in regions of superadiabatic
(\ie,  inefficient) convection.
An improved formulation by \citet[][hereafter \textsl{CM}]{1991ApJ...370..295C} and \citet[][hereafter \textsl{CGM}]{1996apj...473..550c}  
takes into account the multi-spatial scale nature of stellar convection (Full Spectrum of Turbulence, FST). 
It has been implemented in a few stellar codes. As a result, the dependency of the  flux  on the efficiency, \ie,  the function 
$\phi(S)$, differs  between the two descriptions MLT and FST.

%-----------------------------------------------
\begin{figure}[!htb]
\begin{center}
\resizebox{0.85\hsize}{!}{\includegraphics{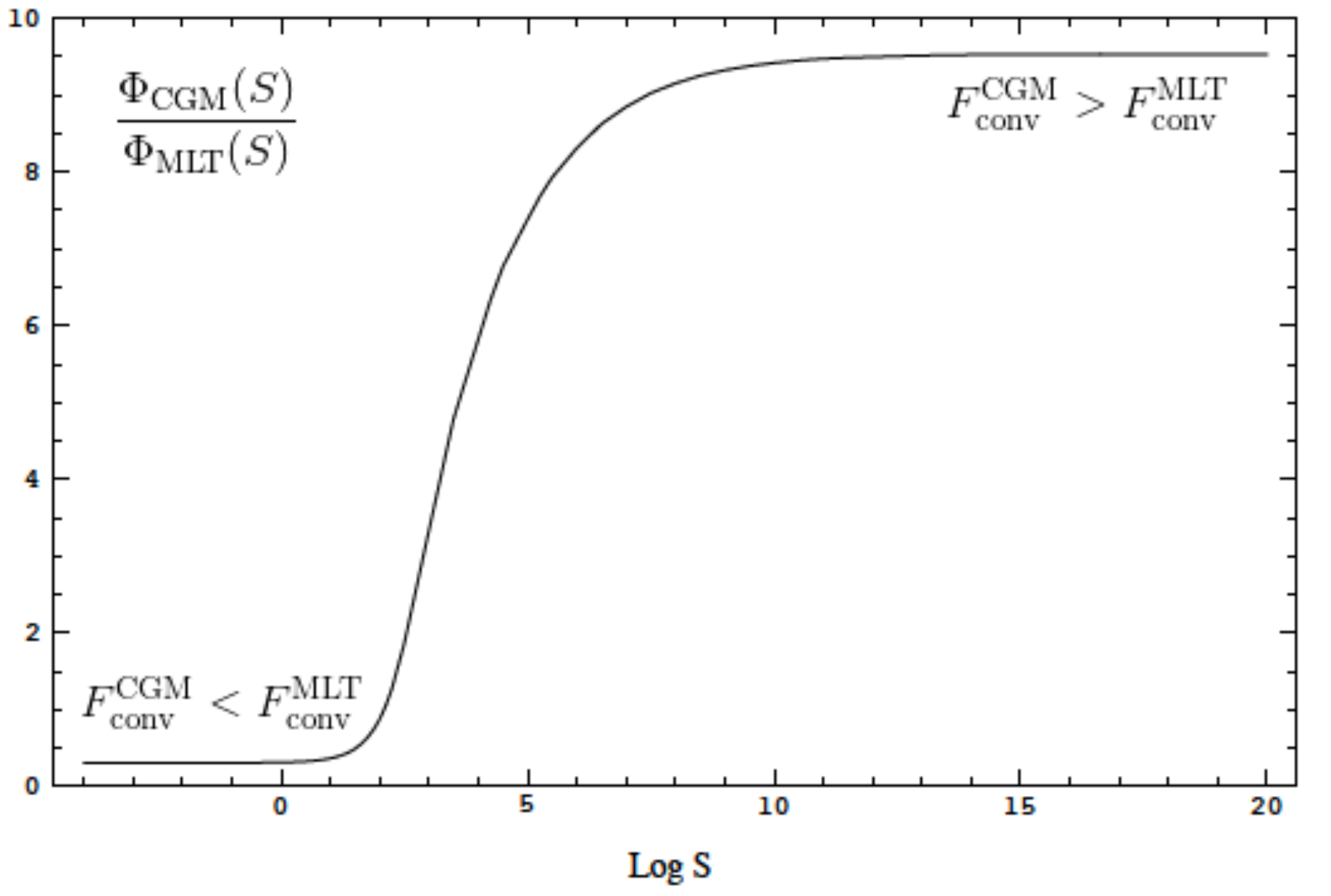}}
\caption{The ratio of the efficiency dependency of the \textsl{CGM}
approach to that of the MLT as a function of the logarithm of the efficiency
$S$ (Eq.~\ref{eff}). [From \citet{1996apj...473..550c}.]
}
\label{CM_S}
\end{center}
\end{figure}
%-----------------------------------------------

Figure \ref{CM_S} from \citet{1996apj...473..550c} shows the ratio of the efficiency dependency of the CGM
approach to that of the MLT as a function of the logarithm of the efficiency $S$. 
The departure from one is the  consequence of including  the whole
 spectrum of kinetic energy in the FST convective flux. The plot   shows that 
the MLT underestimates the convective flux for high efficiency and overestimates
it for small efficiency. 

For an efficient convection ($S\gg1$) , one has:
$$\phi_\mathrm{CM} (S) \sim \phi_\mathrm{CGM} (S)  \sim 10\  \phi_\mathrm{MLT} (S)~;~~ F_\mathrm{conv}^\mathrm{CM} \gg F_\mathrm{conv}^\mathrm{MLT} \; .$$

For an inefficient convection, ($S\ll1$) 
$$\phi_\mathrm{CM} (S) \sim \frac{1}{3} ~\phi_\mathrm{CGM} (S)~~; ~~\phi_\mathrm{CGM} (S)\sim \frac{1}{3}~ \phi_\mathrm{MLT} (S)~~;~~ F_\mathrm{conv}^\mathrm{CM} \ll F_\mathrm{conv}^\mathrm{MLT} \; .$$

%Comparisons with 3D simulations show that the FST represents an improvement over the MLT 
%\citep{2008IAUS..252...75L,2007AIPC..948..141T}. 
Comparisons with  observations show that the FST represents an improvement over the MLT 
\citep{1995A&A...302..271G,1995A&A...302..382M}.
However it suffers from 
 the same other limitations as the MLT, 
particularly this is a local  theory which depends on a free parameter, the mixing length.

\subsubsection{Convection in stellar envelopes}

 In the MLT description, the efficiency is given by 
$$\Gamma \propto \alpha_\mathrm{MLT}~ \kappa ~ \Bigl(\frac{\rho}{T}\Bigr)^2, $$
where $\Gamma$ is  related to the efficiency $S$ as  $\Gamma \approx 0.025 ~ S$.
Because of the opacity peak in partial H ionization regions,  
near the superadiabatic layer (SAL), the opacity  $\kappa $ is   large.  
As a consequence, $(\nabla_\mathrm{rad}-\nabla_\mathrm{ad}) \gg1$ and the instability generates
 a strong driving. But $\rho / T$ is small in the outer
layers and  $\Gamma\ll1$ despite the strong driving. 
Convection is therefore inefficient in envelopes of cool stars.   
The temperature gradient then is intermediate between 
the radiative and the adiabatic gradient.  
For small efficiency, 
the gradient writes
 \begin{equation}
 \label{grad}
 \nabla \; \approx \; \nabla_\mathrm{rad}-\frac{9}{4}~ \Gamma^2 \, \left(\nabla_\mathrm{rad}-\nabla_\mathrm{ad} \right) \; ,
 \end{equation}
see \eg~\citet{1958za.....46..108b,1992isa..book.....B}.
 The convective 
flux carries little energy
$$\frac{F_\mathrm{conv}}{F_\mathrm{tot}} \; \approx \; \frac{9}{4}~ \Gamma^2 \,  
\left(1-\frac{\nabla_\mathrm{ad}}{\nabla_\mathrm{rad}}\right) \approx  \frac{9}{4}~ \Gamma^2 \ll  1 \quad .$$  

%-----------------------------------------------
\begin{figure}[!htbp]
\begin{center}
\resizebox{0.5\hsize}{!}{\includegraphics{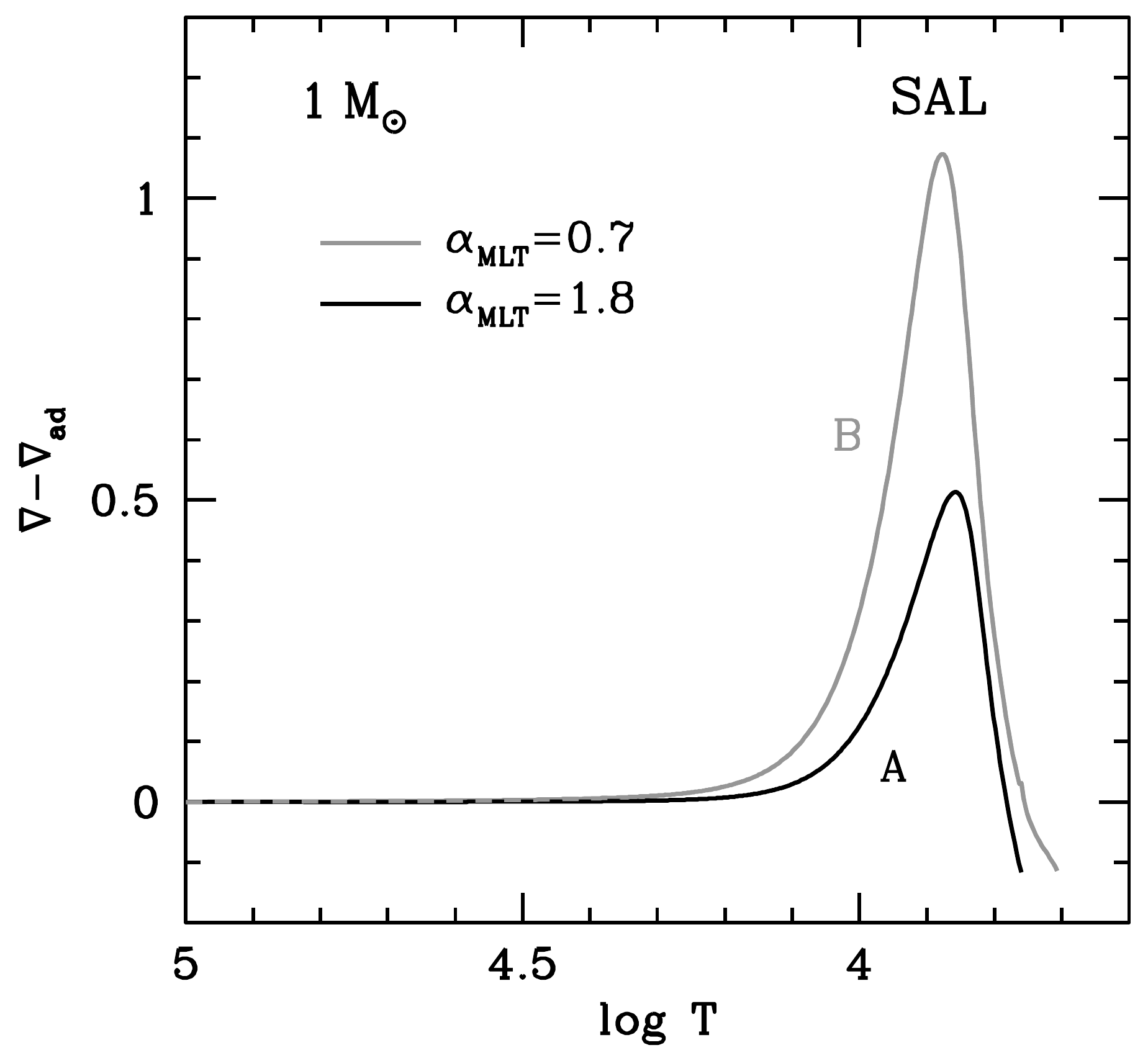}}\resizebox{0.5\hsize}{!}{\includegraphics{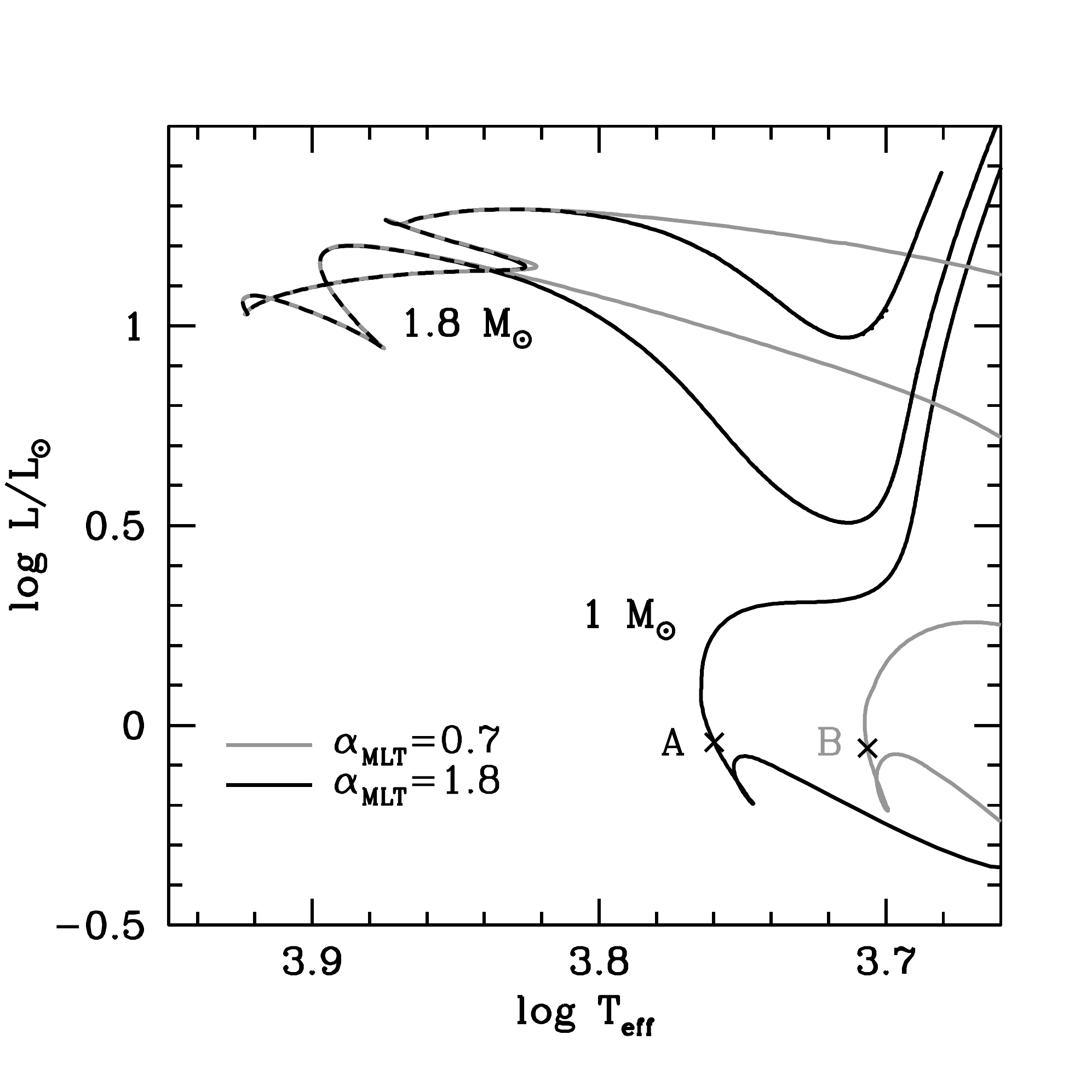}}
\caption{ {\sl {\sl Left:}} profile of 
the superadiabatic gradient ($\nabla -\nabla_\mathrm{ad}$)
 as a function of  temperature $\log T$  in the  vicinity 
of the hydrogen ionisation region ($\log T\sim 4.1$)  in  a $1.0\  M_\odot$ model 
for two values of the
$\alpha_\mathrm{MLT}$ parameter,  $\alpha_\mathrm{MLT} = 0.7$ and  $\alpha_\mathrm{MLT}=1.8$. 
{\sl Right:}  HR diagram showing evolutionary tracks for two values of the mass 
and mixing-length parameter. 
The grey track corresponds to  $\alpha_\mathrm{MLT} = 0.7$ and the
black track is for $\alpha_\mathrm{MLT} =1.8$.
}
\label{sal}
\end{center}
\end{figure}
%-----------------------------------------------

\paragraph{Impact of the mixing-length value on the temperature gradient.} 
Left panel of Fig.~\ref{sal}  shows the  run of 
the superadiabatic gradient $(\nabla -\nabla_\mathrm{ad})$ 
 as a function of the temperature $T$ 
  in the  outer layers of a $1.0\ M_\odot$ model for two values of the
$\alpha_\mathrm{MLT}$ parameter  $\alpha_\mathrm{MLT} = 0.7$ and  $1.8$. 
From Eq.\ \ref{grad}, one obtains: 
$$ \nabla -\nabla_\mathrm{ad} \; \approx \; \left(1-\frac{9}{4}~ \Gamma^2\right) ~  \; \left(\nabla_\mathrm{rad}-\nabla_\mathrm{ad}\right).$$
For a given stratification, the mixing-length value determines the magnitude
 of the efficiency and therefore the gradient:
 the larger $\alpha_\mathrm{MLT}$, the larger the convective  efficiency and 
 the farther  the gradient from  the
 radiative one. The convective efficiency is small but the driving is strong 
$(\nabla_\mathrm{rad}-\nabla_\mathrm{ad})\gg1$ hence the  actual gradient in presence of convection 
is much larger than the adiabatic one  and closer to, although significantly
smaller than the radiative one.  How smaller depends on the value one adopts for
the mixing-length parameter. 

Below the SAL (up to $r/R =0.9 $, 
$\log  T =4.6$ for the $1.0\ M_\odot$ model), the convection becomes  quite efficient, \ie,  $\Gamma >1$, 
 because $\rho/T$ becomes  large.  For a large efficiency, one has
$$ \nabla \approx \nabla_\mathrm{ad} +\frac{9}{4 \Gamma} ~ (\nabla_\mathrm{rad}-\nabla_\mathrm{ad}) \sim \nabla_\mathrm{ad}.$$
The larger $\alpha_\mathrm{MLT}$, the larger the efficiency and the smaller the actual 
gradient compared to the radiative one, the closer to the adiabatic one.

\paragraph{Impact of the mixing-length value on evolutionary tracks.} The right panel of Fig.\ \ref{sal} 
shows the effect of  increasing $\alpha_\mathrm{MLT}$ on 
the evolutionary tracks in the HR diagram. The impact depends on the effective temperature, 
hence on the mass at a given luminosity. 
Stars of $T_\mathrm{eff}$ hotter than $\sim 7000 ~  K$ are not impacted 
 as can be seen for the $1.8 M_\odot$ tracks.  The reason is that the outer
  convection region is too thin and not dense enough 
   to play an important role in the energy transport. 

The comparison of the $1.0\ M_\odot$ tracks 
on the other hand shows that  there is a clear shift of the higher $\alpha_\mathrm{MLT}$ track 
toward the  blue for  models at the same
evolutionary stage, \ie,  with the same value of central hydrogen abundance.
For a larger $\alpha_\mathrm{MLT}$, the star is  more compact, 
the radius is smaller, and $T_\mathrm{eff}$ is higher  at  the same luminosity.
Hence, for low mass stars ($T_\mathrm{eff} <7000K$), an increase of $\alpha_\mathrm{MLT}$  causes
 an increase of $T_\mathrm{eff}$.

%-----------------------------------------------
\begin{figure}[!htb]
\begin{center}
\includegraphics[width=0.9\textwidth]{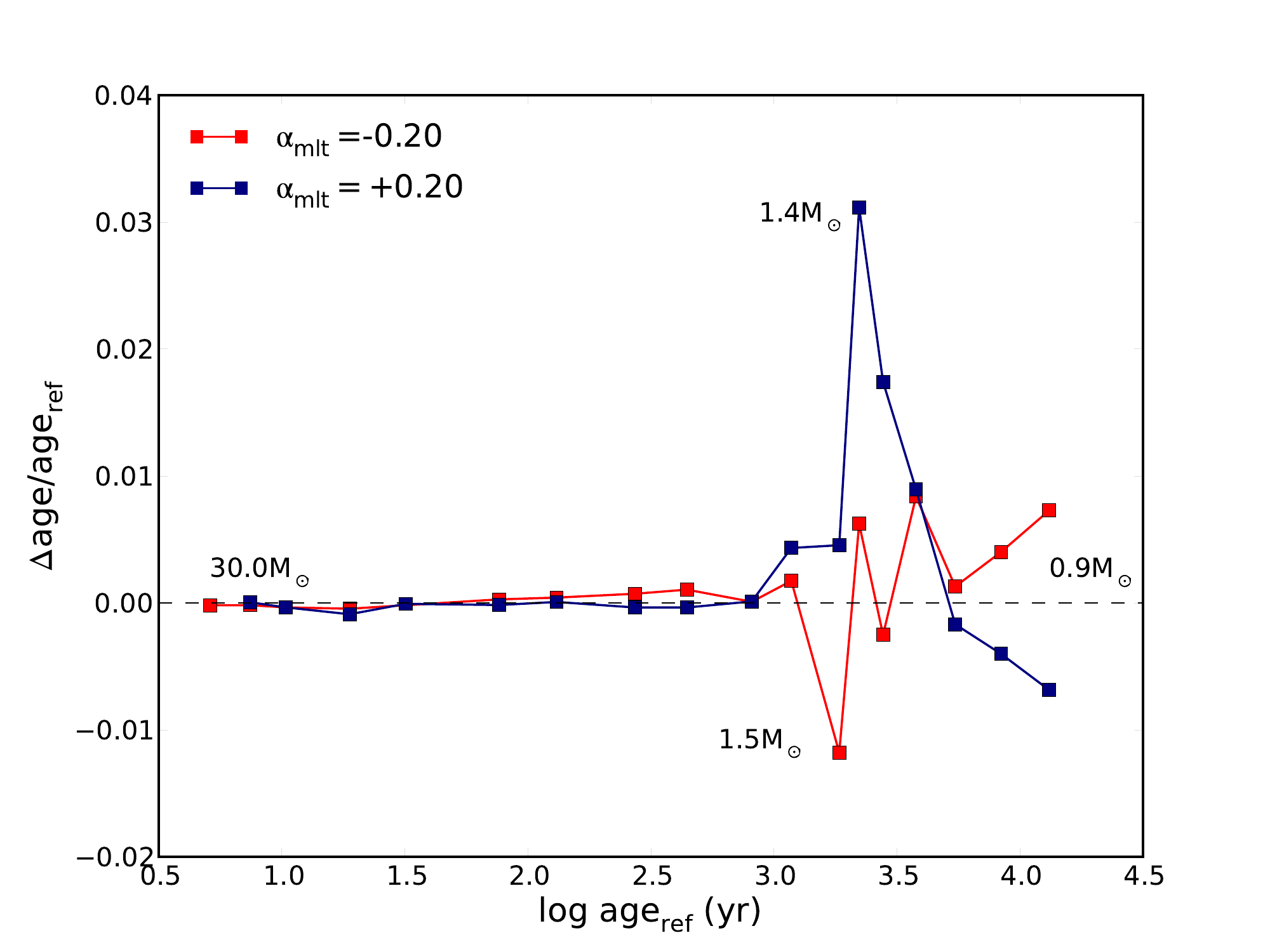}
\caption{Same comparison as in Fig.~\ref{errorpp}. In blue the comparison of TO ages of the
reference model calculated with the 
solar $\alpha_\mathrm{MLT}=1.76$ value and a model calculated with
$\alpha_\mathrm{MLT}=\alpha_\mathrm{MLT,\odot}+0.20$. In red, the reference model 
is compared to a
model calculated with $\alpha_\mathrm{MLT}=\alpha_\mathrm{MLT,\odot}-0.20$.
}
\label{ageMLT}
\end{center}
\end{figure}
%-----------------------------------------------

\paragraph{Uncertainty on the mixing-length value: impact on TO ages.} 
Figure \ref{ageMLT} compares the age at TO of a reference model with solar composition
and solar $\alpha_\mathrm{MLT}$ value  with the TO ages of models 
 computed assuming   $\alpha_\mathrm{MLT,\odot}\pm 0.20$ dex.
The maximum effect of a change of $\alpha_\mathrm{MLT}$ by $\pm 0.20$ dex
 on the TO age occurs in the mass range  $1.2 -1.5\ M_\odot$. 
A maximum difference of 3 per cent is found at $1.4\ M_\odot$.
The impact is therefore small. The small impact on isochrones has been shown by
 \citet{1999MNRAS.303..265C}.

%-----------------------------------------------
\begin{figure}[!hptb]
\begin{center}
%\resizebox{0.5\hsize}{!}{
\includegraphics[width=0.87\textwidth]{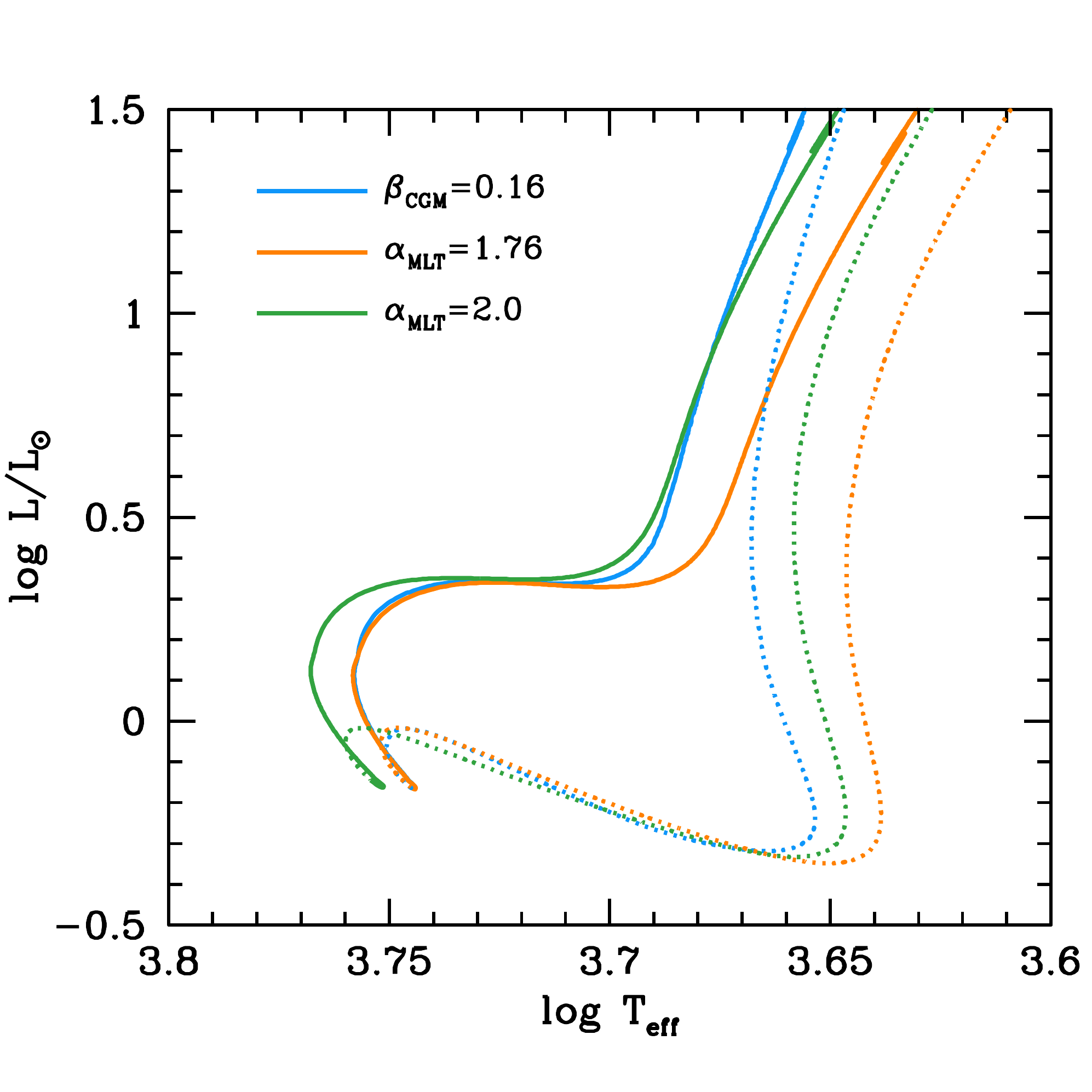}
%}\resizebox{0.5\hsize}{!}{
\includegraphics[width=0.87\textwidth]{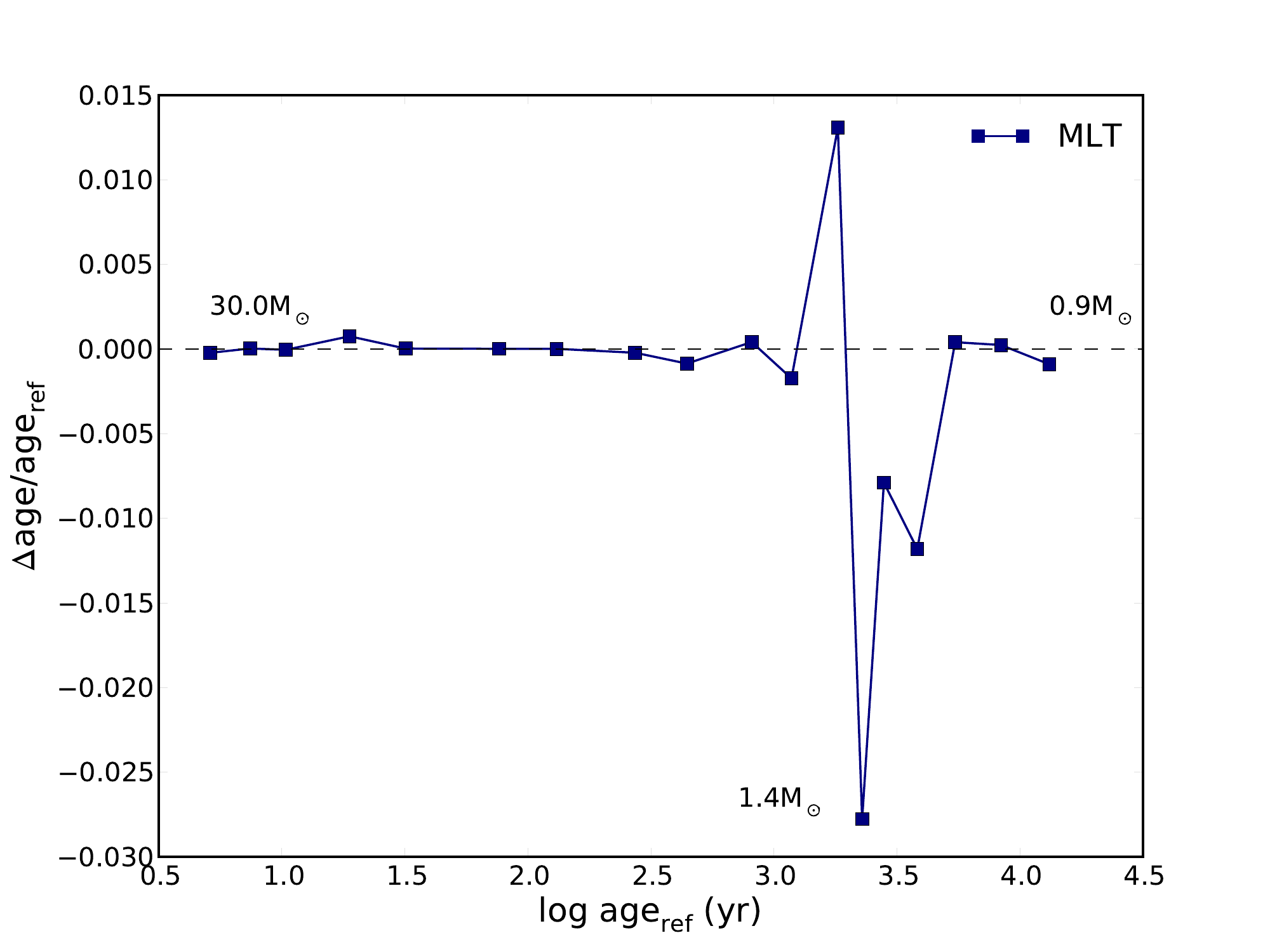}
\caption{{\sl Top:} 
evolutionary  tracks  for $1.0\ M_\odot$  models computed assuming either 
 $\beta_\mathrm{CGM} =0.16$, 
 $\alpha_\mathrm{MLT} = 2.0 $, or $\alpha_\mathrm{MLT, \odot} = 1.76$ (all are solar calibrated values).
{\sl Bottom:} Same comparison as in Fig.~\ref{errorpp} between a model based on the MLT approach
(with $\alpha_\mathrm{MLT,\odot}=1.76$)
and a model based on the CGM approach (with $\alpha_\mathrm{CGM, \odot}=0.688$).
}
\label{FSTMLT}
\end{center}
\end{figure}
%-----------------------------------------------

\paragraph{FST versus MLT: impact on TO ages.}
As mentioned above, the FST theory provides an improved model of turbulent convection. That 
leads to a different prescription of the convective flux with respect 
to the standard MLT one. Nevertheless, the FST remains a local theory, which  
 also requires the definition of a mixing-length scale. Either 
$\Lambda_\mathrm{CGM}=z+\beta_\mathrm{CGM}\times H_{P,\mathrm{top}}$  
(where  $z$ is the distance to the convection boundary 
and $H_{P,\mathrm{top}}$ is the pressure scale-height at the top boundary), 
or $\Lambda_\mathrm{CGM}=\alpha_\mathrm{CGM}\times H_{P}$ are used. 
The free parameters $\beta_\mathrm{CGM}$, $\alpha_\mathrm{CGM}$, or  $\alpha_\mathrm{MLT}$  
are calibrated to fit the solar radius at solar age. 
Their values depend on the convective flux description, 
but also on input physics such as opacity, solar mixture, 
EoS, and atmospheric boundary conditions (BC), see \eg,  
\citet{1998A&A...332..127B,2004A&A...416.1081M,2006A&A...445..233S}.  
For instance, for a given set of microphysics and BC, we could match the current Sun with  either 
$\beta_\mathrm{CGM,\odot}=0.16$, $\alpha_\mathrm{CGM,\odot}=0.688$, or  $\alpha_\mathrm{MLT,\odot}=1.76$. 
However, because of the different dependence of the convective flux on the superadiabaticity, 
evolution
%For a calibrated Sun,  adopting the FST approach  instead of the MLT one 
% has a small effect  if one uses a \citet{1993ycat.6039....0k}'s
%model atmosphere; it is quite large if one uses an Eddington $T-\tau$ law
%\citep[e.g.][]{2002A&A...392..619H,2006A&A...445..233S,1999ApJ...526L..45K}. The evolution 
in the HR diagram with a constant $\alpha_\mathrm{MLT}$ value  is not equivalent 
to the evolution of a model of same mass with a constant value of $\alpha_\mathrm{CGM}$. 
As shown in left panel of Fig.\ref{FSTMLT}, while $\alpha_\mathrm{MLT} = 1.76$ 
yields a similar radius 
than CGM during the MS of a one solar mass model, its value must be increased to 
 $\alpha_\mathrm{MLT} = 2.0 $ during the RGB to mimic the same convective efficiency 
than the CGM treatment. 
 
%-----------------------------------------------
\begin{figure}[!hp]
\begin{center}
%\resizebox{0.45\hsize}{!}{
\includegraphics[width=0.65\textwidth]{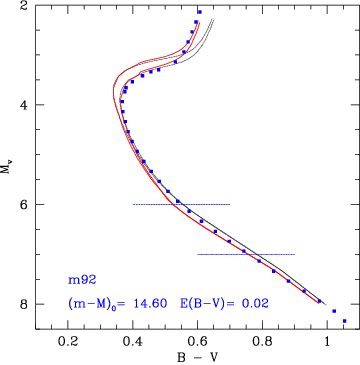}
%}\resizebox{0.45\hsize}{!}{
\includegraphics[width=0.65\textwidth]{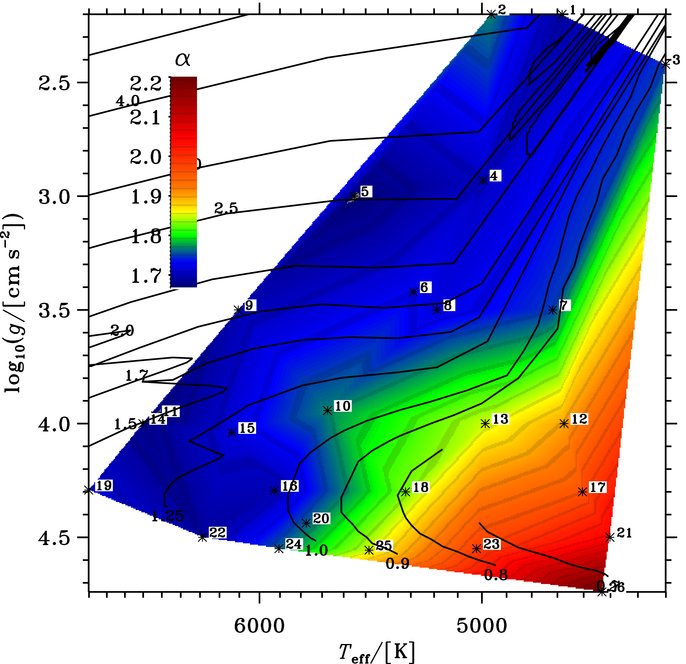}
\caption{
{\sl Top:} isochrones calculated for the metallicity of the globular cluster M92. 
Continuous lines are for the FST treatment of convection and ages of $12$~\Gyr~ (left isochrone) and $14$~\Gyr~(right isochrone).
Dotted lines are isochrones of the same ages but calculated with the MLT treatment of convection.
Blue squares are observational data of M92 from \citet{1996PASP..108..560S}.  [From  
\citet{2001A&A...370..982M}.]  
{\sl Bottom:} variation of the $\alpha_\mathrm{MLT}$ value derived
from 3-D surface convection simulations  with a solar chemical composition 
in a $\log~ g - \log~ T_\mathrm{eff}$ diagram \citep{2011ApJ...731...78T}.
Evolutionary tracks with various masses are  from 
\citet{1992A&AS...96..269S,1999A&AS..135..405C}.
}
\label{MLTisocb}
\end{center}
\end{figure}
%-----------------------------------------------

A comparison assuming the solar calibrated values 
$\alpha_\mathrm{MLT,\odot}=1.76$ and $\alpha_\mathrm{CGM, \odot} =0.688$  (Fig.~\ref{FSTMLT}, bottom panel)
indicates that the impact on age at turn-off is small ($\Delta$age$/$age$ <3$ per cent). 
The maximum impact occurs for masses in the range $1.2-1.5 M_\odot$.
%for a solar composition with a maximum difference for 	a $1.4 M_\odot$ model. 
The impact on isochrones is small, for a constant  free parameter
 ($\alpha_\mathrm{MLT}$ or $\alpha_\mathrm{CGM}$). This remains  true for isochrones with low metallicity used to 
  reproduce  globular clusters. This is illustrated in Fig.~\ref{MLTisocb}, left panel, in the case of M92
 \citep[\eg,][]{1995A&A...302..382M,2001A&A...370..982M}. 

\paragraph{Calibrations of the mixing-length value.}  
 Figure \ref{MLTisocb} (right panel) shows the variation of the $\alpha_\mathrm{MLT}$ value derived
from 3-D surface convection simulations in a $\log~ g - \log~ T_\mathrm{eff}$ diagram
for evolutionary tracks of various masses  and a solar chemical
composition \citep{2011ApJ...731...78T}.  For a one solar mass for instance, the $\alpha_\mathrm{MLT}$  value 
roughly varies 
 from $1.8$ to $1.7$  on the MS. For  ZAMS models with masses decreasing from
 $1.$5 to $0.7\ M_\odot$, $\alpha_\mathrm{MLT}$ increases  from $1.7$ to $2.2$. 
  
One then needs to calibrate the $\alpha_\mathrm{MLT}$ value across the HR diagram. This can be obtained  
with a prescription  for the $\alpha_\mathrm{MLT}$ value derived from 2-D or
3-D numerical simulations 
\citep{1999A&A...346..111L,2008IAUS..252...75L,2011ApJ...731...78T,2014arXiv1403.1062M}. 
An alternative is 
to use patched models, which are built as  a 1-D stellar interior with the outer layers 
originating from a 3-D simulation 
\citep{1999A&A...351..689R,2006ApJ...636.1078S,2010A&A...509A..16S}.
An observational calibration  can also be directly
obtained on a case by case level by performing \`{a} la carte seismic studies (see lecture 2).

\subsection{Overshooting from convective cores}
\label{overshooting}

MS stars of masses $\gtrapprox 1.2\ M_\odot$ develop a 
convective core because of the high temperature dependence of the nuclear CNO cycle. 
Convection in the  dense central layers is very efficient and the temperature gradient is 
 nearly adiabatic. 
 As a consequence, the value of the mixing length has no effect on the properties of the convective core.
 
 %----------------------------------------------------------------
 \begin{figure}[!hptb]
 \begin{center}
 %\resizebox{0.5\hsize}{!}{
\includegraphics[width=0.75\textwidth]{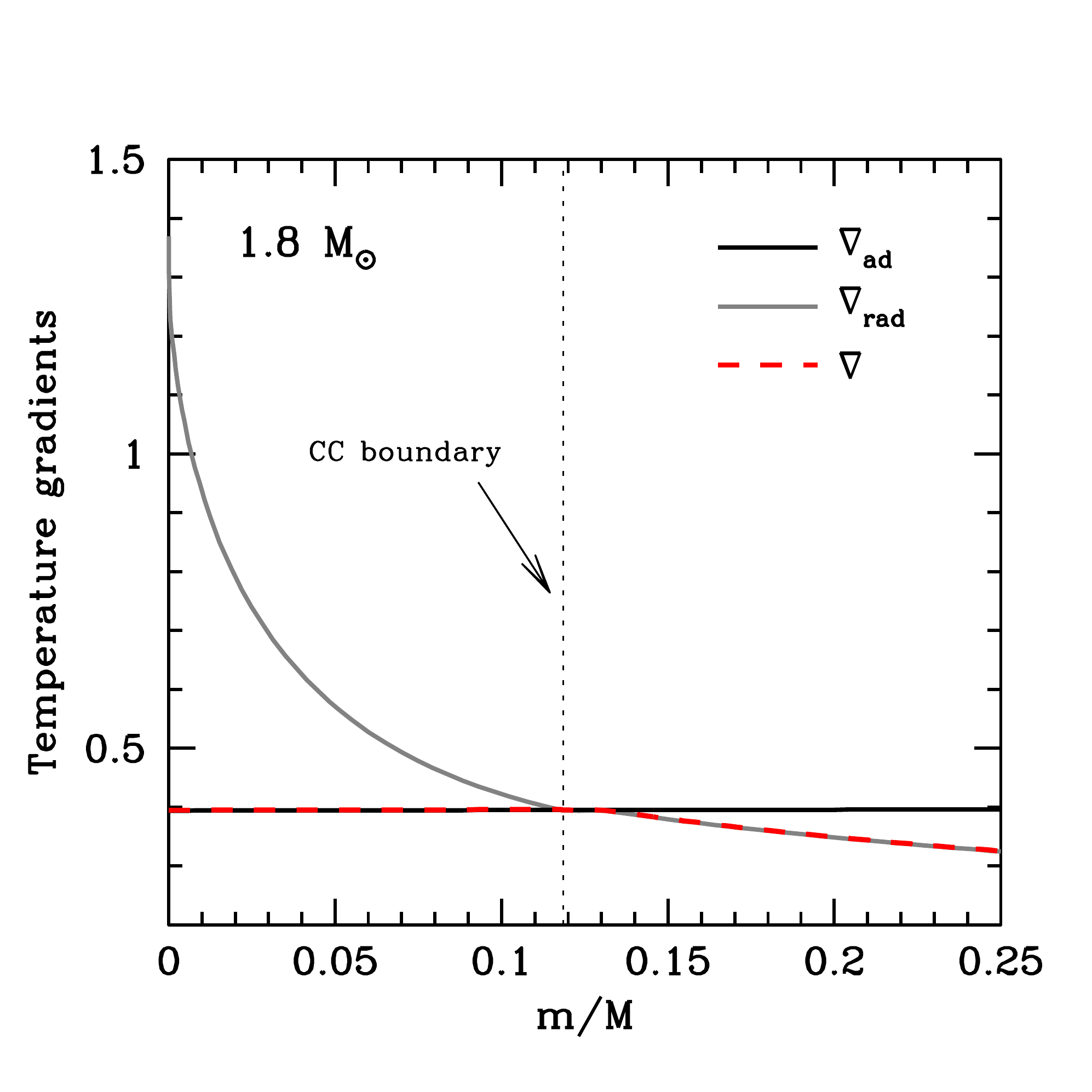}
%}\resizebox{0.5\hsize}{!}{
\includegraphics[width=0.7\textwidth]{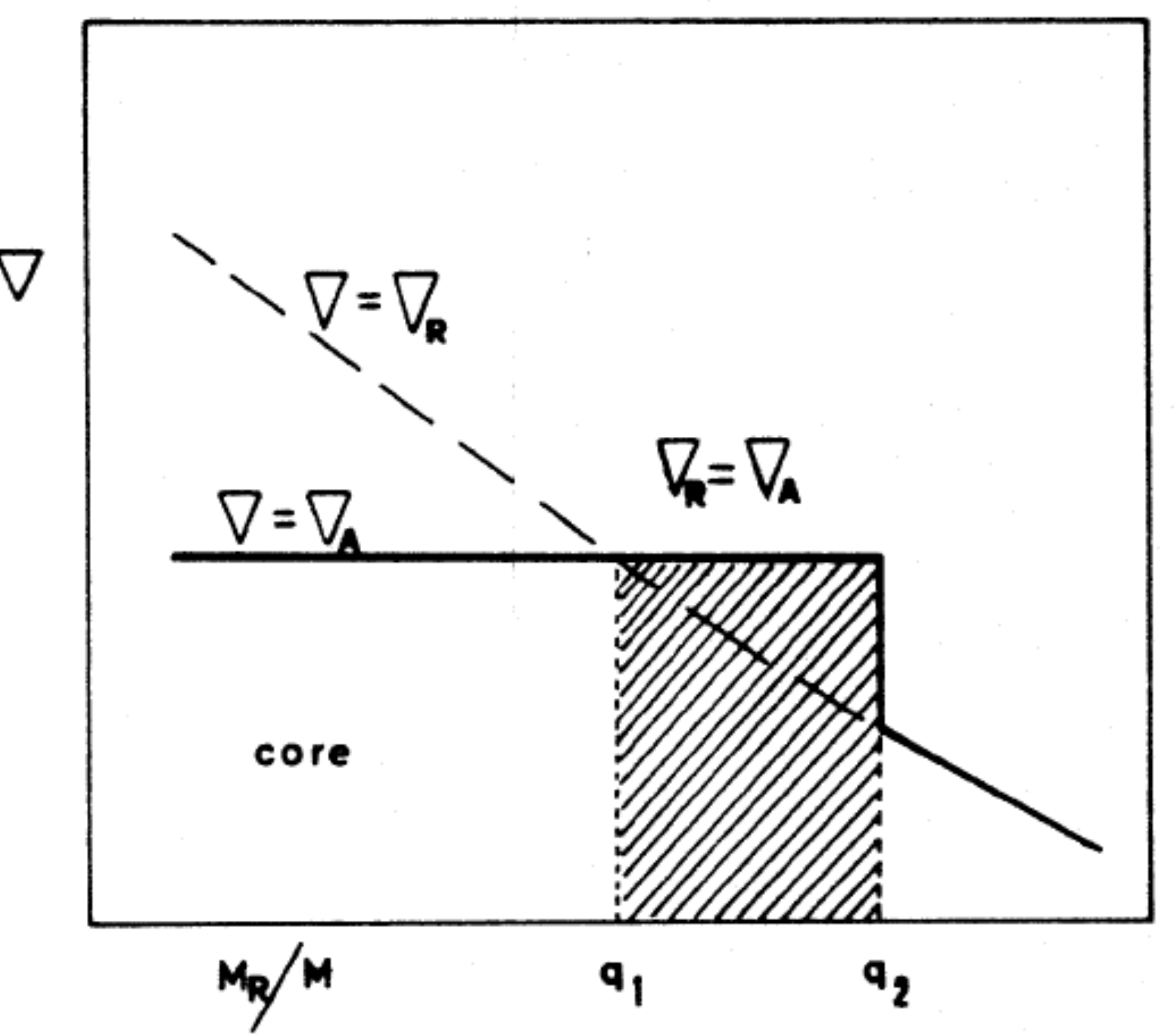}
 \caption{
 {\sl Top:} profiles of the temperature gradients 
 $\nabla_\mathrm{rad}$ (grey) and $\nabla_\mathrm{ad}$ (black) and of the 
 actual gradient $\nabla$  (red)
 as a function of the fractional mass, for the inner 25 per cent in mass,
  in a $1.8\ M_\odot$ model. 
{\sl Bottom:} a schematic view
  of the simplest description of the overshooting impact on the  temperature
  gradient. [From \citet{1981A&A...102...25B}.]}
 \label{ov1}
 \end{center}
 \end{figure}
 %-------------------------------------------------------------------
 
  Figure~\ref{ov1} (left panel) shows the temperature gradients
   $\nabla_\mathrm{ad}$, $\nabla_\mathrm{rad}$, and the actual gradient 
   $\nabla$, as a function of 
  fractional mass, in the central regions
of a $1.8\ M_\odot$ MS model. The
   convective core  extends over the inner 12 per cent in mass.
   The radiative gradient sharply decreases with radius 
  and reaches the adiabatic gradient at a radius defined as the Schwarzschild radius $r_\mathrm{sc}$.
 In the convective region,  the temperature gradient $\nabla$, 
 which is required  for transporting the energy (or luminosity) 
 remains close to $\nabla_\mathrm{ad}$, namely 
    the difference $(\nabla-\nabla_\mathrm{ad})$ is small and remains of the order of 
        $\sim  10^{-8}$ independently of the value of the mixing length.
 On the other hand,  core overshooting is expected to occur in stars. 
 It consists in convective elements moving over into 
  the radiation layers above the convective core \citep[see \eg,][for a review]{2009CoAst.158...45D}.
   However this process is poorly understood and crudely modelled in stellar evolutionary codes 
\citep[see][for a review]{2007IAUS..239..235C}. 
  
   Indeed 
 the transition between the convective core and the radiative region above
 is delimited by  the Schwarzschild criterion 
 (\ie,  $\nabla_\mathrm{ad}=\nabla_\mathrm{rad}$). 
 This  corresponds to  the location where the
 acceleration of the convective motion vanishes. Because of inertia, the moving fluid
 keeps on travelling over some distance into the adjacent  radiative region. During
 its travel, the bubble is decelerated till its velocity vanishes.
This induces  mixing of chemical elements and  heat transport in the overshooting
 region, and a modification of the corresponding gradients in the region above the convective core. 
  To evaluate the overshooting distance properly, a non-local description of
  convection is necessary. Instead, a crude  formulation is often used, 
  which states  that  the overshooting distance simply is some fraction of the pressure
  scale height, $d_\mathrm{ov}=\alpha_\mathrm{ov} H_P$. 
  
  In order to avoid some incoherence when the convective core is quite small
  (for instance in low-mass stars), the overshooting distance is often actually 
  set to be a fraction of  $H_P$, or of the core radius if the latter is lower than $H_P$.
  In addition, in the overshooting region it is often assumed that the matter 
is  fully mixed and that the temperature gradient is the adiabatic one (see Fig.~\ref{ov1}, right panel).
  
 Several important open questions/issues remain: 
 \begin{itemize}
 \item What is the size of the zone of extended mixing? The parameter $\alpha_\mathrm{ov}$
  is a free parameter of models. The question is to know whether it depends on
  the mass, metallicity, or other properties of the star.
 
 \item Is the stratification fully adiabatic in the overshooting region?
 
 \item What kind of chemical mixing 
 does actually occur? Is it instantaneous or diffusive?
 
 \end{itemize}
 
 Convective core overshooting widens the MS, which modifies the shape of isochrones. 
Therefore, one way to quantify overshooting has been to try to fit the observed isochrone MS turn-off of open clusters, and the width of the MS band of groups of stars
 \citep[see \eg,][]{1981A&A....93..136M,1990ApJ...363L..33A,1991ApJ...383..820S,1992A&AS...96..269S,2001A&A...374..540L,2002A&A...392..169C}.
Furthermore, insights on how the overshooting distance varies with stellar mass, metallicity, and evolutionary state were obtained by 
the modelling of samples of binary stars of known mass and/or radius, and chemical composition
  \citep{1990ApJ...363L..33A,2000MNRAS.318L..55R, 2007A&A...475.1019C}.
Different empirical calibrations of the overshooting distance suggest that it roughly covers the range $d_\mathrm{ov} = 0.0 - 0.4\ H_P$.
However, the value depends on the model input physics. For instance, \citet{1992A&AS...96..269S} showed that the improvement of opacities
implies a decrease of $d_\mathrm{ov}$ from $d_\mathrm{ov} = 0.25-0.30\ H_P$ to $d_\mathrm{ov} \lessapprox 0.20\ H_P$.

 %----------------------------------------------------------------
 \begin{figure}[!hptb]
 \begin{center}
 %\resizebox{0.55\hsize}{!}{
\includegraphics[width=0.87\textwidth]{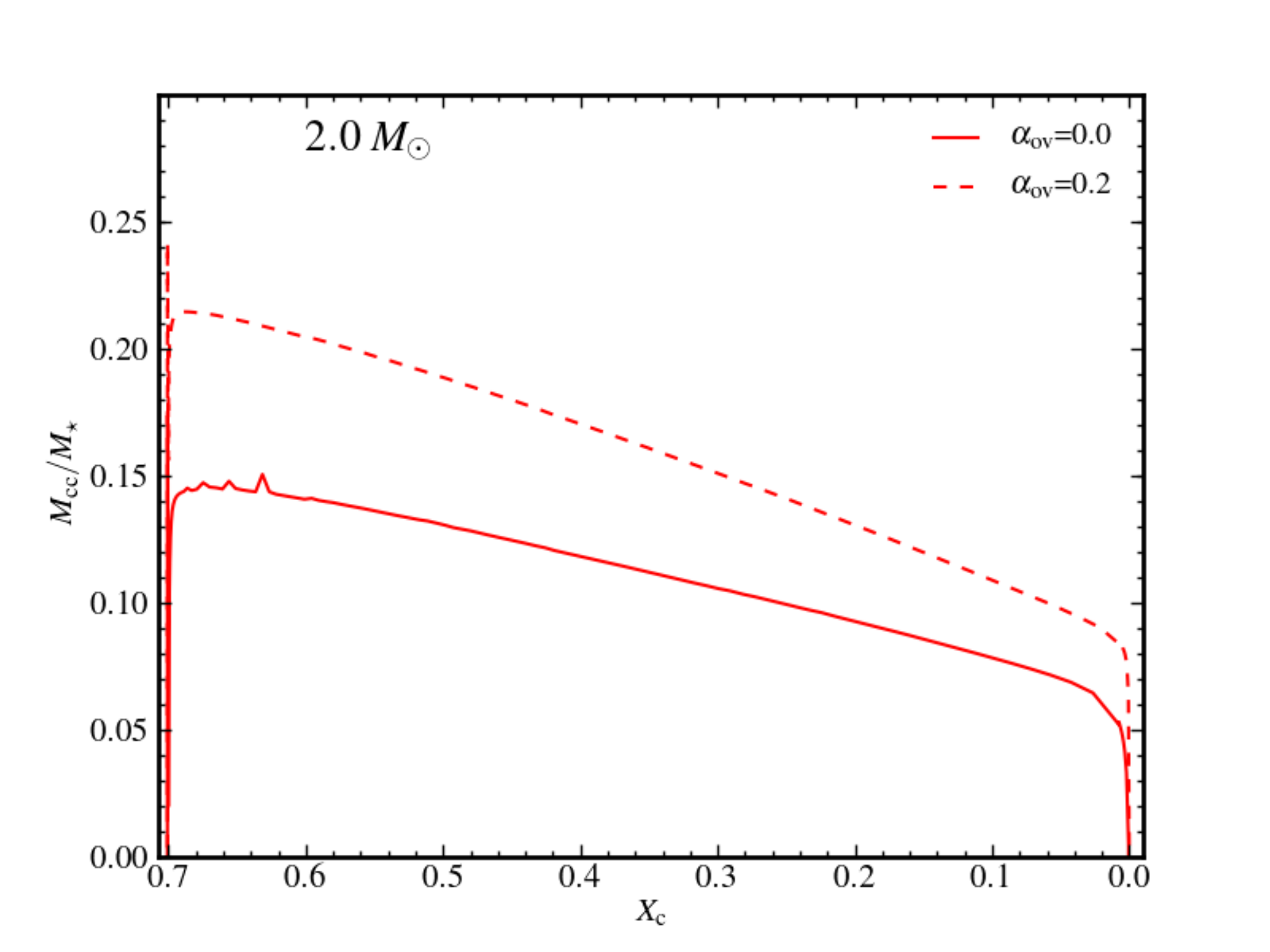}
%}\resizebox{0.55\hsize}{!}{
\includegraphics[width=0.87\textwidth]{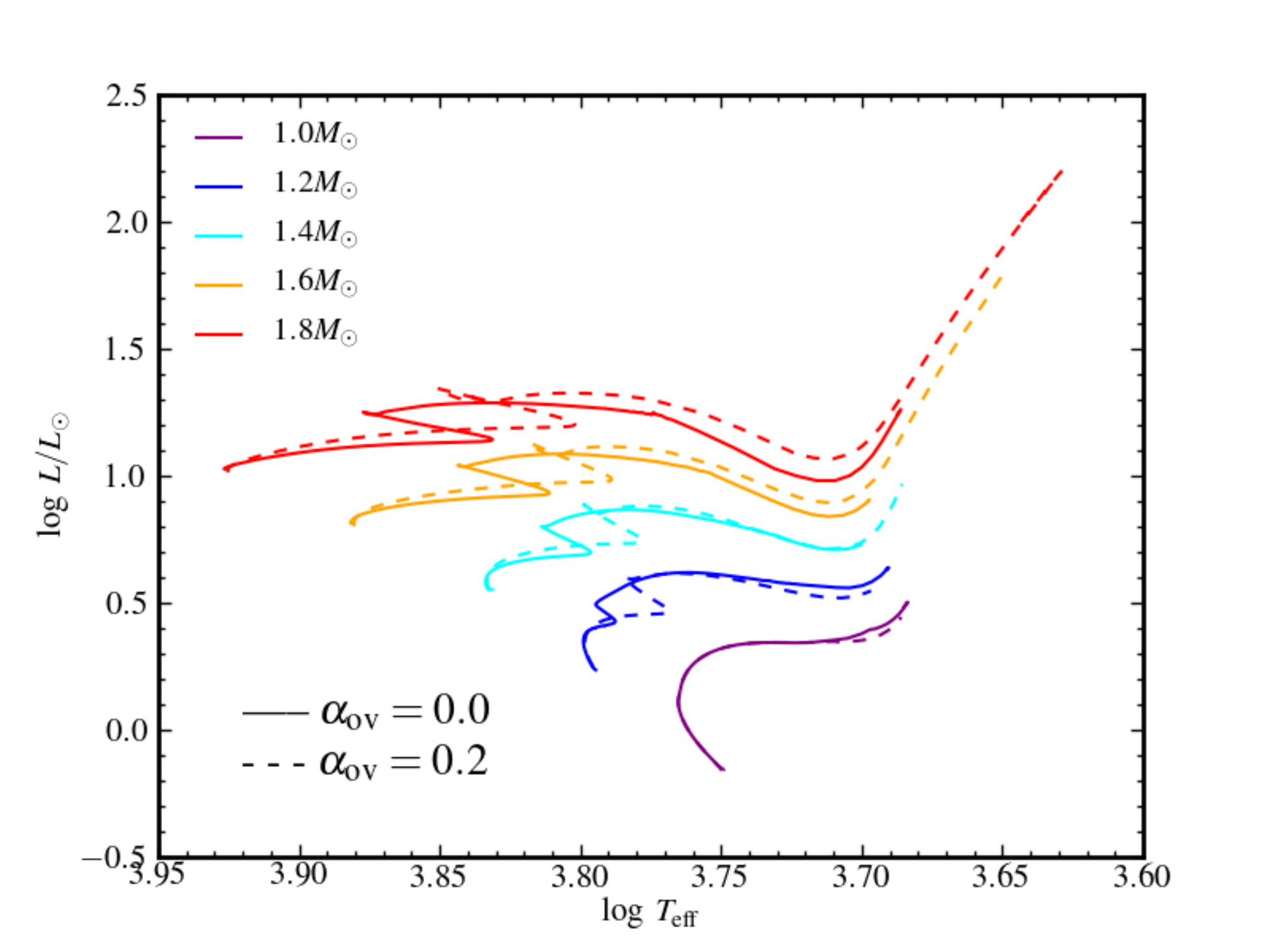}
 \caption{Comparison of models without convective core overshooting (continuous lines) to models
 with an overshooting parameter $\alpha_\mathrm{ov}=0.2$ (dashed lines).
  {\sl Top:} evolution of the  size of the mixed central region (in relative mass) as a function of $X_c${\tiny }, the central H abundance (a proxy for the age) in a star of $2.0\ M_\odot$.
    {\sl Bottom:} 
comparison of the evolutionary tracks in the HR diagram for
masses in the range $1.0\ M_\odot-1.8\ M_\odot$. 
 }
 \label{ov4}
 \end{center}
 \end{figure}
 %-------------------------------------------------------------------
   
 \subsubsection{Impact of overshooting on stellar age}
 
 Including core overshooting in the modelling increases the size of the mixed 
 core. As a result, on the MS, more hydrogen is available for nuclear burning.
This lengthens the MS phase and yields older models at TO. This is clearly illustrated 
  in Fig.~\ref{ov4} (left panel), which shows the evolution of the
  size of the mixed central region (in relative mass) as a function of $X_c$ the
  central hydrogen abundance (a proxy for the age)  for two $2.0\ M_\odot$
  models, one  without core overshooting and the other
   with core overshooting of $0.2 H_P$.
 Fig.~\ref{ov4} (right panel) shows a HR diagram comparing 
 evolutionary tracks  without overshooting to
 tracks including  core overshooting of $0.2 H_P$, for masses in the
range $1.0-1.8\ M_\odot$.  Comparison of the location of the
 TO for the two types of tracks evidences 
 the lengthening of the MS by about 20 per cent when
 a core overshooting  of $0.2\ H_P$ is included. 
 
 %----------------------------------------------------------------
 \begin{figure}[!hptb]
 \begin{center}
 %\resizebox{0.5\hsize}{!}{\includegraphics{Figs/ov5.eps}} \resizebox{0.5)\hsize}{!}{\includegraphics{Figs/ov5.eps}}
%\resizebox{0.5\hsize}{!}{
\includegraphics[width=0.7\textwidth]{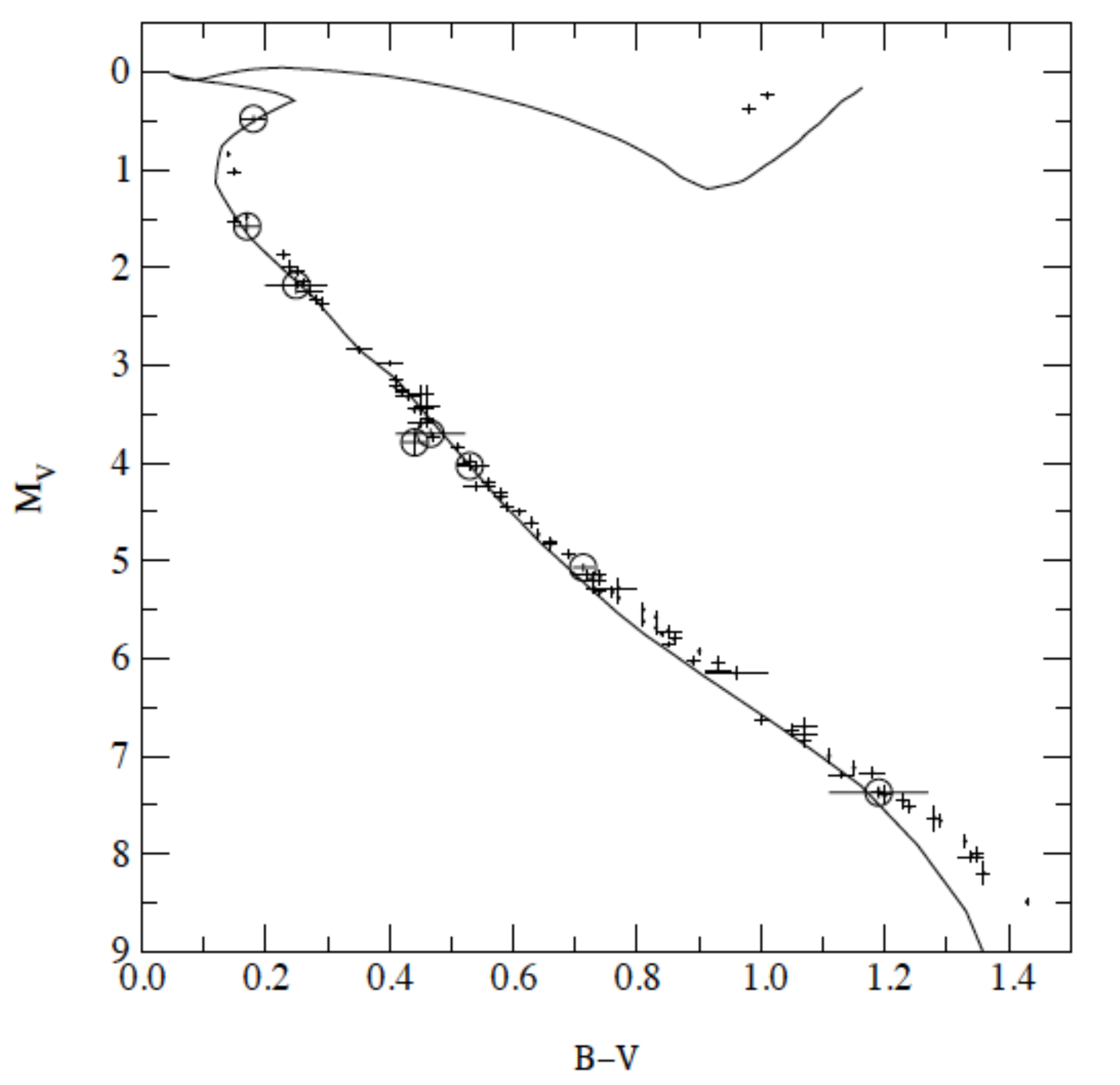}
%}\resizebox{0.5\hsize}{!}{
\includegraphics[width=0.7\textwidth]{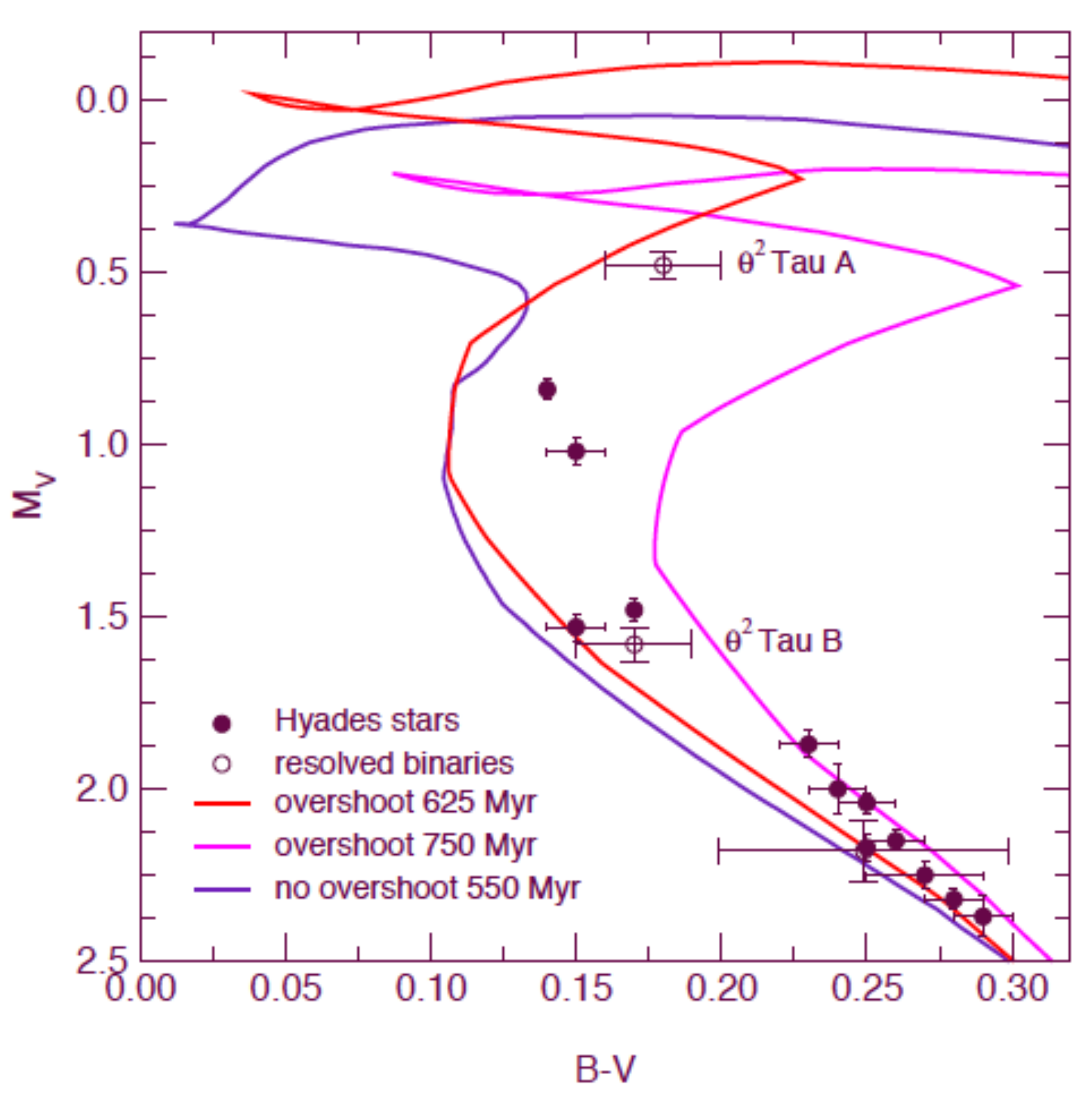}
 \caption{
 {\sl Top:} Observed location of the Hyades cluster stars in a HR
 diagram with the best fitted isochrone (625~\Myr). 
 {\sl Bottom:}  Comparison in a
 HR diagram of the shape of three isochrones (one without overshooting, two with overshooting but
 different ages) in the vicinity of the turn-off of the Hyades
 cluster (observations are also reported). [From  \citet{2001A&A...374..540L}.] 
 }
 \label{ov5}
 \end{center}
 \end{figure}
 %-------------------------------------------------------------------
 
 \subsubsection{Overshooting of convective cores: the age of the Hyades}
 As  just discussed, the end of the MS for a given stellar mass occurs at 
 lower effective temperature and  higher luminosity when core overshooting is
 included. Accordingly, the  shapes of the isochrones, 
 which have turn-off masses larger than $\sim 1.2\ M_\odot$ are significantly 
  modified.
   This modifies the age of rather young open clusters.
   This is the case of the Hyades, the closest ($26$ pc) and well-studied 
   open cluster  which has turn-off masses in the range $2.0-2.5\ M_\odot$ (see 
   Fig.\ \ref{ov5}). With an  overshooting amount 
   of $d_\mathrm{ov}=0.20 ~ H_P$, the isochrone fit gives
 an age of $625$~\Myr, whereas when no overshooting is assumed, the derived age is 
 $550$~\Myr~(Fig.~\ref{ov5} right panel).  The relative age difference in 
 this case amounts to  13.6 per cent.  
 
Note that seismic studies of stars at TO could provide constraints for core mixing.
Interesting candidates are the  $\theta^2$ Tau binary system components, which are $\delta$-Scuti stars of known mass located in the vicinity of the Hyades TO.
However their fast rotation will make the task difficult.

 %----------------------------------------------------------------
 \begin{figure}[!hptb]
 \begin{center}
 %\resizebox{0.55\hsize}{!}{
\includegraphics[width=0.87\textwidth]{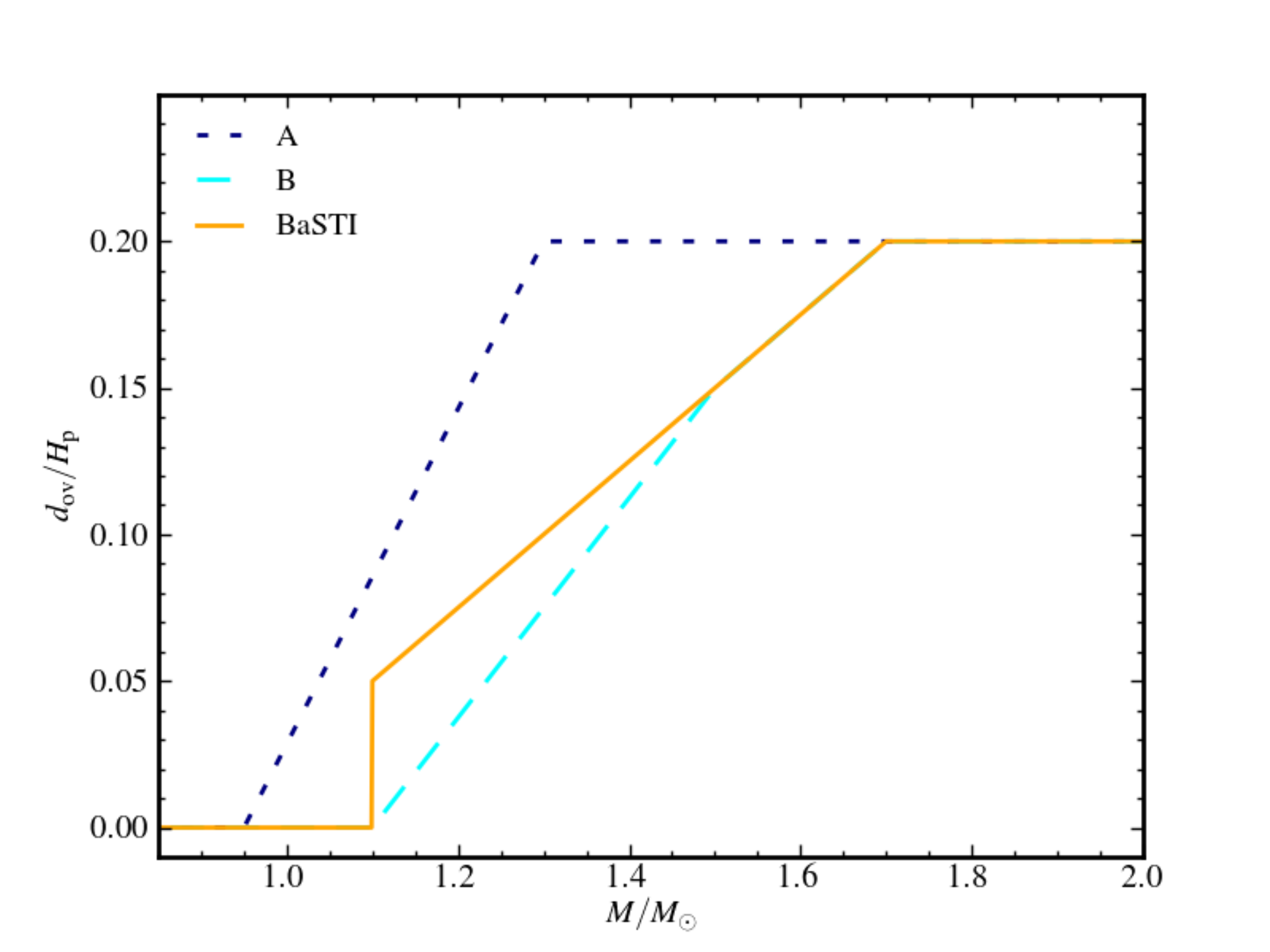}
%}\resizebox{0.45\hsize}{!}{
\includegraphics[width=0.75\textwidth]{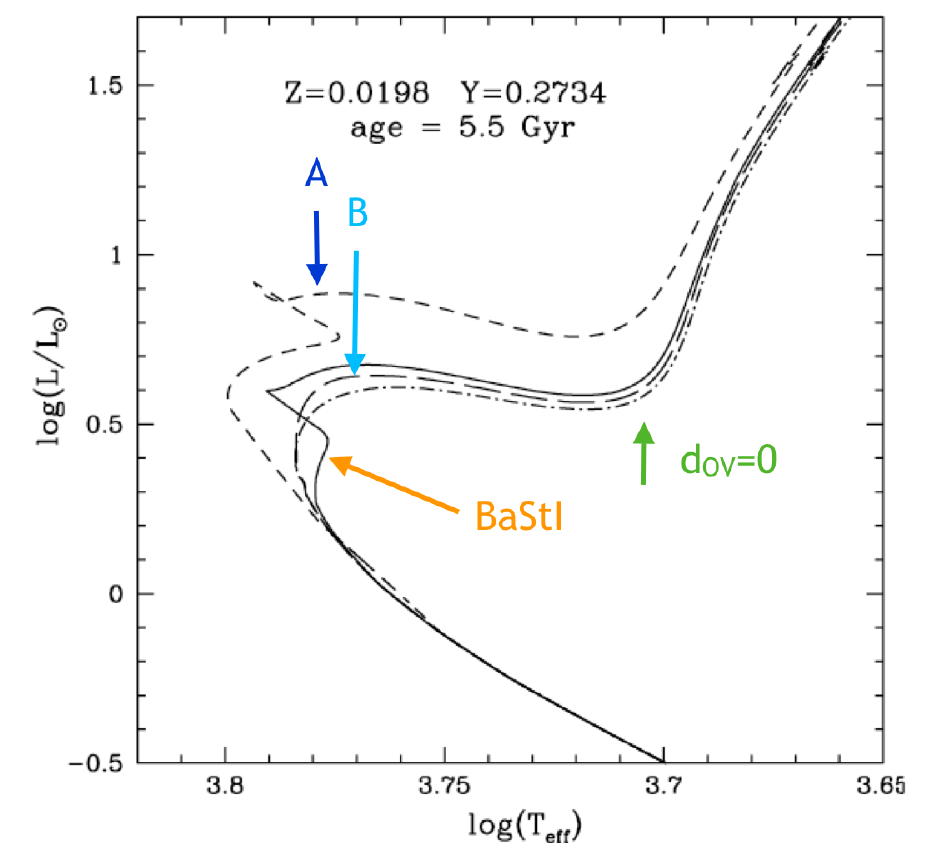}
 \caption{
 {\sl Top:} three options for the  variation of $\alpha_\mathrm{ov}$ 
  for small convective core overshooting 
  as a function of the stellar  mass. {\sl Bottom:} 
 impact of the different options shown on the left
  on a $5.5$~\Gyr~isochrone. [From  \citet{2004apj...612..168p}.]
 }
 \label{ov6}
 \end{center}
 \end{figure}
 %-------------------------------------------------------------------
 
 \subsubsection{Vanishing convective cores}
 \label{smallcores}
 
 As discussed above, for masses larger than $\sim 1.4-1.5 M_\odot$ (the exact mass depending  on the chemical composition)
 the convective core is well developed. 
 On the other hand, convective cores begin to
 form at  $M_\mathrm{min}\approx 1.1\ M_\odot$. 
In the intermediate mass range, \ie,  $\sim 1.1 - 1.4  M_\odot$, the problem is how 
 to treat overshooting of very small convective cores. 
 
The simplest  case is to assume no overshooting  for
  masses $M\leq M_\mathrm{min}$ and an overshooting of  $\alpha_\mathrm{ov}\times H_P$
  for masses $M> M_\mathrm{min}$. A typical case is to take $M_\mathrm{min}=1.1\ M_\odot$ and
 $\alpha_\mathrm{ov} =0.20$.  More sophisticated prescriptions have been proposed by, for instance,
\citet{2004apj...612..168p}, but see also \citet{2012MNRAS.427..127B},  
who considered variations of the overshooting parameter
$\alpha_\mathrm{ov}$ with mass, as schematically represented  in Fig.~\ref{ov6}, right panel.
  In their first option (case A), $\alpha_\mathrm{ov}$  linearly increases 
  with the model mass from $\alpha_\mathrm{ov} =0.0$ at $0.95\ M_\odot$ to $0.2$ at $1.7\ M_\odot$. 
  In their case B, $\alpha_\mathrm{ov}$ also linearly increases 
   with the model mass from $0.0$ at $1.1\ M_\odot$ to $0.2$ at $1.7\ M_\odot$,  but with a change of slope at $1.5\ M_\odot$.
The third option, adopted in the BaSTI  code, is intermediate between case A and B,  \ie, 
no overshooting for masses lower  than  $1.1\ M_\odot$,  then
 an overshooting  $\alpha_\mathrm{ov} =0.05$ at $1.1\ M_\odot$,
  followed by a linear increase of $\alpha_\mathrm{ov}$ 
  until it reaches $\alpha_\mathrm{ov} =0.20$  at $1.7\ M_\odot$. 
    The impact of these different options on a 5.5~\Gyr~isochrone
    is shown in Fig.~\ref{ov6} (right panel). There are little differences between the 
  BaSTI case, case B, and the case with no overshooting. On the other hand, for the case A option,
  overshooting is larger and the maximum amount of overshooting is reached earlier, at $1.3\ M_\odot$. 
   To match the same TO location, in case A, an isochrone would have to be older than in the other cases.
   
 \subsubsection{Impact on ages at MS turn-off}
 
\begin{figure}[!htb]
 \begin{center}
 \resizebox{0.9\hsize}{!}{\includegraphics{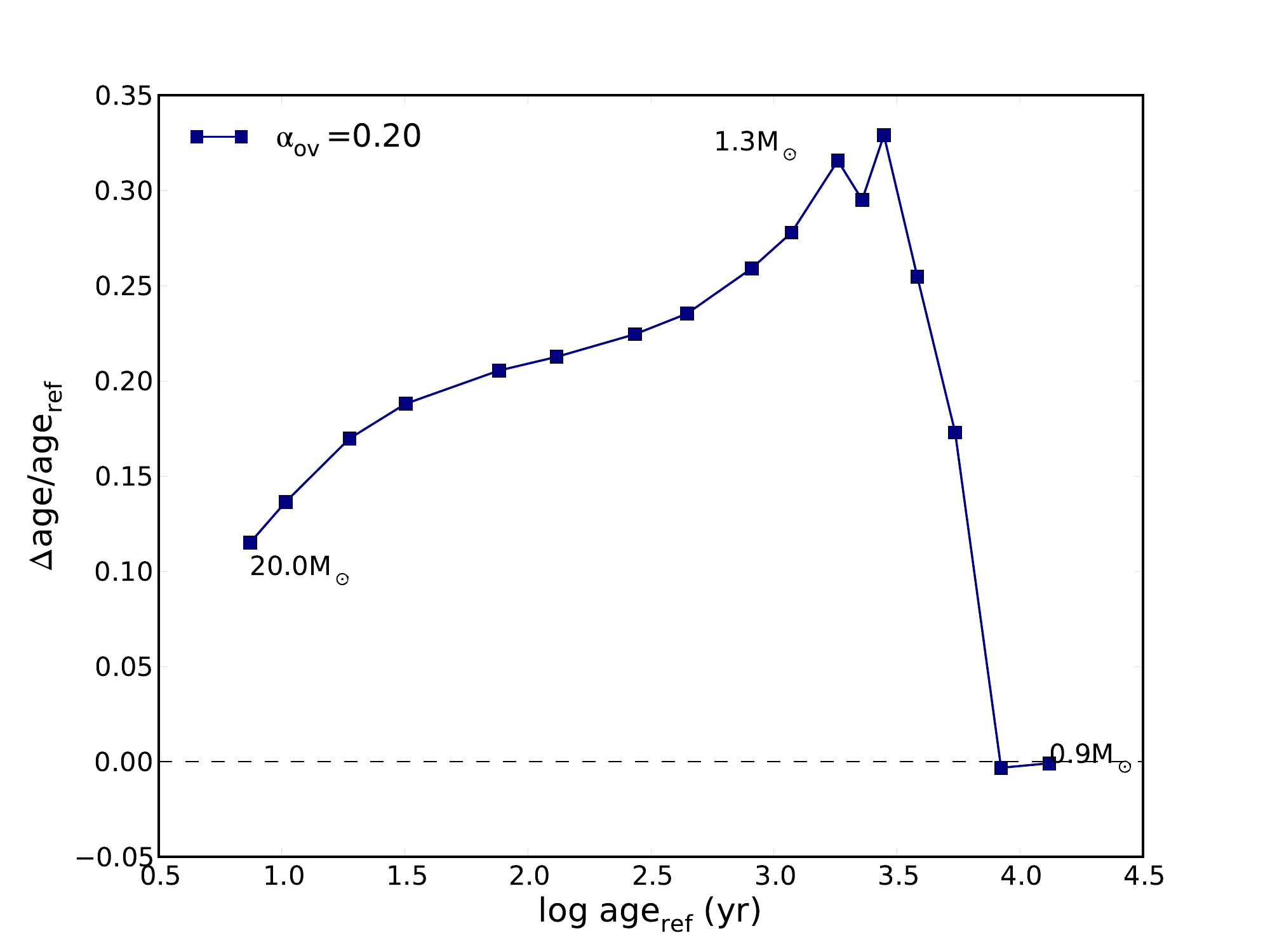}}
 \caption{
  Relative differences between the age of the model with an overshooting of $0.2 H_P$
 and the reference model (no overshooting) as a function of the age of the reference model.
 }
 \label{ov7}
 \end{center}
 \end{figure}

 Figure\ \ref{ov7} shows the relative differences in age at turn-off between models including
 a core overshooting of $0.2\ H_P$ and the reference models (no overshooting). 
  For models with masses below $\sim 1.1\ M_\odot$, the convective core  is quite small, hence the core 
  overshooting distance and its impact on the age are  negligible.  For models with masses higher than  
 $\sim 1.2\ M_\odot$,  the core overshooting is no longer negligible. The impact on age amounts to about 30
  per cent  in a $1.5\ M_\odot$ model. It decreases
 with the mass of the model but still equals $\sim 10$ per cent at $20 ~ M_\odot$. 
 The decrease of the impact with the mass of the turn-off models is due to a
 decrease  of the pressure scale height  $H_P \propto T/g$ 
 of the central layers with the model mass. 
 
 %----------------------------------------------------------------
 %-------------------------------------------------------------------
 
 \subsubsection{Modelling overshooting of convective cores: alternatives}
 
 The large uncertainty of the core overshooting extent is a severe flaw for stellar models at least when the age-dating is concerned. 
This is why other ways to model core overshooting 
 have been proposed as more realistic alternatives. One can mention for instance:
 
 \begin{itemize}
 \item Diffusive overshooting.
This approach assumes that convective velocities decrease exponentially from a level located inside the Schwarzschild 
convective core, beneath its upper boundary, to a level located up in the radiative region 
\citep[][]{1996A&A...313..145D,1998A&A...334..953V,2012ApJ...750...11Z,2013ApJS..205...18Z}.
 This adds a diffusion coefficient in Eq.~\ref{dXidtconv}.
When the classical non diffusive overshooting is assumed, models of low-mass stars (in particular the solar model) develop 
a convective core at the end of the PMS, which may remain on the MS. Small cores then require a special treatment (Sect.~\ref{smallcores}), 
 not necessary in the diffusive approach.
For more massive stars,  there are no differences between the diffusive and non diffusive treatments on the MS, but 
evolved stages like He-burning may be affected \citep{2005A&A...440..623V}.
  
 \item A prescription derived from energy conservation principles has been 
 derived by \citet{1978A&A....65..281R,1989A&A...211..361R}. 
 This prescription allows the estimate of an upper limit  to the extent of  overshooting from
  convective cores. 
  Numerical estimations \citep{1992A&A...266..291R} suggest  that  
  the mass of the extended convective core is proportional
   to the mass of the Schwarzschild convective core ($M_\mathrm{ov} = 1.7\ M_\mathrm{Schwarzschild}$).  
   This approach does not provide the right
   amount of overshooting (as estimated from observations) unless 
    the dissipation of energy is included but dissipation is a major unknown 
  \citep[\eg,][]{1992A&A...266..291R,1993A&A...277...93R,Maeder09,2006ApJS..162..375V}.    

\item Overshooting due to plumes, based on an assumption of universal 
  turbulent entrainment. A theoretical prescription can be  derived, 
 which involves several parameters 
\citep[\eg,][]{1991A&A...252..179Z,1995A&A...296..127R,1997A&A...322..545L}.
 
\item Several variants of 
 a non-local description (Reynolds stress model) 
 using moment equations and closure models
\citep[\eg,][]{1978ChA.....2..118X,1985A&A...150..133X,1996MNRAS.279..305G,1998ApJ...493..834C,1999ApJ...518L.119C,2012ApJ...750...11Z}.

 \end{itemize}
 Unfortunately,  all these formulations involve  one (or more) free parameters. 
 Some progress will come from prescription and/or calibration with 3-D simulations, although this currently 
 remains somewhat difficult 
 \citep[\eg,][]{2007ApJ...667..448M,2013ApJ...773..137G,2013ARep...57..380S,2013ApJ...769....1V}.

\subsection{Rotation }
\label{rotation} 

Stars rotate and their 
 surface rotation velocity is known to change with time (see R. Jeffries' lecture). 
 That can be the consequence of angular momentum loss by external torques 
 (for instance magnetic braking, coupling with an accretion disc), and of structural 
 changes such as core contraction and envelope expansion during MS and post-MS evolution. 
 The physical processes linked with rotation and its evolution are manifold and interconnected. 
 This makes the treatment of rotation in stellar evolution modelling very complex 
 \citep[\eg,][]{2013A&A...553A...1M,2013LNP...865....3M,2013LNP...865...23M}.
  Several books and reviews deal with the effect of rotation on stellar structure and evolution 
 \citep[see \eg,][and references therein]{2000ARA&A..38..143M,2007stro.book.....T,Maeder09,2013EAS....62..227P,2014IAUS..301..161G}.
 
First of all, rotation breaks the spherical symmetry of stars and therefore creates a thermal imbalance. 
As a result, large-scale circulation (meridional circulation) takes place that transports chemicals and angular momentum (AM).  
Since the resulting rotation regime is not uniform ($\Omega=\Omega(r,\theta)$), the vertical and horizontal shear induced by 
differential rotation give rise to various hydrodynamical instabilities, which generate turbulence and hence transport  
 angular momentum and chemicals.  The turbulent transport  (mainly the horizontal one) in turn modifies 
the efficiency of large-scale transport and the final rotation profile.

The effects of the  interaction between magnetic fields and rotation are also manifold.  
The interaction between convection and rotation leads to the generation, by a dynamo mechanism, of a magnetic field, 
whose intensity seems to be linked with rotation. On the other hand,  this magnetic field may channel stellar winds to large 
distances (as in solar-like stars), increasing the loss of angular momentum from the stellar surface and braking the star.  
In radiative zones, magnetic fields can freeze plasma motions and also induce magnetohydrodynamical instabilities 
(for instance the Taylor-Spruit instability) that may affect the transport of angular momentum and chemical elements.

Finally, the propagation of internal gravity waves\footnote{These waves are excited at the boundaries of convective zones
 and propagate in the gravitationally stratified radiative zone where they extract or deposit AM.}
 (IGW) in a rotating medium can also lead to AM transport and modify the internal rotation, and in turn chemical composition profiles.

To summarize, rotation interacts with many physical processes which may transport AM and matter, in a complex way.
These interactions are currently addressed in many theoretical studies \citep[see \eg,][]{2013A&A...553A...1M,2013LNP...865....3M,2013LNP...865...23M} 
and benefit from the results of 3-D numerical simulations \citep{2004ApJ...601..512B}. 
Resulting modifications of  internal angular velocity and chemical composition profiles may strongly affect age-dating. 
 
\subsubsection{Angular momentum transport and rotational mixing in stellar models}

Here, we only briefly summarize the main aspects related to model calculation.

Concerning the AM transport, two main approaches have been followed. 
The first one, proposed by \citet{1976ApJ...210..184E}, 
consists in treating  both AM and  chemical transport
as diffusive processes; the problem is then reduced to 1-D \citep[see also][]{1989ApJ...338..424P}.
In the second approach, the AM transport is treated as 
an advective-diffusive process, while the transport of chemical elements obeys a diffusion equation.
In that context, \citet{1992A&A...265..115Z} assumed that the turbulence induced by differential rotation is stronger in the
horizontal direction than in the vertical direction, which implies that the angular 
velocity $\Omega$ is about constant on isobars. This behavior justifies the hypothesis
of so-called shellular rotation, which leads to express any quantity as a function of pressure only, 
or of radius provided the rotation is slow \citep[see also][]{1998A&A...334.1000M}.

In the framework of shellular rotation, the transport of angular momentum in radiative zones
obeys an advection-diffusion equation that reads,
  \begin{eqnarray}\label{AM}
  \frac{\partial j}{\partial t}+ \dot r ~ \frac{\partial j}{\partial r}
 &=& -\frac{1}{\rho r^2} \frac{\partial (r^2 {\cal F})}{\partial r} +
  \Bigl(\frac{dj}{dt}\Bigr)_\mathrm{ext}, 
 \end{eqnarray} 
 where $j = r^2 \Omega$ is  the local specific AM and   
${\cal F}$ is the AM flux, both at level $r$, and
$\dot r$ is the time derivative of the radius.
The AM flux ${\cal F}$ is the sum of several contributions to be evaluated.
In convective zones the angular momentum is assumed to be constant, that is 
convective zones are assumed to rotate like solid bodies. This assumption has been investigated
 in some specific cases \citep{2006A&A...453..261P}.

In order to solve Eq.~\ref{AM}, one has to specify  the
 surface AM losses  $(dj/dt)_\mathrm{ext}$. 
 For stars with convective envelopes, the surface AM losses are assumed to result from
 magnetic braking by stellar winds \citep{1962AnAp...25...18S,1968MNRAS.138..359M}. 
 %(Mestel 84, Kawaler 88, Reiners \& Mohanty12).  
 One has also to specify the initial AM profile across the star. %condition  $j_0(r) = j(r, t = 0)$, 
 It is commonly assumed that stars rotate like solid bodies at the beginning of the PMS and as long as they remain 
 entirely convective. 
 On the other hand, the surface rotation is assumed to remain
 constant and equal to the rotation of the protostellar disc as long as disc locking occurs.
 
 The total AM flux, ${\cal F}$ results from several AM transport processes 
 \citep[\eg,][]{2008EAS....32...81T,Maeder09}. 
 It is given by 
 ${\cal F} = {\cal F}_\mathrm{MC} + {\cal F}_\mathrm{turb} + {\cal F}_\mathrm{IGW} + {\cal F}_\mathrm{B}$, 
 where the currently identified contributions are:
 
\begin{itemize}
\item $ {\cal F}_\mathrm{MC} $: the AM transport by meridional circulation.
reads, ${\cal F}_\mathrm{MC}= -\frac{1}{5} \rho r^2 ~ \Omega ~ U_r$,
 where $U_r$ is the vertical velocity of the meridional circulation.

\item $ {\cal F}_\mathrm{turb} $:   the AM transport by the turbulence generated by different kinds 
of instabilities taking place in the  radiative regions. The combined effects are modelled as
 a diffusive process and contribute to a total turbulent viscosity $\nu_\mathrm{v}$. 
 The AM flux then reads  ${\cal F}_\mathrm{turb}= -\rho r^2 \nu_\mathrm{v} (\partial \Omega/\partial r)$. 
 %This process transports AM from inner fast rotating layers to
 %outer slower regions and can be dominant in regions with sharp rotation gradient
 %s. Turbulent transport as a diffusion process cannot account for of all
 %possible types of AM transport.
 
\item $ {\cal F}_\mathrm{IGW} $:   the AM transport by IGW in stellar radiative regions 
 has been proven to be efficient to transport AM and to influence the chemical mixing 
\citep[\eg,][]{1993A&A...279..431S,2005Sci...309.2189C}. 
The determination of  an accurate  expression for the IGW, AM flux  ${\cal F}_\mathrm{IGW}$ is nowadays 
 the object of intense theoretical research
\citep[see \eg,][]{2008EAS....32...81T,2013EAS....62..227P}.
 
\item $ {\cal F}_\mathrm{B} $:   the AM transport in presence of a magnetic field. 
The  torque of the Lorentz force, magneto-hydrodynamic instabilities (as the Taylor-Spruit instability), and multi-diffusive magnetic instabilities may slow down the star and lead to rigid rotation. The efficiency of these mechanisms to slow down 
   the radiative regions of  solar-like stars and giant stars depends 
   on its modelling, which remains debated 
  \citep{2011A&A...532A..34S,2013ApJ...764..146G,2014arXiv1405.1419C}.

\end{itemize}  

The turbulence induced by shear instabilities is strongly anisotropic.
The vertical transport of chemicals elements due to rotation (rotational mixing) in radiative zones 
can then be modelled by  a diffusion process resulting from the interaction between the
 meridional circulation and the shear turbulence 
 \citep{1992A&A...253..173C}. Therefore, a diffusion
 coefficient $D_\Omega$ is added to the total diffusion coefficient $D$ in 
  the equations of evolution of the chemical abundances (Eq.~\ref{dXidtconv}). 
Convection zones are assumed to be homogenized on very short time-scales (\ie,  instantaneously).

  The transport coefficients $\nu_\mathrm{v}$, $\nu_\mathrm{h}$, $U_r$, and $D_\Omega$  
  couple up the evolution of AM with the evolution of chemical elements. For instance, 
  $D_\Omega$ explicitly depends both on the vertical velocity of the meridional circulation   and of the turbulent
  viscosity. While prescriptions for these coefficients  exist
\citep[for a review, see][]{2013LNP...865...23M}, they however
 suffer from several uncertainties \citep{2013LNP...865....3M, 2013A&A...553A...1M}.
 
  \subsubsection{Impact of rotationally-induced mixing on stellar structure and isochrones}
  
\begin{figure}[!htb]
  \begin{center}
  %\resizebox{0.5\hsize}{!}{
\includegraphics[width=0.75\textwidth]{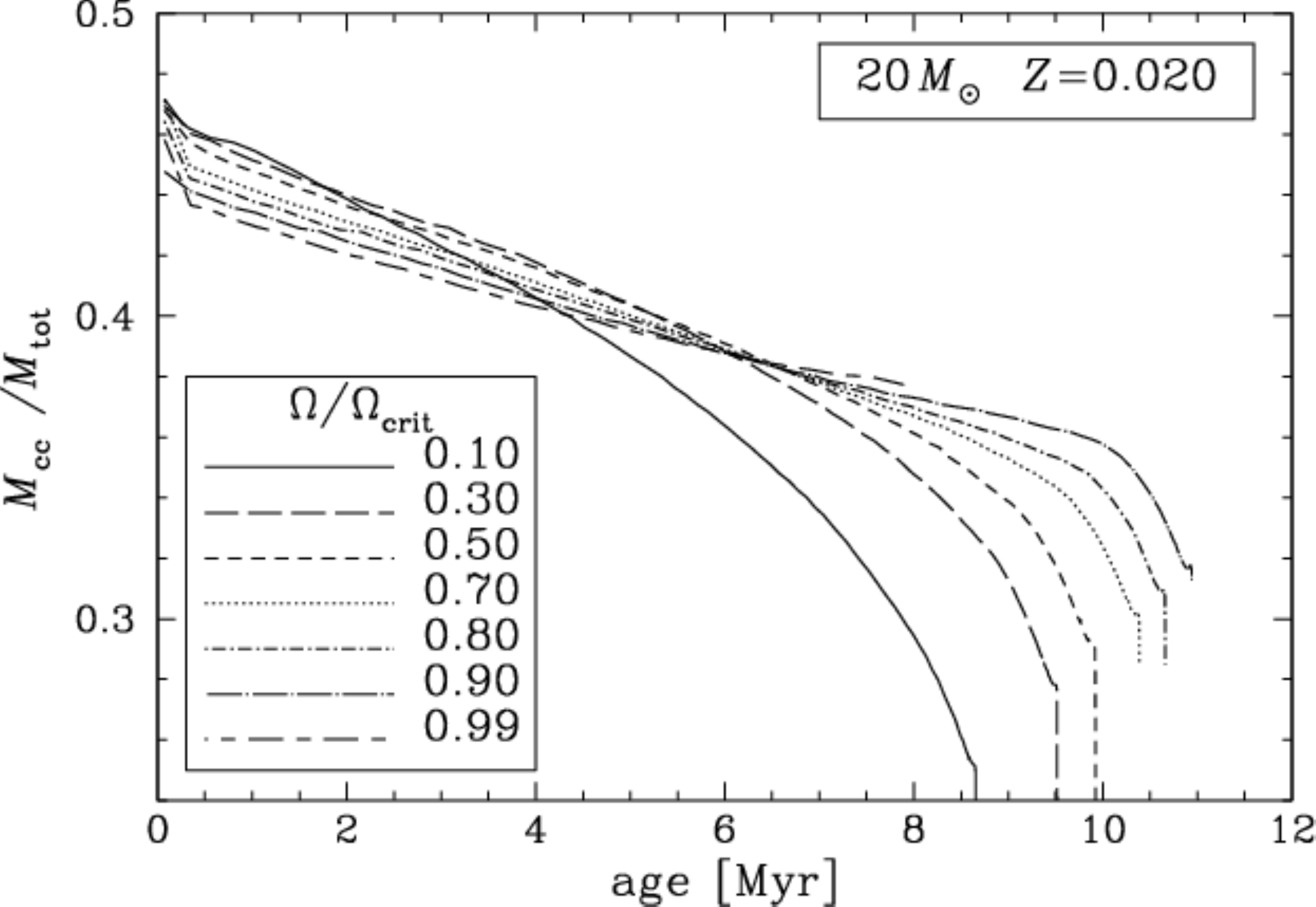}
%}\resizebox{0.5\hsize}{!}{
\includegraphics[width=0.7\textwidth]{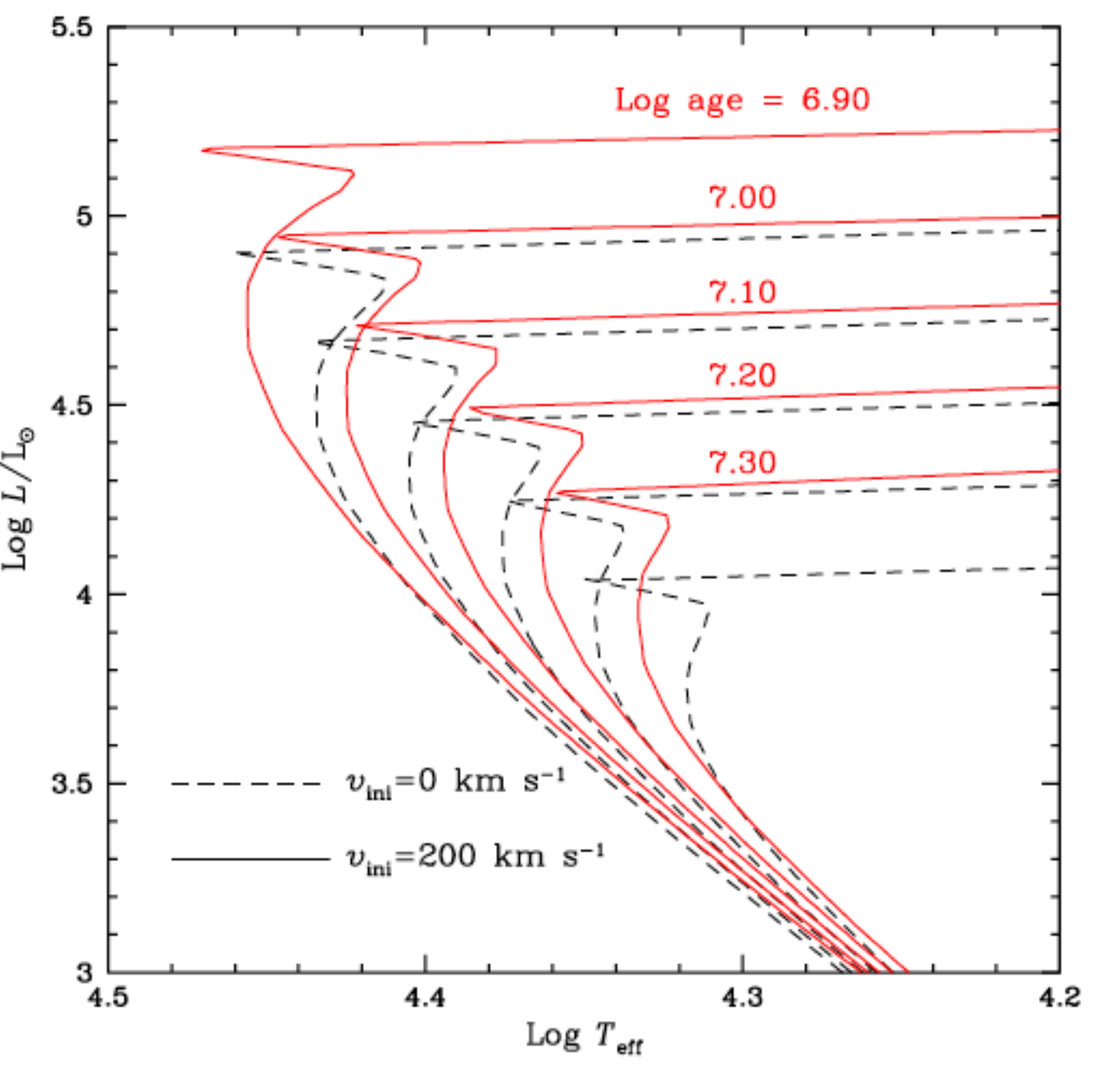}
  \caption{{\sl Top:} temporal evolution of the mass of the convective core in the presence of rotation. [From  \citet{2008A&A...478..467E}.] 
   {\sl Bottom:} impact of shellular rotation on isochrones in the HR diagram. 
   [From \citet{2000A&A...361..101M}.] 
  }
  \label{rot1}
  \end{center}
  \end{figure}

  As mentioned above, rotationally-induced mixing refuels the core with fresh hydrogen. Therefore, at a given evolutionary state, that is 
  for a given value of $X_c$, the mass of the mixed core is larger. As a result, as illustrated in left panel of Fig.\ \ref{rot1}, the higher the rotation rate, the longer the MS duration. The effect of rotationally-induced mixing on isochrones can be seen in Fig.\ \ref{rot1}, right panel: at a given age, when rotation is included, the TO mass is higher and the TO sits at higher effective temperature, which affects the
  age-dating.
  
   %----------------------------------------------------------------
  %-------------------------------------------------------------------
   
  \subsubsection{Rotationally-induced mixing versus convective core overshooting}
  \label{rotvsov}
  
\begin{figure}[!hptb]
  \begin{center}
  %\resizebox{0.5\hsize}{!}{
\includegraphics[width=0.75\textwidth]{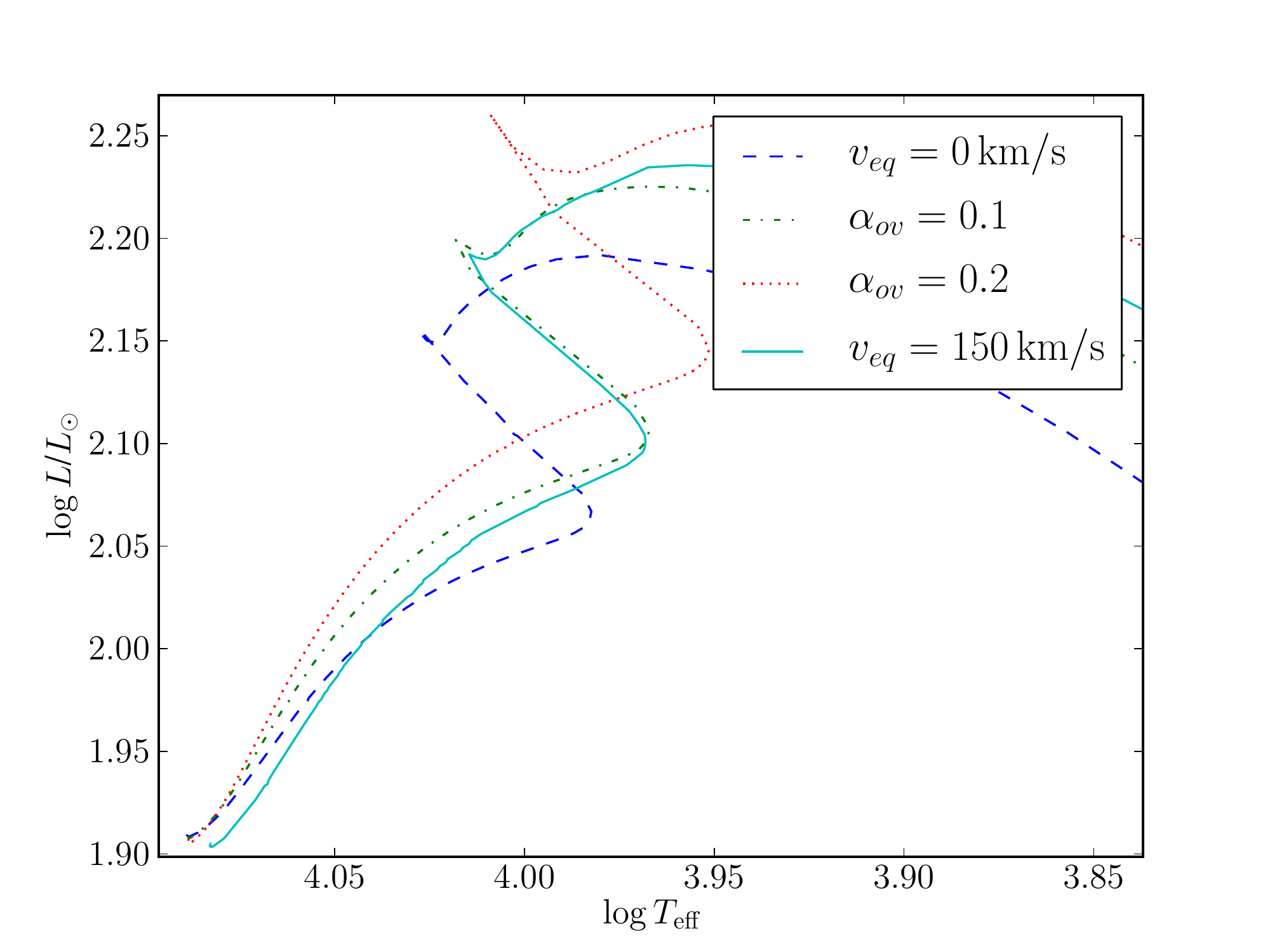}
%}\resizebox{0.5\hsize}{!}{
\includegraphics[width=0.7\textwidth]{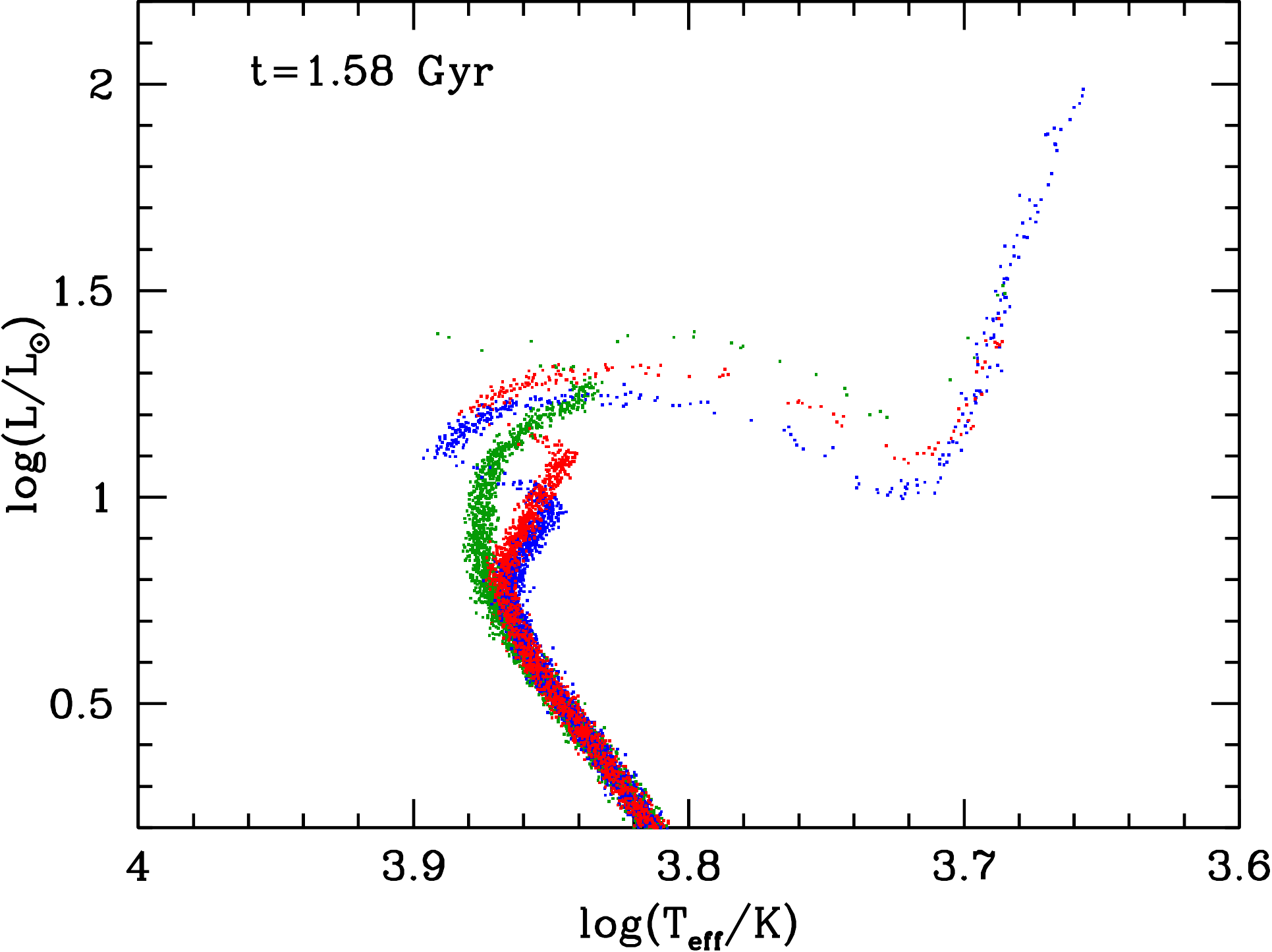}
  \caption{{\sl Top:} evolutionary 
  tracks for a $3 ~  M_\odot$ model 
  including either \textsl{(i)} no rotation, no overshooting (blue dashed line), 
  \textsl{(ii)} shellular rotation with no overshoot (continuous green line), or \textsl{(iii)} no rotation but 
  overshooting with $d_\mathrm{ov}=0.1~ H_P$ (green, small dashed) or $d_\mathrm{ov}=0.2~ H_P$ (red dotted).
  [After \citet{2013A&A...549A..74M}, but see also \citet{2010A&A...519L...2E}.]
  {\sl Bottom:} synthetic HR diagram populated at 1.58~\Gyr~by $10^4 ~ M_\odot$. Stellar masses are distributed according to a chosen IMF and isochrones correspond to either 
 $v_\mathrm{ini} = 0$; $\alpha_\mathrm{ov}=0$ (blue),
  $v_\mathrm{ini} = 150\ \mathrm{km.s^{-1}}$; $\alpha_\mathrm{ov}=0$ (red), and 
  $v_\mathrm{ini}   = 0$; $\alpha_\mathrm{ov}=0.25$ (green). [From  \citet{2011MNRAS.412L.103G}.] 
  }
  \label{rot3}
  \end{center}
  \end{figure}

 Several sets of evolutionary tracks calculated with different options, \ie,  shellular rotation, overshooting or both, can be 
  found in the literature: for instance for a $9\ M_\odot$ model \citep{1997A&A...322..209T}, 
  a  $3\ M_\odot$ model \citep{2010A&A...519L...2E,2013A&A...549A..74M}, 
  or a $1.8\ M_\odot$ model \citep{2002ASPC..259..306G}, but see also the book by \citet{Maeder09}.
 The conclusion of these studies, illustrated in  Fig.\ \ref{rot3}, is that for 
  masses above $1.8 ~M_\odot$, the effect on the MS of rotationally-induced mixing is roughly equivalent to the
  effect of a core overshooting of  $d_\mathrm{ov} \approx 0.1\ H_P$.
  This result is consistent with the result of a 3-D numerical simulation of a rotating
  convective core  for a $2 ~ M_\odot$ A-type star \citep{2004ApJ...601..512B}.

  %----------------------------------------------------------------
  %-------------------------------------------------------------------
 
 However, the confrontation of models to observations indicates that rotational mixing is not sufficient 
 to reproduce the MS width, which makes additional core mixing necessary.
 Therefore stellar models have to include both shellular rotation and core overshooting.
 As discussed by \citet{2012A&A...537A.146E}, the 
 comparison with rotation velocity measurements in young B-stars by \citet{2010ApJ...722..605H} provides a
 prescription for  the initial rotation velocity, that is $v_\mathrm{ini}= 0.4 ~ v_\mathrm{br}$ where
     $v_\mathrm{br}$ is the break-up velocity (velocity corresponding to a balance between gravitational and centrifugal accelerations).
 Moreover, \citet{2012A&A...537A.146E} proposed a prescription for the amount 
 of overshooting necessary to match the observed MS-width:
  \begin{itemize}
  \item no overshoot for $M\leq 1.25\ M_\odot$, 
  \item $d_\mathrm{ov}=0.05\ H_P$ for $M < 1.7\ M_\odot$
  \item $d_\mathrm{ov}=0.10\ H_P$ for $M > 1.7\ M_\odot$.
  \end{itemize}
 Therefore, in presence of rotationally-induced mixing with $v_\mathrm{ini}= 0.4 ~ v_\mathrm{br}$ the necessary amount of core overshooting on the MS ($d_\mathrm{ov}\lessapprox 0.10\ H_P$) is smaller than in the case without rotation 
 ($d_\mathrm{ov}\approx 0.15-0.20\ H_P$, see Sect.~\ref{overshooting}), 
 while the theoretical and observed MS-widths better agree  when rotationally-induced mixing is accounted for. 
 Note also that  in order to reproduce the effects of rotationally-induced mixing both on the MS and post-MS 
 evolutionary tracks, the value of the overshooting parameter must vary 
  in models accounting for overshooting only  \citep{2010A&A...509A..72E}. 
   
  A similar comparison has been carried out for 1.58~\Gyr~isochrones by \citet{2011MNRAS.412L.103G}:
  a synthetic HR diagram was populated by $10^4$ stars distributed according to an IMF and
  isochrones assuming either 
  $v_\mathrm{ini} = 0$ and $\alpha_\mathrm{ov}=0$,  $v_\mathrm{ini}   = 0$ and $\alpha_\mathrm{ov}=0.25$, or
 $v_\mathrm{ini} = 150\ \mathrm{km.s^{-1}}$ and $\alpha_\mathrm{ov}=0$.
  The impact of the rotation and/or overshooting is visible on the shape of the
  isochrone at the turn-off.
  
   \subsubsection{Impact on TO ages}
  
  %----------------------------------------------------------------
  \begin{figure}[!htb]
  \begin{center}
  \resizebox{0.95\hsize}{!}{\includegraphics{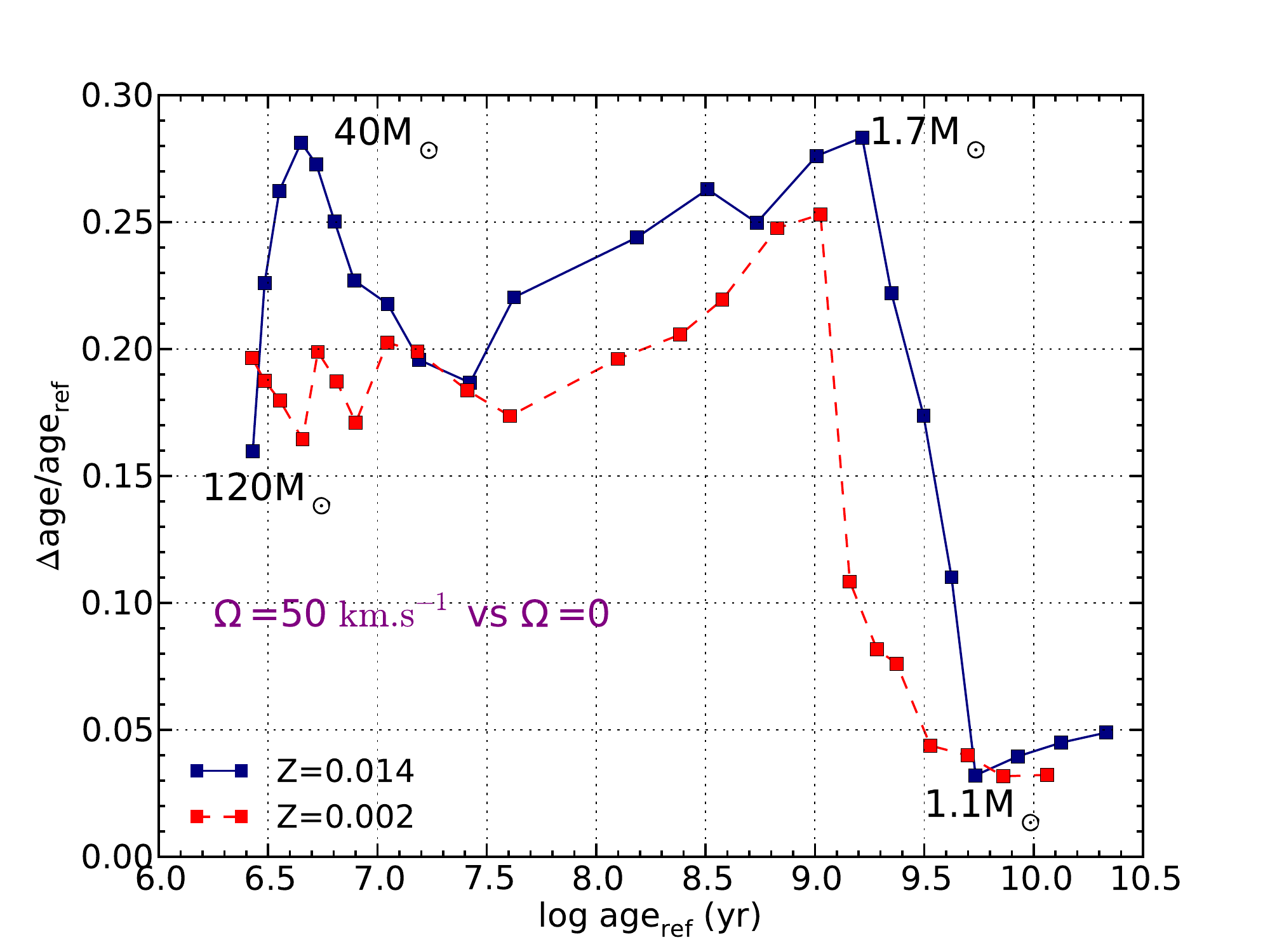}}
   \caption{Relative age differences at TO between models including 
 rotationally induced mixing and models without rotation. Differences calculated
 from the data provided in the Geneva model grids 
  \citep{2012A&A...537A.146E}. 
  }
  \label{rot5}
  \end{center}
  \end{figure}
  %-------------------------------------------------------------------
  
  To estimate how rotationally induced mixing affects stellar ages, 
  we used the data provided by the  
 Geneva team \citep{2012A&A...537A.146E} to compare the ages at TO of models including 
  shellular rotation with an initial rotational velocity of 
  $v_\mathrm{ini}=110~\mathrm{km.s^{-1}}$ 
 with the ages of models without rotation. In both cases, the Geneva grids include 
 overshooting according to the prescription given in Sect.~\ref{rotvsov}. 
 The results are shown in Fig.\ \ref{rot5}. The TO ages 
 differ by up to 30 (25) per cent for $Z=0.014\ (0.002)$. 
 For $M \gtrapprox 1.7\ M_\odot$, the MS-lifetime increase remains 
  the same for any mass, \ie,  scales as $v_\mathrm{ini}/v_\mathrm{br}$. 
 The impact is smaller (a few per cent) for stellar masses below 
  $1.3\ M_\odot $. The large increase of the age difference
 between $1.3\ M_\odot$ and $1.7\  M_\odot$  is attributed to the fact that the
  convective core does not appear at same age for a given mass whether
  rotation is included or not. 
 We obtained similar results using the \textsl{STAREVOL} grids \citep{2012A&A...543A.108L}.
 
Finally, let us add that MS stars of mass  $\gtrapprox 2.0\ M_\odot$ use to rotate fast.
Fast rotation may change the aspect ($T_\mathrm{eff}$ and $L$) 
of the star, depending on rotational velocity and inclination angle \citep[see \eg,][]{1999A&A...346..586P}.
In turn, the shape of the isochrones can be modified (independently of rotational mixing), which may further affect age-dating 
\citep[see \eg~Fig.~13 in][]{2001A&A...374..540L}.
%################################################################# 
 \section{Conclusion}
\label{conclusions}
%\input conclusion.tex

% ----------------------------------------------------------------
\begin{figure}[!htb]
\begin{center}
\resizebox{0.85\hsize}{!}{\includegraphics{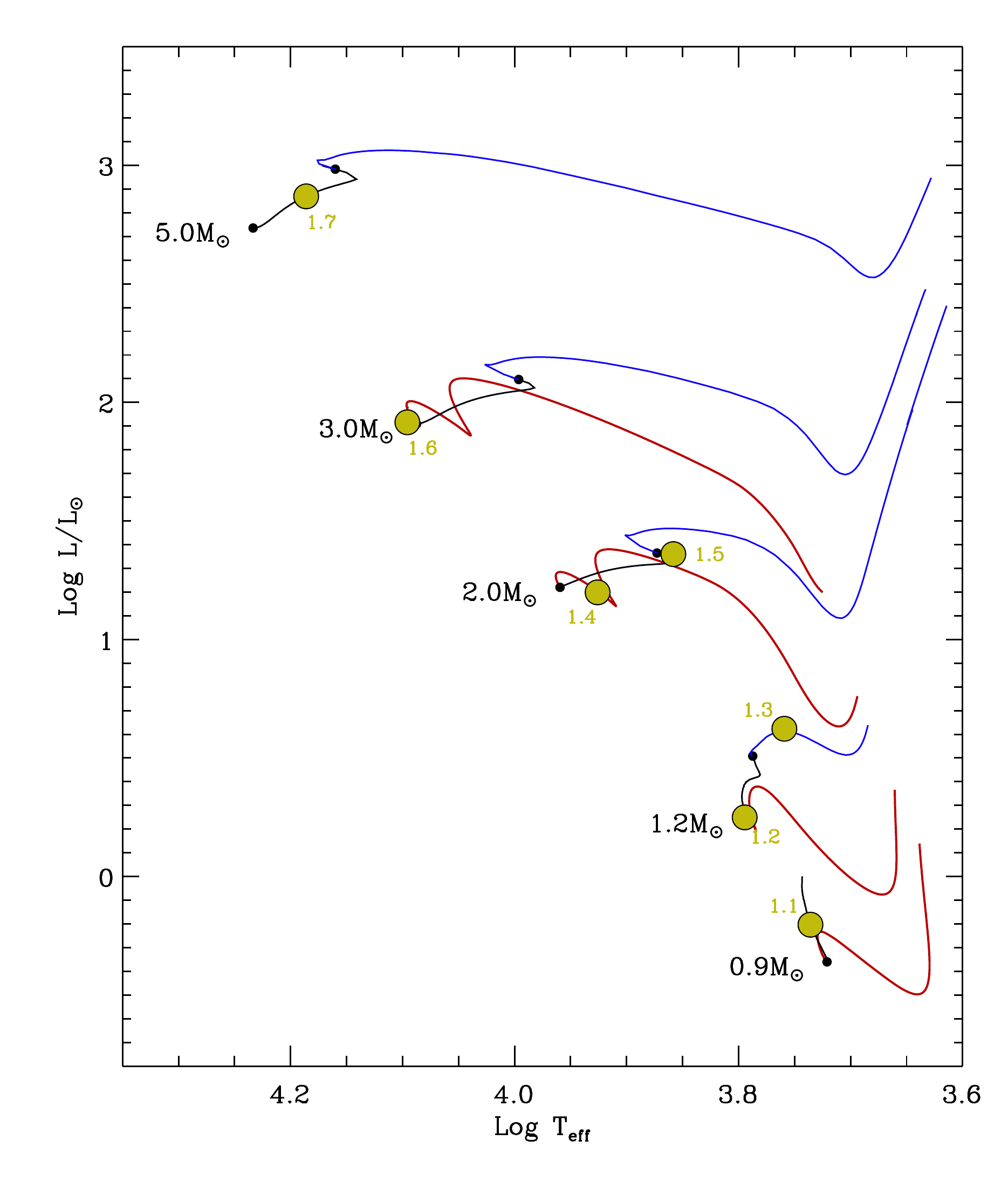}}
\caption{ESTA-CoRoT comparisons. Evolutionary tracks in a HR diagram. 
Filled circles indicate the location of the stellar models (targets) chosen for the comparison.
Red lines correspond to the PMS, black lines to
the MS, and blue lines to the post MS evolution.
%The percentages indicate the larger relative age differences between the
%models from different evolutionary codes and the reference \texttt{cesam2k} model.
[From \citet{2006ESASP1306..363M}.]}
\label{esta1}
\end{center}
\end{figure}
%-------------------------------------------------------------------

% ----------------------------------------------------------------
\begin{figure}[!htb]
\begin{center}
\resizebox{0.5\hsize}{!}{\includegraphics{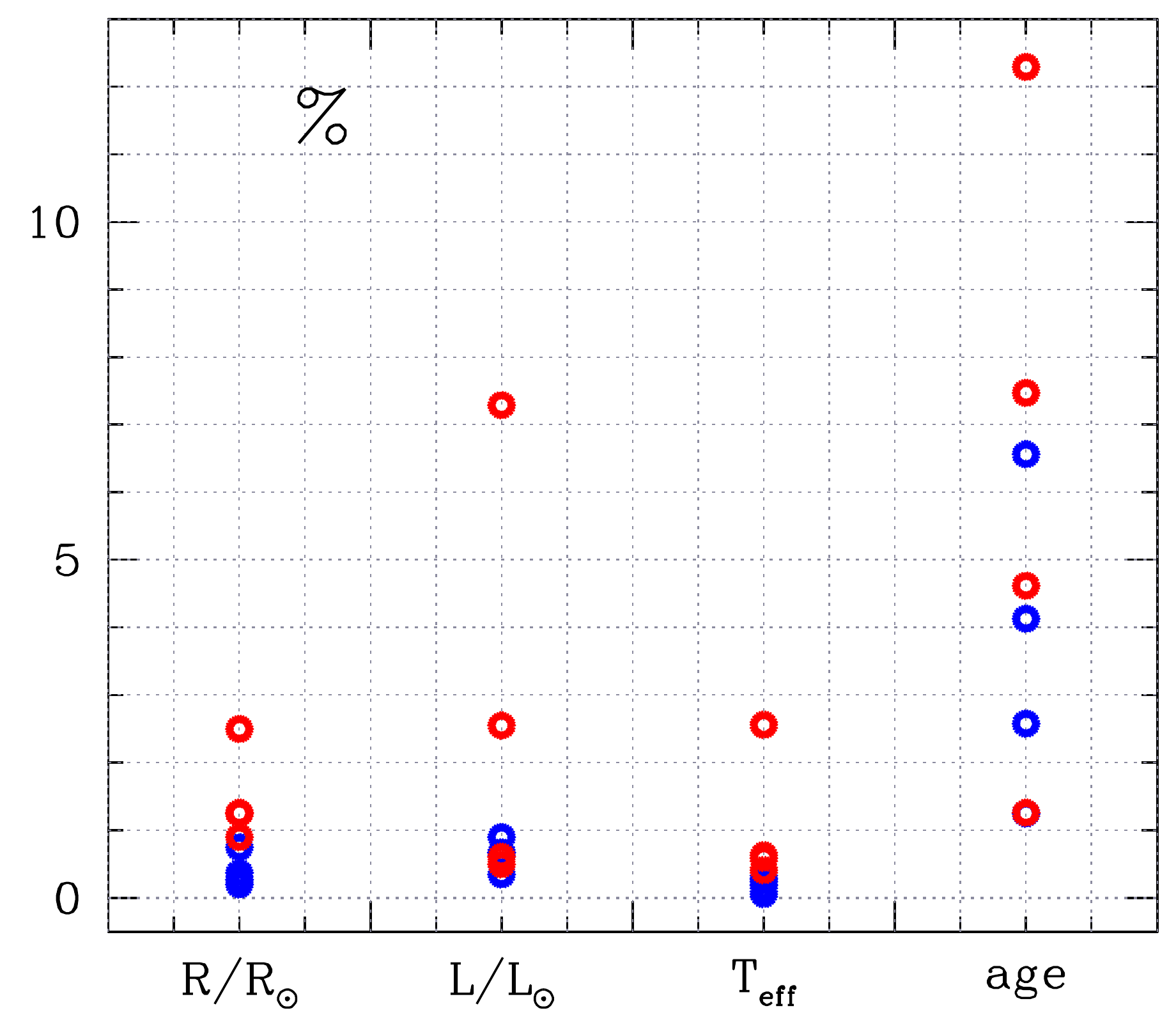}}\resizebox{0.5\hsize}{!}{\includegraphics{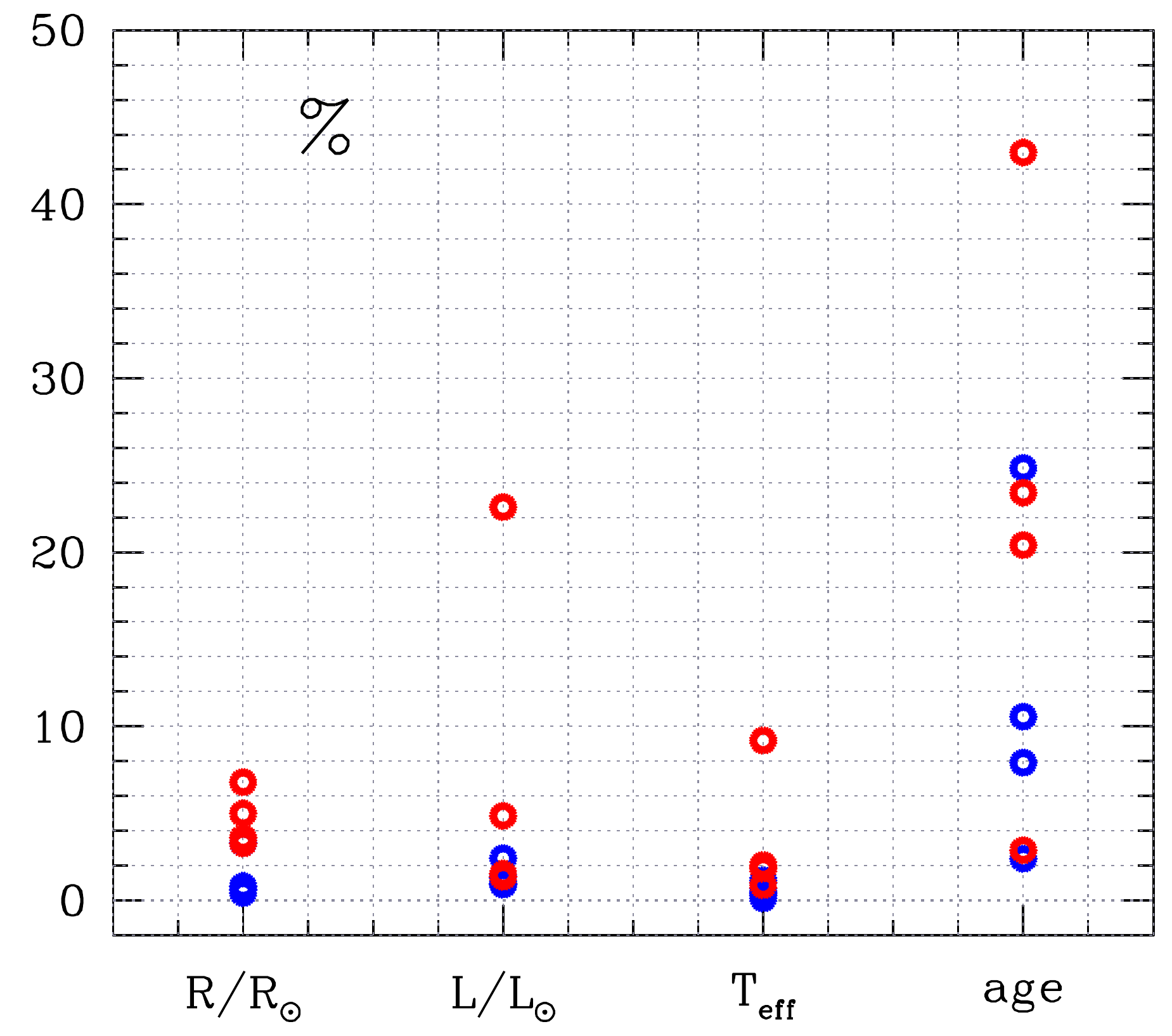}}
\caption{ESTA-CoRoT comparisons. {\sl Left:} mean differences in the classical parameters obtained in models
calculated with different codes and \texttt{cesam2k} models. We distinguish in blue symbols the results of 
codes that have strictly
followed the prescription of the model calculation from the others in red  (see text).
{\sl Right:} same comparison showing the maximum differences obtained for each code.
[From  \citet{2010Ap&SS.328...29L}.]}
\label{esta2}
\end{center}
\end{figure}
%-------------------------------------------------------------------
% ----------------------------------------------------------------
\begin{figure}[!htpb]
\begin{center}
\resizebox{0.995\hsize}{!}{ \rotatebox{270}{\includegraphics{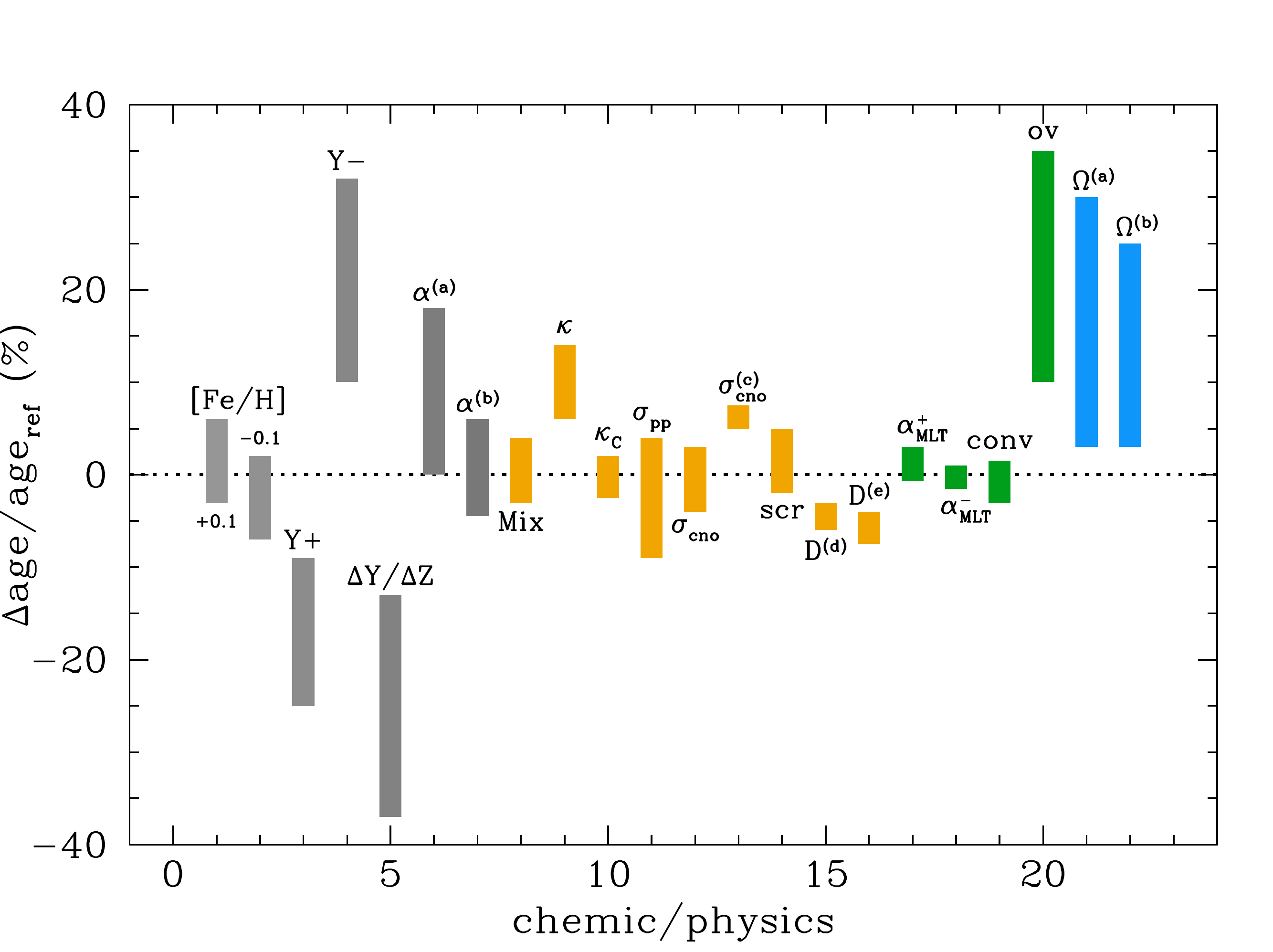}}}
%\resizebox{0.47\hsize}{!}{\includegraphics{Figs/fig_synth_2.eps}}
\caption{
%\orange{\sl Left:} 
Synthesis of the ranges of relative age differences at TO, as obtained when changing one of the inputs
of the  reference  model (defined in Sect.~\ref{grids}). 
From left to right, the case labels on the abscissae correspond to the following changes: 
[1, 2] [Fe/H] abundance by $\pm 0.1$ dex with respect to solar,
[3, 4]  initial helium abundance by $\pm 0.03$ with respect to solar,
[5] $\Delta Y/\Delta Z$ by $+3$ with respect to solar,
[6,7]  $\alpha$-elements enhancement of $+0.4$ dex at [Fe/H]=0.0 (a) and $-$1.0 dex (b),
[8] solar mixture (\textsl{AGSS09} \textit{vs} \textsl{GN93} mixture),
[9] opacity (increased by $10$ per cent),
[10] conductive opacity, \citet{1975ApJ...196..525I} \textit{vs} \citet{2007ApJ...661.1094C} formalism,
[11] $\sigma_{pp}$ reaction rate (decreased by $15$ per cent), 
[12, 13] $\sigma_\mathrm{CNO}$ (\textsl{LUNA} \textit{vs} \textsl{NACRE} rate for 
the $^{14}\mathrm{N} (p, \gamma) ^{15}\mathrm{O}$ rate) at [Fe/H]$=0.0$ dex (a) and $-2.0$ dex (c),
[14] screening factor in nuclear reaction rates (no screening \textit{vs} screening),
[15, 16]  atomic diffusion for (d) diffusion \textit{vs} no diffusion and (e) 
no diffusion \textit{vs} diffusion with diffusion velocities increased by $20$ per cent,
[17, 18] $\alpha_\mathrm{MLT}$ value by $\pm 0.20$ dex with respect to solar,
[19]  prescription for convection (MLT \textit{vs} FST), 
[20]  convective core overshooting ($\alpha_\mathrm{ov}=0.20$ \textit{vs} no overshooting),
[21, 22] rotation ($\Omega=50\ \mathrm{km \, s^{-1}}$ \textit{vs} no rotation), at [Fe/H]=0.0 dex (a) and $\sim -1.0$ dex (b).
}
\label{synthesis}
\end{center}
\end{figure}
%-------------------------------------------------------------------

The most appropriate method to obtain accurate stellar ages relies on the 
computation of stellar models. However these models are far from being perfect,
they are affected by several sources of uncertainties.
Some of them are well-known, some are difficult to identify. 
It is then not easy to provide ages with realistic error bars.  
With the nowadays, high-quality, ground and space data, efforts are made  to identify and 
quantify the  most important  biases affecting stellar age-dating. The ultimate goal is to eliminate them. 

% ----------------------------------------------------------------
\begin{figure}[!ht]
\begin{center}
\resizebox{0.80\hsize}{!}{\includegraphics{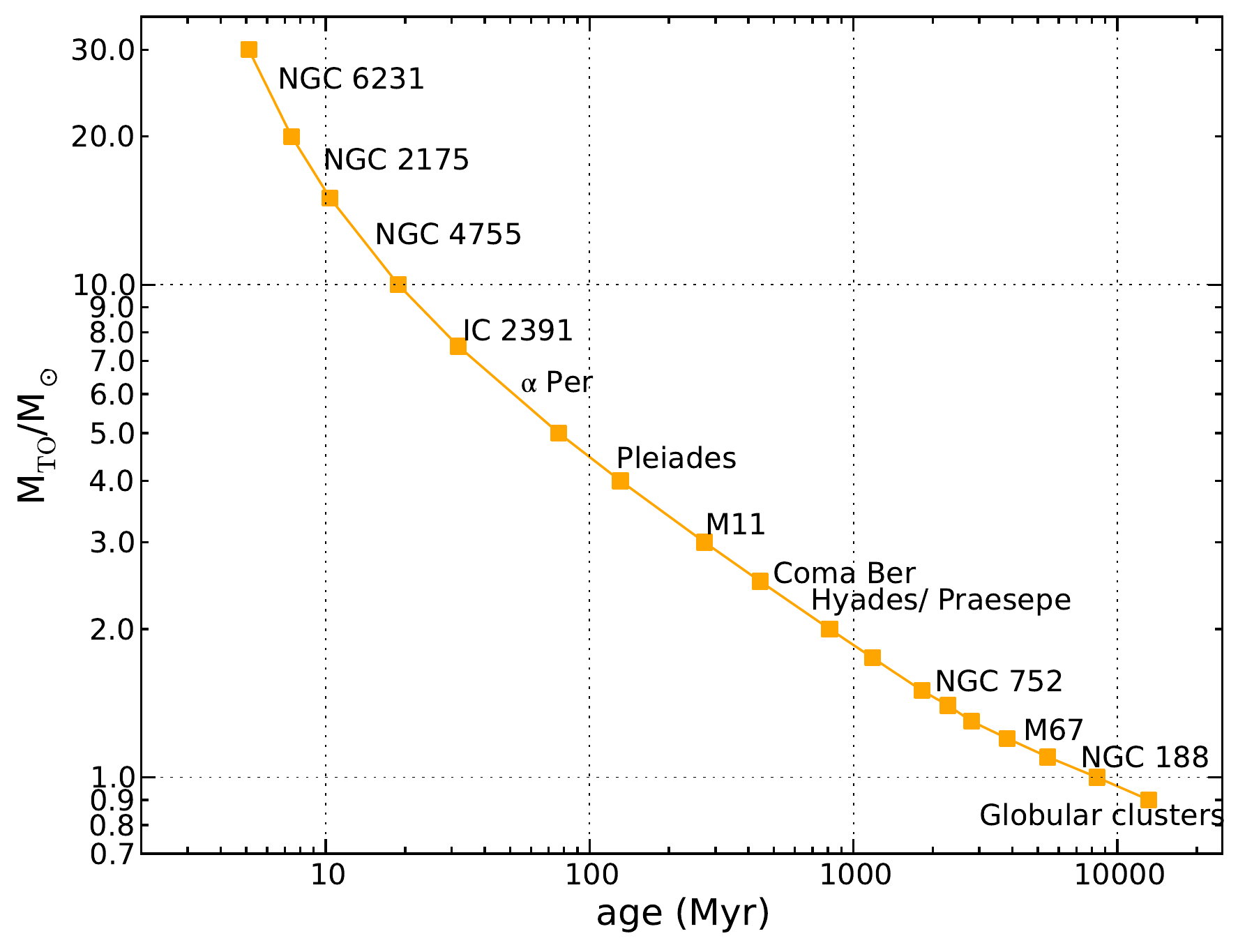}}
\caption{
Value of the mass at turn-off as a function of age for
our reference models (solar [Fe/H]). The correspondence with observed stellar clusters
is indicated.
}
\label{synthTO}
\end{center}
\end{figure}
%-------------------------------------------------------------------
% ----------------------------------------------------------------
\begin{figure}[!hb]
 \begin{center}
 \resizebox{0.95\hsize}{!}{ \rotatebox{270}{\includegraphics{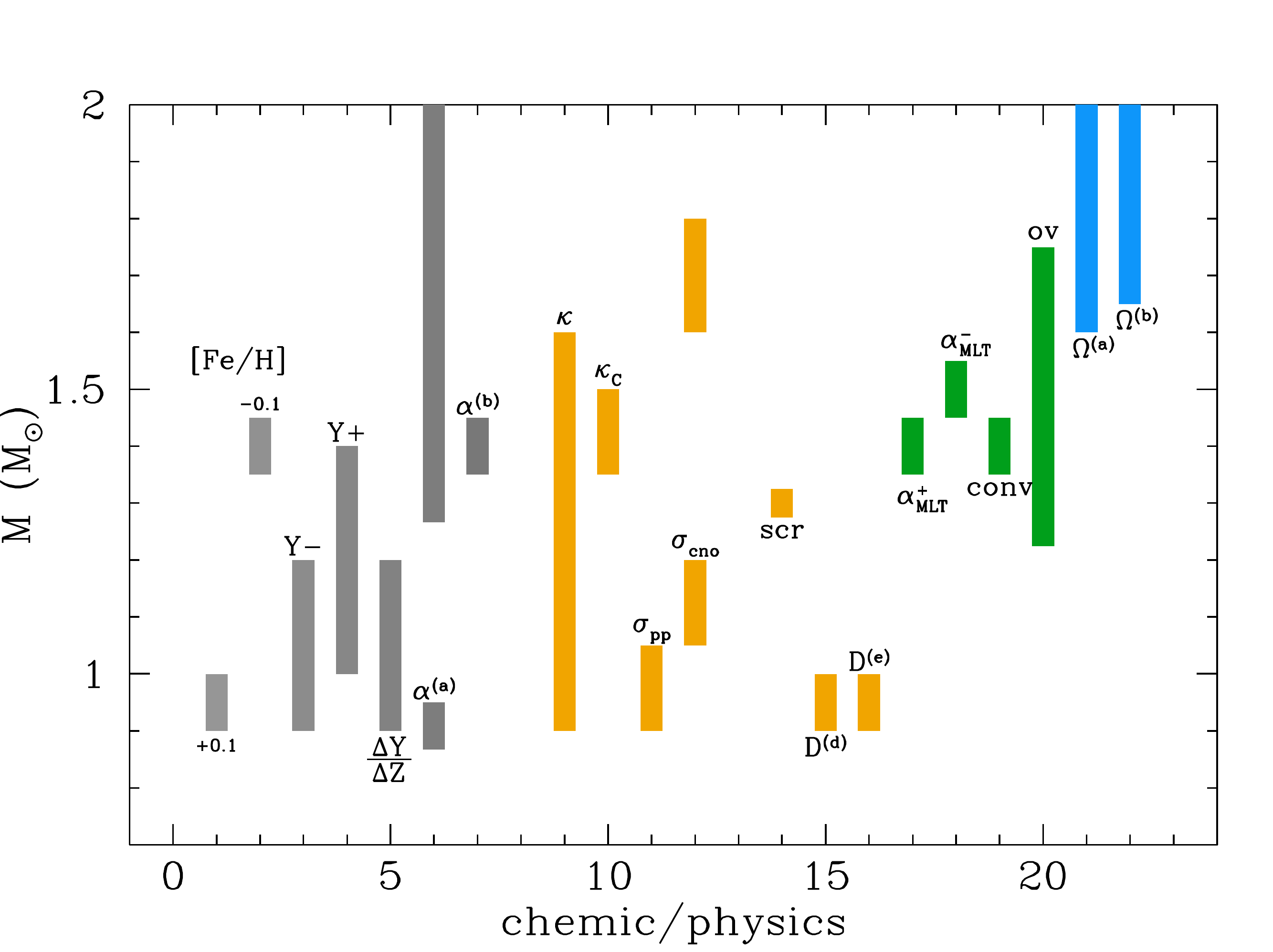}}}
 \caption{
 The range of mass for which the age is affected by the different uncertainties in the model inputs considered here
 (see Fig.~\ref{synthesis}). The effect of the chosen solar mixture is not reported because it affects the whole range of possible stellar
 masses.
 }
 \label{synthmass}
 \end{center}
 \end{figure}
 %-------------------------------------------------------------------

  %% ----------------------------------------------------------------
\begin{figure}[!htb]
 \begin{center}
 \resizebox{0.5\hsize}{!}{\includegraphics{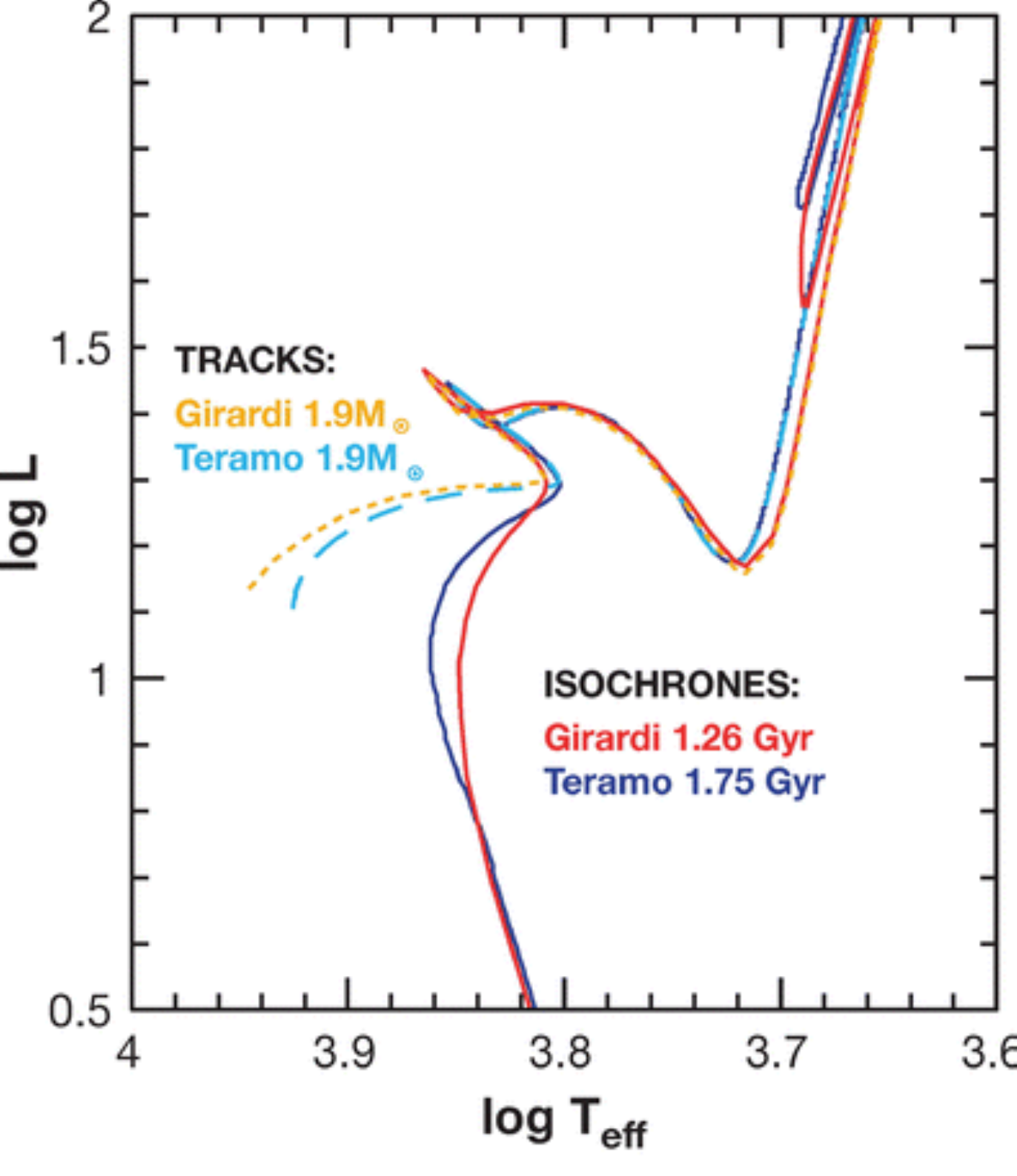}}\resizebox{0.5\hsize}{!}{\includegraphics{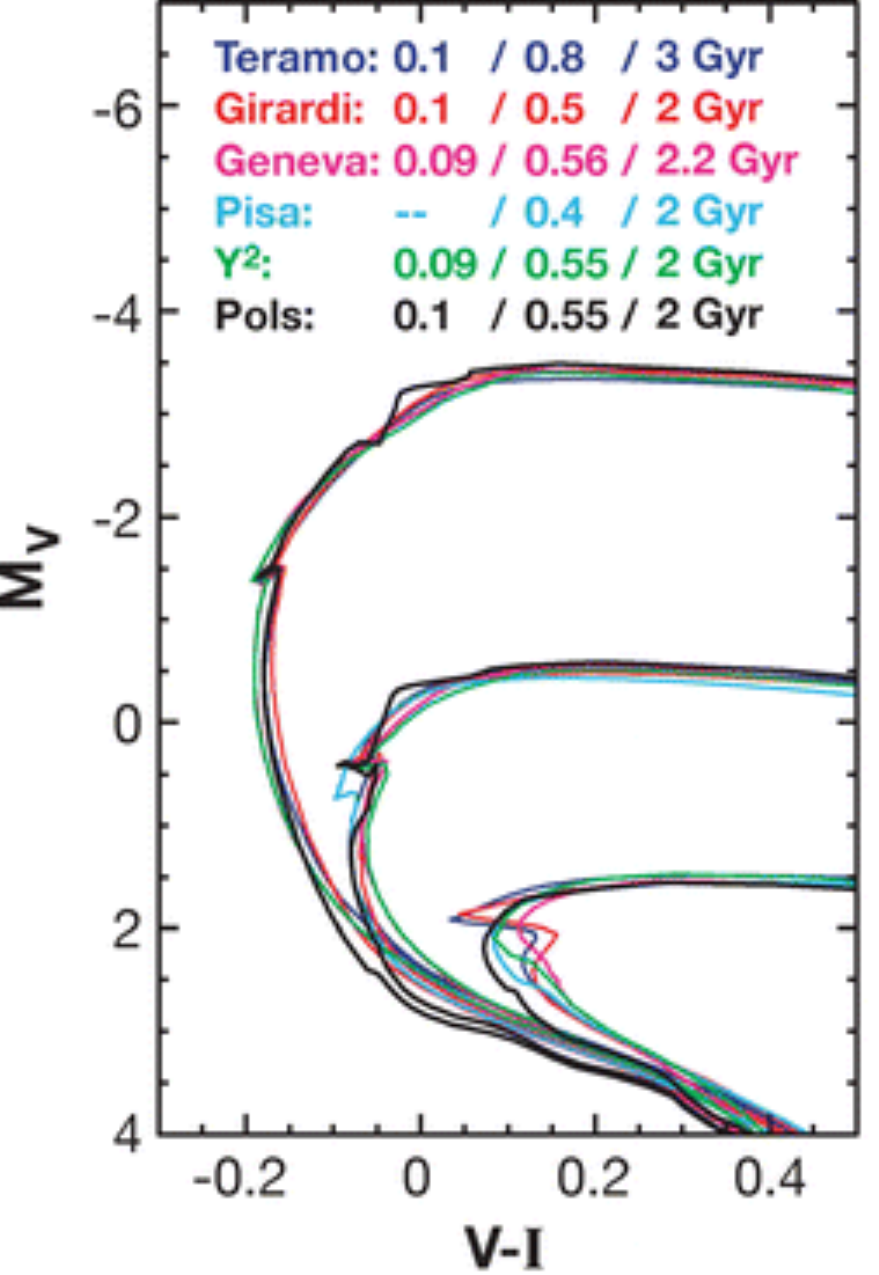}}
 %\resizebox{!}{0.5\vsize}{\includegraphics{Figs/Gallart1.png}}\resizebox{!}{0.5\vsize}{\includegraphics{Figs/Gallart2.png}}
 \caption{{\sl Left: } 
 tracks of $1.9 M_\odot$ models from two sets of grids (see text) plotted in the HR diagram together with 
 related isochrones providing the same TO location.
  {\sl Right:} three groups of isochrones
  providing the same TO and SGB position (same metallicity  $Z=0.001$).
  The  upper curves correspond to ages in the range $\sim  0.09-0.1$~\Gyr,
 the middle curves to ages in the range $\sim 0.4-0.8$~\Gyr, and the lower curves 
 correspond to ages in the range $\sim 2-3$~\Gyr. [Both figures from \citet{2005ARA&A..43..387G}.]
 }
 \label{conc5}
 \end{center}
 \end{figure}
 %-------------------------------------------------------------------

In that respect, in the framework of the scientific preparation of the CoRoT mission, 
a European working group, ESTA-CoRoT, was in charge of carrying out in-depth comparisons of stellar models
\citep[see \eg,][]{2006ESASP1306..363M,2008Ap&SS.316....1L,2008ap&ss.316..187l,2008Ap&SS.316..219M}.  
The group has compared models calculated with ten different stellar evolutionary codes. 
Prescriptions had been given to have, as far as possible, the same input physics, physical constants, and astronomical constants 
in all codes. The model comparisons were made for several cases corresponding to different choices of the mass, 
initial chemical composition, and evolutionary stage \citep[for more details see][]{2006ESASP1306..363M,2008Ap&SS.316....1L}.
These cases were chosen to be representative of the CoRoT seismic targets.
The position in the HR diagram of one of the sets of models that have been compared is shown in Fig.~\ref{esta1}.
In Fig.~\ref{esta2} we show the comparisons of the radius, luminosity, effective temperature, and age
of the models calculated by each participating code with the results obtained with the \texttt{cesam2k} code \citep{2008ap&ss.316...61m}. 
The left panel shows the mean relative differences in these quantities, while the right panel displays the
maximum differences. Concerning the age, the mean differences are in the range $1-12$ per cent, 
while the maximum differences
are in the range $2-43$ per cent, which is quite high. 
However, if we only consider the codes that strictly followed the prescriptions for the
comparison, that is adopting exactly the prescribed input physics and constants (blue symbols in Fig.~\ref{esta2}), 
the differences are reduced by a factor of about two. In this case, we can consider that 
the differences between the code results are only due to differences in numerical treatments, that is  
the handling of table interpolation (to get the opacity, EoS outputs, etc.),
the methods used to solve the equations, and hidden 
numerical mistreatments  
(time steps and mesh points, convective boundaries, etc.). The thorny problem of the numerical determination
of convective boundaries has been addressed recently by \citet{2014arXiv1405.0128G}.
More generally, numerical treatments are discussed in detail in \eg, \citet[][]{2008ap&ss.316..187l,2008Ap&SS.316..219M,2010Ap&SS.328...29L}, and references therein.

In the  present lecture, our approach has been to use the same evolutionary 
code (\texttt{cesam2k}) to estimate
the impact of different physical inputs on the TO age value. 
The results are synthesized in Fig.~\ref{synthesis}, which highlights the huge impact of, on the one hand, the 
uncertain value of the initial helium abundance, and, on the other hand, the uncertain amount of mixing induced by overshooting  
and rotation, which determines the quantity of fuel available on the MS.

To better visualize the domain of mass and age concerned by model uncertainties, we show in
Fig.~\ref{synthTO} the TO mass of observed stellar clusters of different ages, while in Fig.~\ref{synthmass}
 we indicate the domain of mass associated to the different uncertainties in the model inputs.
We stress that the  model uncertainties do not add up linearly,
 but correspond to the variation of one process at a time. Monte Carlo techniques can be used to
estimate the cumulative effect  of all sources of uncertainties
\citep[\eg,][]{1992ApJ...388..372C,2014A&A...561A.125V}.

 Actually, different input physics are used in different stellar evolution codes. 
 As a result, different stellar ages are inferred from different grids of models. 
 \citet[][]{2005ARA&A..43..387G} compared evolution tracks of a $1.9\ M_\odot$ star in the HR diagram provided by
 the  BaSTI group \citep{2004apj...612..168p} and by the Padova group \citep{2002a&a...391..195g} and found that
 the same TO position is reached by isochrones of ages differing by 30 per cent (see Fig.~\ref{conc5}, left panel).
 This quite significant difference 
  is attributed to  several differences in the model inputs, in particular overshooting
  \citep{2004apj...612..168p}.
  \citet[][]{2005ARA&A..43..387G} also compared the ages of metal-deficient stellar populations that would be 
  predicted by different sets of model isochrones for three given choices of the pair (TO location, subgiant position).
  As shown in Fig.~\ref{conc5} {right panel}, the differences in age are more important for older stars: 
  at ages of ${\sim}0.1$~\Gyr~the predicted ages differ by ${\sim}10$ per cent, 
  while for older stars (ages of $0.4-0.8$ or $2-3$~\Gyr) the ages differ by $50-100$ per cent.
The differences are attributed to different model input physics 
 (atomic diffusion, overshooting, microscopic physics).
 
 To summarize, the present lecture showed that the physical description of stellar models  
 must still be significantly improved in
 order to provide accurate ages of MS stars at a precision level better than 20 per cent. 
 We could not discuss here all the processes which can affect the age determination.
 While the impact of numerics, input chemical composition, and microphysics can be estimated 
 rather easily, the impact of macrophysics is much more difficult to estimate because its description
 involves processes which are not well described, imply many parameters, 
 and are sometimes even unidentified.
 Furthermore, the present lecture focused on MS stars, therefore the sources of uncertainties
 on the time elapsed during advanced stages of stellar evolution have not been evaluated.
  
 Accurate ages at the level of 20 per cent or less for low mass MS stars are required in many
 astrophysical fields like the formation and evolution of planetary systems or the evolution of the Galaxy. 
 This need will be even more crucial when high-quality space data provided by, \eg, the 
 PLATO-ESA mission \citep{2013arxiv1310.0696r} and the Gaia-ESA mission \citep{2001a&a...369..339p}  will be available.
 Improving the physical description of stellar models is then certainly worth the effort.  
 The observational progress will have to be supported by theoretical and experimental developments, and by
 numerical simulations in 2-D and 3-D. For instance, recent progress came from the 
  the building of patched 1-D stellar models which include the convective outer layers obtained from 3-D simulations
  \citep{2012A&A...543A.120S}, and from 2-D rotating models such as those calculated with the \texttt{ROTORC} code \citep{1995ApJ...439..357D} and the 
  \texttt{ESTER} code  \citep{2013sf2a.conf..101R}.
%################################################################# 
\section*{Acknowledgements}
The authors would like to thank the ``Formation permanente du CNRS'' for
 financial support.  The preparation and writing of these lectures 
largely benefited from the use of the SIMBAD database, operated at
 CDS, Strasbourg, France and of the NASA's Astrophysics Data System. 
Elements of these lectures have been presented in 2007 at the ESA 
ELSA School on the Science of Gaia in Leiden, The Netherlands, by 
Y. Lebreton (unpublished), in 2008, at the International Young
 Astronomer School on CoRoT Astrophysics in Meudon, France, by  
Lebreton \& Montalb{\'a}n (unpublished), and, in 2009, at the 
EES 2009 CNRS School in Aussois, France (Lebreton \& Montalb{\'a}n, 
unpublished).

%\bibliographystyle{aaees2013}
%\bibliography{lebreton1}{}

%\bibliographystyle{aa}
%\bibliography{yl}

\end{document}